\documentclass[trackchanges,twocolumn]{aastex7}
\usepackage{graphicx}
\usepackage{gensymb}
\usepackage{longtable}
\usepackage{amsmath}

\def\O3{[\ion{O}{3}]}

\begin{document}

\title{Optical Discovery of New and Candidate Galactic Supernova Remnants \\ Plus Optical Imaging of the Large Monogem Supernova Remnant}

\author[0000-0003-3829-2056]{Robert A.\ Fesen}
\email{robert.fesen@dartmouth.edu}
\affil{6127 Wilder Lab, Department of Physics and Astronomy, Dartmouth
       College, Hanover, NH 03755 USA}

\author[0000-0002-7855-3292]{Marcel Drechsler}
\email{epost@marcel-drechsler.de}
\affiliation{\'Equipe StDr, B{\"a}renstein, Feldstraße 17, 09471 B{\"a}renstein, Germany}

\author{Bray Falls}
\email{brayfalls@gmail.com}
\affiliation{Sierra Remote Observatories, 42120 Bald Mountain Road, Auberry, CA, 93602, USA}
\affiliation{Starfront Observatories, 1724 Co Rd 244, Rockwood, TX 76873, USA}

\author[0009-0009-7869-0762]{Tim Schaeffer}
\email{tim.schaeffer30@gmail.com}
\affil{The Deep Sky Collective, Luxembourg }

\author[0009-0008-5838-8147]{Justin Anderson}
\email{anjustin@umich.edu}
\affil{MDM Observatory, Kitt Peak National Observatory, 950 N. Cherry Ave., Tucson, AZ 85719, USA}

\author[]{Deepanshu Arora}
\email{arora.deepanshu@gmail.com}
\affil{The Deep Sky Collective, Luxembourg }

\author[0000-0003-1173-6964]{Werner Becker}
\email{web@mpe.mpg.de}
\affil{Max-Planck-Institut für extraterrestrische Physik, Giessenbachstraße, 85748 Garching, Germany}
\affil{Max-Planck-Institut für Radioastronomie, Auf dem Hügel 69, 53121 Bonn, Germany}

\author[0009-0002-8962-0806]{Carl Bj\"ork}
\email{carl.bjork@gmail.com}
\affil{The Deep Sky Collective, Luxembourg }
\affil{Elveteek Sàrl, Lausanne, Switzerland}

\author[0009-0007-4704-1466]{Steeve Body}
\email{sbody@collarts,edu.au}
\affil{The Deep Sky Collective, Luxembourg }
\affil{Griffith University, Brisbane, Queensland, Australia}


\author[]{Laurent Huet}
\email{l-huet@wanadoo.fr}
\affil{Pond Observatory, 28120 Illiers-Combray, France}

\author[]{Tarun Kottary}
\email{tkottary@gmail.com}
\affil{The Deep Sky Collective, Luxembourg }

\author[]{Mathew Ludgate}
\email{matludgate@me.com}
\affil{Ross Creek Observatory, Dunedin 9010, New Zealand}

\author{Nicolas Martino}
\email{n.martino@hotmail.fr}
\affiliation{Various Amateur Observatory Sites, Lorraine, France}

\author[0000-0001-5324-359X]{Manuel C.\ Peitsch}
\email{manuelcp.astro@gmail.com}
\affiliation{Roboscopes at e-EyE Entre Encinas y Estrellas Observatory, 
Camino de los Molinos 06340 Fregenal de la Sierra, Spain}

\author[]{Nicolas Puig}
\email{deepskycollective@gmail.com}
\affil{The Deep Sky Collective, Luxembourg }

\author[0000-0002-5313-6409]{Wolfgang Reich}
\email{wreich@mpifr-bonn.mpg.de}
\affil{Max-Planck-Institut  f\"ur Radioastronomie,
Auf dem Hugel 69, 53121 Boon, Germany}

\author[0000-0001-5164-3602]{J.\ Rodrigues} 
\email{contact@joserodrigues.space}
\affil{ Instituto de Astrofísica e Ciencias do Espaço, Universidade do
     Porto, CAUP, Rua das Estrelas, PT4150-762 Porto, Portugal}
\affil{Departamento de Fisica e Astronomia, Faculdade de Ciencias, Universidade do Porto, Rua Campo Alegre, 4169-007 Porto, Portugal}
\affil{Observatoire de Genève, Département d’Astronomie, Université de Genève, Chemin Pegasi 51b, 1290 Versoix, Switzerland}

\author{Yann Sainty}
\email{yann.sainty@gmail.com}
\affiliation{YSTY Astronomy, 54000 Nancy, Lorraine, France}

\author[]{Patrick Sparkman}
\email{psparkman@gmail.com}
\affil{The Deep Sky Collective, Luxembourg }

\author[0000-0003-2692-2321]{Sean Walker}
\email{astrowalker535@yahoo.com}
\affil{MDW Sky Survey, New Mexico Skies Observatory, Mayhill, NM 88339, USA}



\begin{abstract}

\noindent
We report the discovery of new and candidate Galactic supernova remnants made through over 2000 hours of H$\alpha$ and [\ion{O}{3}] imaging using amateur class instrumentation.  These nebulae 
range from 0.5 to over 8 degrees in size. Five of these, namely
G27.8-17.1, G115.6+9.2, G190.5+25.3, G195.9+2.2 and G239.9+7.0 exhibit low-dispersion  optical spectra indicating shock emission by the presence of strong [\ion{S}{2}] $\lambda\lambda$6716,6731 and  [\ion{O}{1}] 
$\lambda$6300 line emissions relative to H$\alpha$, or exhibit a nonradiative shock generated Balmer dominated spectrum. Their filamentary emission morphology also supports the presence of interstellar shock fronts. Three other optical nebulae,
G191.3+11.1, G205.7-1.7, and G305.4-0.7 appear likely to be remnants and are proposed as new SNR candidates. 
We also present \O3 and H$\alpha$ images of optical filaments 
associated with the 25 degree diameter Monogem Ring SNR which reveal extensive emission-line filaments around most of the remnant's X-ray boundaries.
Deep images of the X-ray suspected remnant G190.4+12.5 and the large and \O3 emission dominated SNR G206.6+6.1 seen toward or near the Monogem SNR are also included.
Key findings of this work include: 1)  optical surveys can sometimes detect SNRs better than radio and X-ray SNR searches, 2) Balmer nonradiative shocks appear not uncommon in high Galactic latitude SNRs, and 3) faint \O3 emission dominated SNRs may have been missed in prior H$\alpha$ emission only surveys.

\end{abstract}

\bigskip


\section{Introduction}


Although some of the most famous Galactic supernova remnants (SNRs) like the Crab Nebula, Kepler's and Tycho's SNRs, and the Cygnus Loop (aka the Veil Nebula) were first discovered optically,
most of the currently $\sim$300 known Galactic SNRs were discovered through radio observations due to the characteristic synchrotron nonthermal emission associated with shocked gas. 
The process of high-velocity interstellar shocks leads to a power law of radio flux density, S, with S $\varpropto$  $\nu^{-\alpha}$ where $\alpha$ is the emission spectral index with typical values between  
$-0.3$ and $-0.7$ \citep{Reynolds2011,Dubner2015}. The discovery of several new or suspected SNRs via radio observations are reported every few years, with the total number of confirmed or suspected SNRs now numbering nearly 700
\citep{Green2025}.   

The majority of Galactic SNRs are less than a degree in angular size, more than 1 kpc distant, and well evolved with estimated ages between 10$^{4}$ and 10$^{5}$ yr. 
Some 45 remnants or roughly 15\% of the radio or optically confirmed Galactic SNRs have angular dimensions greater than one degree, with more than a dozen larger than 2 degrees. 

Only about 100 or roughly 30\% of the currently cataloged Galactic remnants 
exhibit any appreciable associated optical emission \citep{Green2025}. 
The ratio of [\ion{S}{2}] $\lambda\lambda$6716,6731  vs.\ H$\alpha$ flux above $\sim$0.4 has proven to be an especially  useful criteria in identifying shocked emission for both the Milky Way and nearby local group galaxies SNRs \citep{MC1973,Dordico1978,Dordico1980,Fesen1985,Long2017,Kop2021}.

Recent discoveries of new Galactic SNRs showing optical emission have largely been made first through deep H$\alpha$ surveys of the Galactic plane
(\citealt{Mav2001,Mav2005,Mav2009,Boumis2002,Boumis2009,Sezer2012, How2018, Fesen2019,Fesen2020,Bakis2023}).
Such surveys include the Virginia Tech Spectral-line Survey (VTSS; \citealt{Dennison1998}), the Southern H$\alpha$ Sky Survey Atlas (SHASSA; \citealt{Gaustad2001}),
the Wisconsin H$\alpha$ Mapper (WHAM; \citealt{Haffner2003}),
the AAO/UKST SuperCOSMOS H$\alpha$ survey \citep{Parker2005, Stupar2008}, and the
Issac Newton Telescope Photometric H$\alpha$ Survey of the Northern Galactic Plane 
(IPHAS; \citealt{Drew2005}).

\begin{deluxetable}{lccc}[ht]
\tablecolumns{4}
\tablecaption{Basic Data on the Program SNRs}
\tablewidth{0pt}
\tablehead{ 
\colhead{SNR} & \multicolumn{2}{c}{\underline {Approx. Center (J2000) } }& \colhead{Angular Size}  \\
\colhead{ID}  & \colhead{RA} & \colhead{Dec}  & \colhead {(degrees)}  } 
\startdata
 \underline{New SNRs} &      &          &                     \\   
G027.8-17.1    & ~ 19:44 & $-12$:10   & $5.0 \times 4.0$  \\ 
G115.6+9.2     & ~ 23:21 & +70:45   & $1.1 \times 0.9$   \\
G190.5$-25.3$  & ~ 04:40 & +06:26   & $8.0 \times 9.0$  \\
G195.9+2.2     &  ~ 06:27 & +16:10   & $1.0 \times 1.0$ \\
G239.9+7.0     &  ~ 08:09 & $-20$:11 & $2.1 \times 2.1$  \\
 \underline{Candidate SNRs} &      &          &                     \\
G190.4+12.5    &  ~06:57 & +25:34 &     $6.1 \times 6.1$ \\
G191.4+11.1    & ~ 06:53 & +24:12   & $3.2 \times 2.2$  \\ 
G205.7-1.7     &  ~ 06:32 & +05:42   & $1.4 \times 1.4$ \\
G305.4-0.7     &  ~ 13:14 & $-63$:29 &  $0.5 \times 0.5$ \\
\underline{Known SNRs} &     &          &                   \\
G203 +12\tablenotemark{a}       &  ~ 06:59 & +13:56  & $25 \times 25$  \\
G206.6+6.1     & ~ 07:02 & +08:23   & $3.6 \times 3.6$  \\ 
\enddata
\tablenotetext{a}{The Monogem Ring SNR}
\end{deluxetable}

These surveys have led to both the discovery of additional optical emissions from already known SNRs (e.g., \citealt{Walker2001, Stupar2008,Stupar2011}) and the discovery of a few new remnants 
\citep{Stupar2007, Stupar2012, Stupar2018, Sabin2013, Fesen2015, Fesen2021}.




Recently, narrow passband emission-line sky surveys conducted by amateurs 
using relatively small optical telescopes has led to several additional Galactic SNR discoveries.
Developments in optical CMOS image detectors plus 
affordable high transmission (T $\geq95$ \%) narrow passband 
filters (FWHM $\approx$ 30 \AA) has created a revolution in deep emission-line imaging
of Galactic nebulae by amateur astrophotographers (see
\citealt{Fesen2024} and references therein). 

Not restricted by telescope allocation committees or outside funding, amateurs taking hundreds of exposures with remote and robotically operated telescopes equipped
with high-throughput emission-line filters and sensitive
digital detectors are capable of detecting extremely faint and previously unknown  Galactic emission line nebulae. 
Moreover, unlike previous emission sky surveys, amateurs have employed a variety of nebular emission line filters besides H$\alpha$, most importantly \O3 $\lambda$5007 and [\ion{S}{2}] $\lambda\lambda$6716,6731, and have surveyed regions both along and far from the Galactic plane \citep{Ziegenbalg2025}.

Here we present discovery optical images plus follow-up low-dispersion spectra of five previously unknown Galactic nebula which appear to be true SNRs.
All were recently discovered through deep emission-line images taken by individual or groups of amateur astronomers who obtained,
processed, and assembled hundreds of hours of exposures
using small to modest size telescopes to reveal these new SNRs. 

The outline of the paper is as follows: 
Section 2 describes the imaging data 
along with the limited optical spectra taken of the new SNRs, with
the imaging and spectral analysis
described in Section 3.
Sections 4, 5, and 6 present our data on confirmed SNRs, 
candidate SNRs, and two already known SNRs including the Monogem Ring SNR,
respectively.
Finally, in Section 7 we briefly
discuss some interesting properties of these new remnants, 
the comparison of new SNR searches in the optical vs radio, 
along with some key findings and conclusions of this imaging program.

\section{Observations}

\subsection{Amateur Instrumentation and Exposure Details}

The wide field emission-line images presented below are the work of over a dozen amateurs taking hundreds of short exposures using small or modest telescopes and reducing these data using a variety of shared reduction software.
These images are the result of focused imaging campaigns often spanning many months.
Taken together, the total exposure time
spent on these five new, three candidate, and two known SNRs  
exceeds 2150 hours which is roughly
three months of open shutter time.

A table of the observers, observing sites, optical equipment used, imaging field-of-view (FOV) and detector pixel size, various filters used,
and total exposure times for each of the SNR imaged in this study can be found in Appendix A. This is not a full list of all data taken on each object but is meant to give an overview of the instruments and exposures used in the development of the  SNR images presented here. 

\subsection{MDM Optical Spectra  and Imaging}

Follow-up narrow passband emission line imaging and low-dispersion optical spectra of these remnants were obtained with the MDM 2.4m Hiltner telescope at Kitt Peak using the Ohio State Multi-Object Spectrograph (OSMOS; \citealt{Martini2011}) during a series of observing runs starting in late 2023 and ending in May 2026. 
The OSMOS instrument employs 
a $4096 \times 4096$ CCD for both imaging and spectra.  In imaging mode, this telescope/camera system yields a clear FOV of $18' \times 18'$.  On-chip $2 \times 2$  pixel binning gave a spatial resolution scale of $0.55''$ per pixel. Deep H$\alpha$ images were of relatively small ($\sim15' \times 15'$) regions of a few of the new SNRs reported here were obtained using a filter with a FWHM bandpass of 80 \AA\ and exposures of between 600 and 2000 s. Data reduction included flat field, bias corrections and cosmic ray removal using 
L.A. Cosmic software \citep{vanDokkum01}.
 
Low-dispersion spectra were taken employing a blue VPH grism (R = 1600) with $1.2''$ or $3.0''$ wide slits and exposure times
of 900 to 3000 s covering 4500--6950 \AA \ with a spectral resolution $\simeq$ 2 \AA \ pixel$^{-1}$ and a FWHM = 3.8 \AA. Spectra were extracted from regions along the $15'$ long slits.  Data were taken under mainly photometric conditions but with highly variable seeing (FWHM = $1'' - 3''$).

Spectra were reduced using using standard IRAF software reduction procedures and OSMOS reduction pipelines in PYRAF\footnote{PYRAF is a product of the Space Telescope Science Institute, which is operated by AURA for NASA.}. L.A. Cosmic \citep{vanDokkum01} was used to remove cosmic rays. Spectral fluxes were calibrated using Ne, Hg, and  Ar lamps and bright spectroscopic standard stars \citep{Oke1974A,Massey1990} and are believed accurate to
$\pm10\%$ for strong lines and up to $\pm25\%$ for weak lines.


\begin{deluxetable*}{llcccccccccccr}[htp]
\tiny
\centerwidetable
\tablecolumns{14}
\tablecaption{Observed Relative Emission Line Strengths }
\tablewidth{0pt}
\tablehead{\colhead{SNR} & \colhead{Slit} & \colhead{H$\beta$ } & \colhead{[\ion{O}{3}]} & \colhead{[\ion{O}{1}]} & 
           \colhead{H$\alpha$} & \colhead{[\ion{N}{2}] } & \colhead{[\ion{S}{2}]} & \colhead{[\ion{S}{2}]} & 
           \colhead{[S II]}  & \colhead{[\ion{S}{2}]/H$\alpha$} & \colhead{H$\alpha$/H$\beta$}  & \colhead{$E(B - V$)} &  \colhead{H$\alpha$} \\
 \colhead{ID}& \colhead{Position} & \colhead{4861} & \colhead{5007} & \colhead{6300} & \colhead{6563} & 
\colhead{6583} & \colhead{6716} & \colhead{6731} &  \colhead{6716/6731} & \colhead{} & \colhead{} & \colhead{} &
\colhead{Flux\tablenotemark{a}}  }                                                                
\startdata
 G27.8-17.1 & NE       &  (9)     & \nodata & \nodata & 100 & \nodata & \nodata  & \nodata &  \nodata & \nodata  & \nodata &  \nodata   & 5.5 ~ \\
            & SW       &  (20)    & \nodata &  (10)   & 100 & 133     & 117      & 67      &  1.75    &  1.84    & \nodata &  (0.24)   & 3.6 ~ \\
 G115.6+9.2 & North P1 &  \nodata & \nodata & \nodata & 100 & \nodata & \nodata  & \nodata &  \nodata & \nodata  & \nodata & \nodata & 1.6 ~ \\ 
            & North P2 &  \nodata & (12)    & \nodata & 100 & (5)     & (7)      & (4)     &  \nodata & \nodata  & \nodata & \nodata & 4.3 ~ \\
            & South P1 &  \nodata & \nodata &  (12)   & 100 &  20     & 51       & 41      &  1.24    &    0.92  & \nodata & \nodata & 7.3 ~ \\   
            & South P2 &  \nodata & \nodata & \nodata & 100 & \nodata & \nodata  & \nodata & \nodata  & \nodata  & \nodata & \nodata & 2.7 ~ \\ 
 G190.5-25.3& East     &  32      & \nodata & 24      & 100 & 68      & 70       & 47      & 1.48     &    1.17  & 3.13    & 0.09    & 14.3 ~ \\
            & Center   &  32      & \nodata & (13)    & 100 & 30      & 42       & 27      & 1.55     &    0.69  & 3.12    & 0.09    & 11.5 ~ \\
            & SE 1A    &  34      & \nodata & 22      & 100 & 38      & 54       & 40      & 1.35     &    1.04  & 2.94    & 0.04    & 28.3 ~ \\
            & SE 1B    &  30      & \nodata & 21      & 100 & 72      & 71       & 48      & 1.49     &    1.19  & 3.12    & 0.09    & 48.6 ~ \\
            & SE 2     &  30      & \nodata & 13      & 100 & 52      & 70       & 46      & 1.52     &    1.16  & 3.33    & 0.15    & 12.7 ~ \\
            & West     &  36      & \nodata & 22      & 100 & 44      & 60       & 41      & 1.46     &    1.01  & 2.78    & 0.00    & 60.6 ~ \\
            & NW       &  32      & \nodata & 26      & 100 & 19      & 38       & 25      & 1.52     &    0.63  & 2.94    & 0.04    & 18.7 ~ \\
G195.9+2.2  & P1 North & \nodata  & \nodata & 76      & 100 & 37      & 31       & 28      & 1.01     &    0.59  & \nodata & \nodata & 2.9 ~ \\  
            & P1 South & \nodata  &   152   & 46      & 100 & 59      & 29       & 34      & 0.85     &    0.63  & \nodata & \nodata & 1.8 ~ \\
            & P2       & \nodata  & \nodata & (31)    & 100 & \nodata & 32       & 28      & 1.14     &    0.60  & \nodata & \nodata & 2.3 ~ \\
G206.6+6.1  & P1       & \nodata  & \nodata & 41      & 100 & 78      & 44       & 33      & 1.30     &    0.70  & \nodata & \nodata & 3.9 ~ \\
            & P2       & 31       & 200     & 40      & 100 & 103     & 76       & 52      & 1.46     &    1.28  & 3.22    & 0.12    & 4.1 ~ \\
G239.9+7.0  & P1       & \nodata  &   130   & (28)    & 100 & 73      & 78       & 54      & 1.44     &    1.28  & \nodata & \nodata &  1.9 ~ \\
            & P2       & \nodata  & \nodata & \nodata & 100 & (27)    & 44       & 23      & 1.91     &    0.67  & \nodata & \nodata &  2.3 ~ \\
            & P4       & \nodata  &   40    & (29)      & 100 & (27)    & 41       & 34      & 1.21     &    0.75  & \nodata & \nodata & 2.1 ~ \\
\enddata
\tablenotetext{a}{Values are in units of 10$^{-16}$ erg cm$^{-2}$ s$^{-1}$. }
\tablecomments{Listed line strengths are relative to H$\alpha$ = 100 and are uncorrected for reddening.}
\label{Table_2}
\end{deluxetable*} 

\section{Data Description and Analysis}

Below we present deep optical imaging and low-dispersion spectra on five newly identified and spectroscopically confirmed Galactic SNRs, plus   emission-line images of four optical SNR candidates.
In addition, we present and discuss optical images of two already known remnants, namely the large and \O3 bright remnant G206.7+6.1, and the huge 
Monogem Ring SNR which significantly
clarify both their fine scale and overall structure and optical emission properties.

In eight cases of optically confirmed or suspected SNRs discussed below, there are no known or suspected coincident SNRs listed in published X-ray or radio SNR catalogs 
(\citep{Safi-Harb2019, Green2025}. Nor are there any obvious radio emissions in either GLEAM 170 MHz or EB 21 cm data
(see Appendix B for radio maps and details). 
Nonetheless, five of these  with optical spectra are clear cases of new and
previously unrecognized Galactic SNRs, while the three candidate objects seem worthy of multi-wavelength follow-up.

To establish that five of these new optical nebulae are truly SNRs, we have employed standard shock criteria, namely emission line ratio diagnostics
where [\ion{S}{2}]/H$\alpha$ $\geq$ 0.4 is an indicator of
shock emission. This [\ion{S}{2}]/H$\alpha$ ratio criteria has been used successfully for several decades in both Galactic and extragalactic SNR surveys 
\citep{Mathewson1972,Mathewson1973, Fesen1985, Long1990, Smith1993, Kop2020}.

The physics behind this is that in photoionized nebula such as H~II regions and planetary nebulae (PNe), sulfur
exists mainly in the form of S$^{+2}$, yielding weak [\ion{S}{2}] emissions relative to that of H$\alpha$. In contrast, radiative shocks have an extended postshock cooling region which generates abundant low ionization species resulting in strong
[\ion{S}{2}] $\lambda\lambda$6716,6731 emissions. The presence of appreciable [\ion{O}{1}] 
$\lambda\lambda$6300,6364 in the spectrum can often provide additional
confirmation for the presence of shock since these emission lines are also not strong in photoionized nebulae
\citep{Fesen1985,Kop2020}.

Morphological evidence for identifying an optical nebulosity as a SNR is the detection of sharp, thin filaments which can be seen in narrow passband images. Although this is not as strong a SNR identification marker as emission-line ratios, relatively high resolution optical images can aid in distinguishing SNRs from some other types of nebulae.

Because all remnant or remnant candidates plus the Monogem SNR discussed below are relatively large in angular size and often show considerable fine-scale structure, we present many of these images in large formats so that a reader can better appreciate each remnant's morphology and fine detail. In addition, in many cases we also present 
versions where the stars have been digitally removed via software to enhance the visibility of faint nebular structures.

However, we note that some caution needs to be exercised when examining the starless images presented here. It is not uncommon that in the process of digitally removing stars from the raw images, some very narrow emission-line filaments are sometimes shown significantly decreased in brightness or even almost removed entirely from the final starless image. Consequently, one should be careful in using the starless images presented below and treat the version with stars as the being ground truth as to relative filament brightness. 

\begin{figure*}[ht!]
\begin{center}
\includegraphics[angle=0,width=14.0cm]{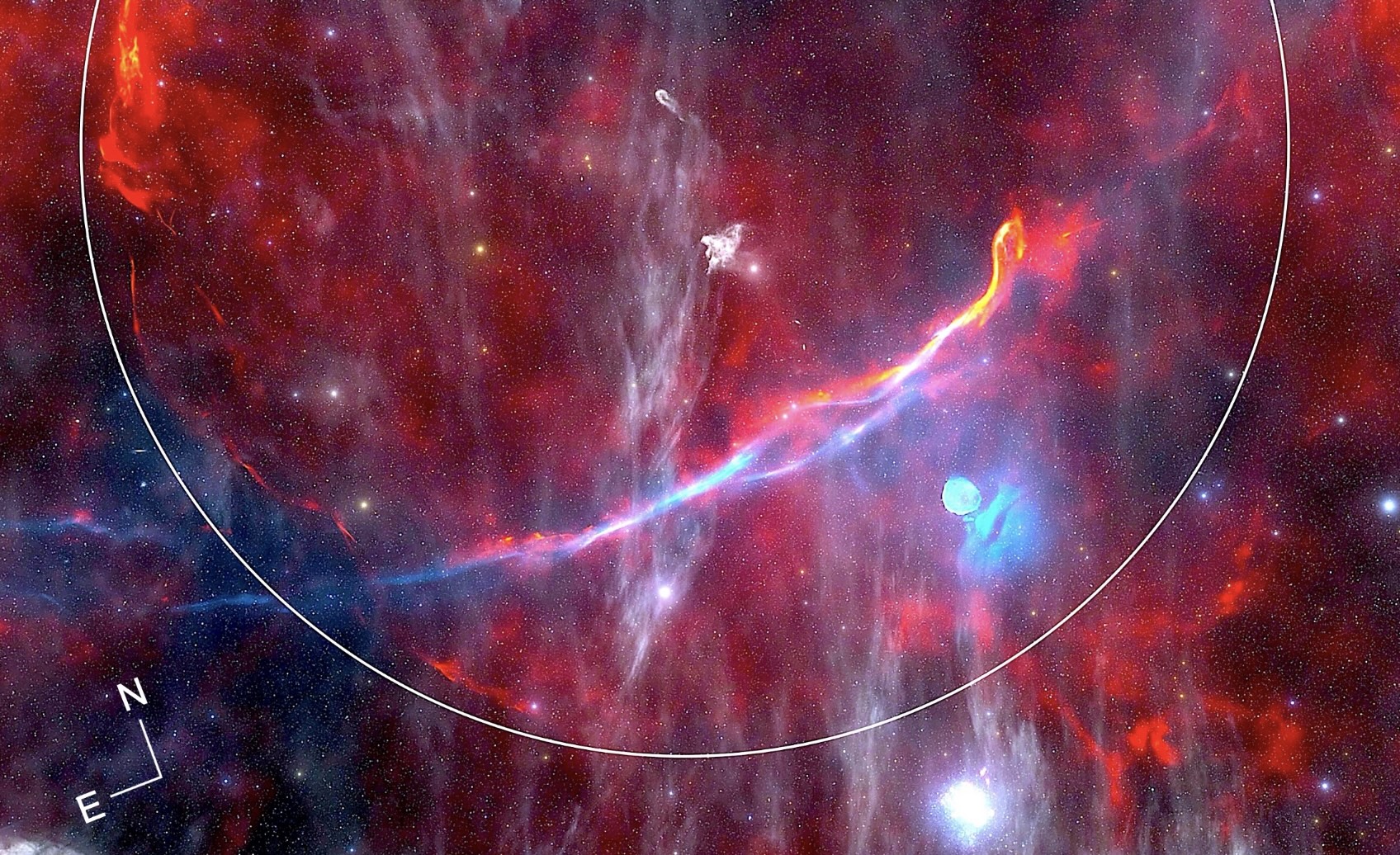} 
\caption{Color composite of  H$\alpha$ (red), [\ion{O}{3}] (blue), and R,G,B continuum (white) images of the southeastern
section of the G27.8-17.1 remnant. White circle indicates the remnant's approximate 5$\degr$ diameter and center. Besides the bright SE-NW filament, note the curved line of eastern H$\alpha$ emission features extending from the southern edge of image to the top left. 
\label{G27_1} 
} 
\end{center}
\end{figure*}

\begin{figure*}[ht]
\begin{center}
\includegraphics[angle=0,width=17.0cm]{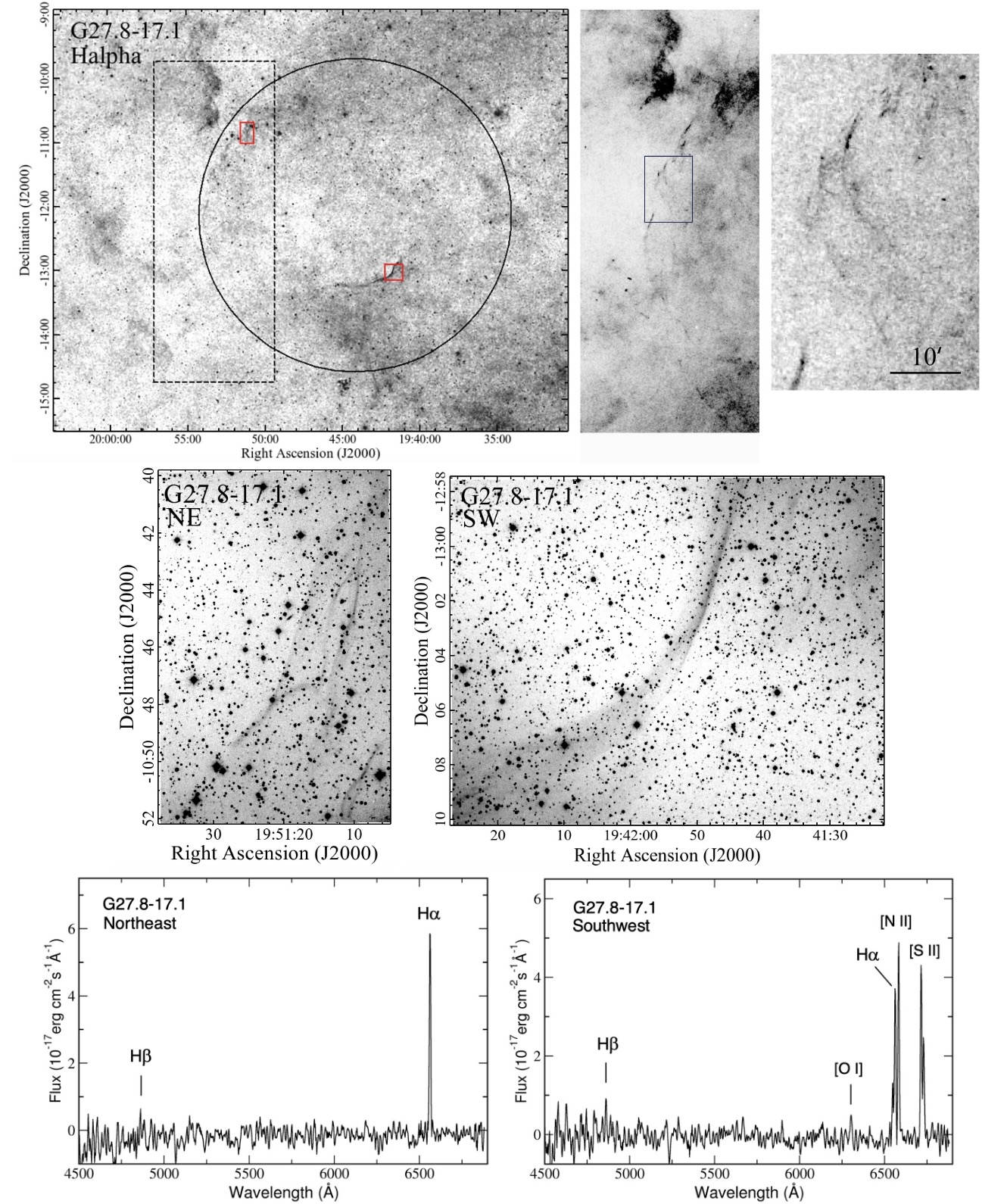} 
\caption{
Top: NSNS H$\alpha$ image of the G27.8-17.1 region. Circle is 5$\degr$ in diameter. Red boxes indicate regions where higher resolution H$\alpha$ images were taken, while dashed rectangle is area of our 
H$\alpha$ image shown in adjacent middle panel. Right panel shows enlarged section of the eastern filaments showing overlapping thin filaments.
Middle: Higher resolution MDM images of NE and SW regions (red boxes).
Bottom: Low-dispersion spectra of G27.9-17.6 filaments in the NE and SW.
\label{G27_2} 
} 
\end{center}
\end{figure*} 

\section{New SNRs}

\subsection{G27.8-17.1}

In 2023, M.\ Drechsler noticed a faint filament in on-line H$\alpha$ sky survey images 
in a region in Sagittarius located some 17 degrees off the Galactic plane.
The filament's westernmost portion is faintly visible on the DSS2 red image but quite obvious in the H$\alpha$ images in the Northern Sky Narrowband Survey (NSNS; \citealt{Ziegenbalg2025}).
Deep follow-up H$\alpha$ and \O3  images of this region obtained by B.\ Falls showed the filament to be nearly $3\degr$ long and bright in \O3 in certain sections (see Fig.~\ref{G27_1}).
These images also revealed a long curved line of very faint H$\alpha$ emission patches and filaments located further to the east from the original filament which extended more than 3$\degr$ in north and south directions.

The top panel of Figure~\ref{G27_2} shows what we designate as the G27.8-17.1 remnant on the NSNS's H$\alpha$ image. While the black circle is meant to mark the remnant's approximate 5 degree diameter size and center, 
the remnant's currently known emission consists of just the bright E-W filament plus a long curved, broken line of eastern emission filaments (see top right panels of Fig.~\ref{G27_2}). Given the limited imaging data currently in hand, both the remnant's size and center are uncertain. Consequently, the remnant's G27.8-17.1 identification name is subject to some uncertainty in both Galactic latitude and longitude. 

There is no sign of a remnant in the radio in this region (see Appendix B).
Despite this, evidence for the existence of a previously unrecognized Galactic remnant  here is compelling. First, the detection of filamentary emission in both the object's southwest and northeast is consistent for the presence of ISM shocks.

More importantly, the spectral properties exhibited by these filaments clearly indicate interstellar shocks.
Low-dispersion spectra of the 
bright SW filament showed very strong [\ion{S}{2}] $\lambda\lambda$6716,6731 line emission relative to H$\alpha$, a classic signature of shocks (see right hand spectrum in the bottom panels of Fig.~\ref{G27_2}). 

The complete absence of \O3 in the more western portion of this filament compared to the strong \O3 emission seen in its eastern section indicates a range of velocities of around 80 km s$^{-1}$ to over 100 km s$^{-1}$ indicating a gradient of electron densities in this filament of at least a factor of around two.
The very weak [\ion{O}{1}] $\lambda$6300 emission in the filament section observed is a bit surprising but does not negate a shock emission identification here given the emission's filamentary morphology and the presence of strong \O3 elsewhere along the filament.

\begin{figure*}[ht!]
\begin{center}
\includegraphics[angle=0,width=17.7cm]{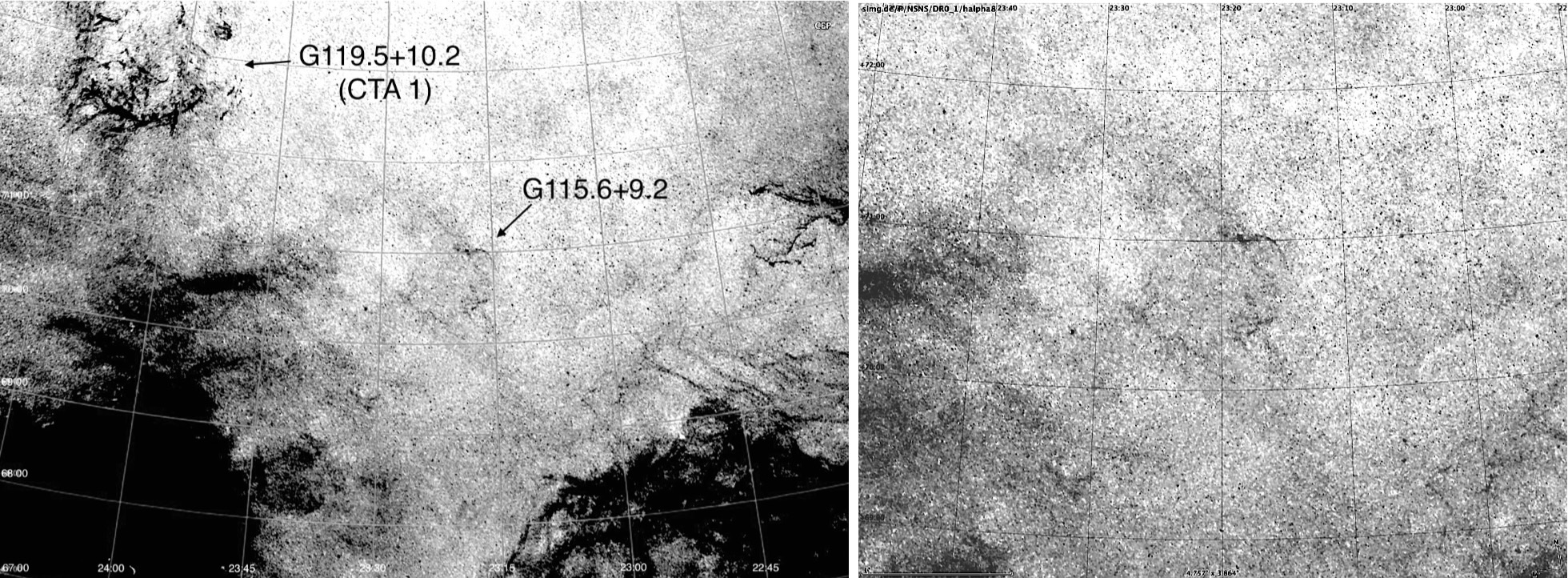}
\caption{Left: A $6.5\degr \times 9.6\degr$ H$\alpha$ image from the Northern Sky Narrowband Survey (NSNS)  
showing the area around SNR G115.6+9.2 along with the well known SNR G119.5+10.2 (CTA 1). Right: A closer view of  G115.6+9.2 showing
the presence of faint diffuse 
H$\alpha$
emission within the nebula's filamentary boundaries and faint emission
extending farther to the northeast.
\label{G115_NSNS} 
} 
\end{center}
\end{figure*}

Surprisingly, the spectrum taken of one of the remnant's NE filaments revealed a pure hydrogen Balmer type shock spectrum, with no detected \O3, [\ion{N}{2}], or [\ion{S}{2}] emission lines at levels above a few percent of that H$\alpha$ (bottom left panel of Fig.~\ref{G27_2}). Such a so-called Balmer-dominated or nonradiative spectrum is like than seen in some SNRs with interstellar shock velocities 
$\sim200$ km s$^{-1}$. 

In fact, based on morphology we suspect that much of the H$\alpha$ emission seen in the NSNS image along the remnant's northeast region as well as the line of eastern emission filaments (top right panels of Fig.~\ref{G27_2}) may also be  nonradiative shock filaments.  If that is the case, then this remnant is unusual in that much of its emission is nonradiative, possibly related to a relatively low ambient ISM density related to it location some 17$\degr$ off the Galactic plane.

\begin{figure*}[ht!]
\begin{center}
\includegraphics[angle=0,width=16.0cm]{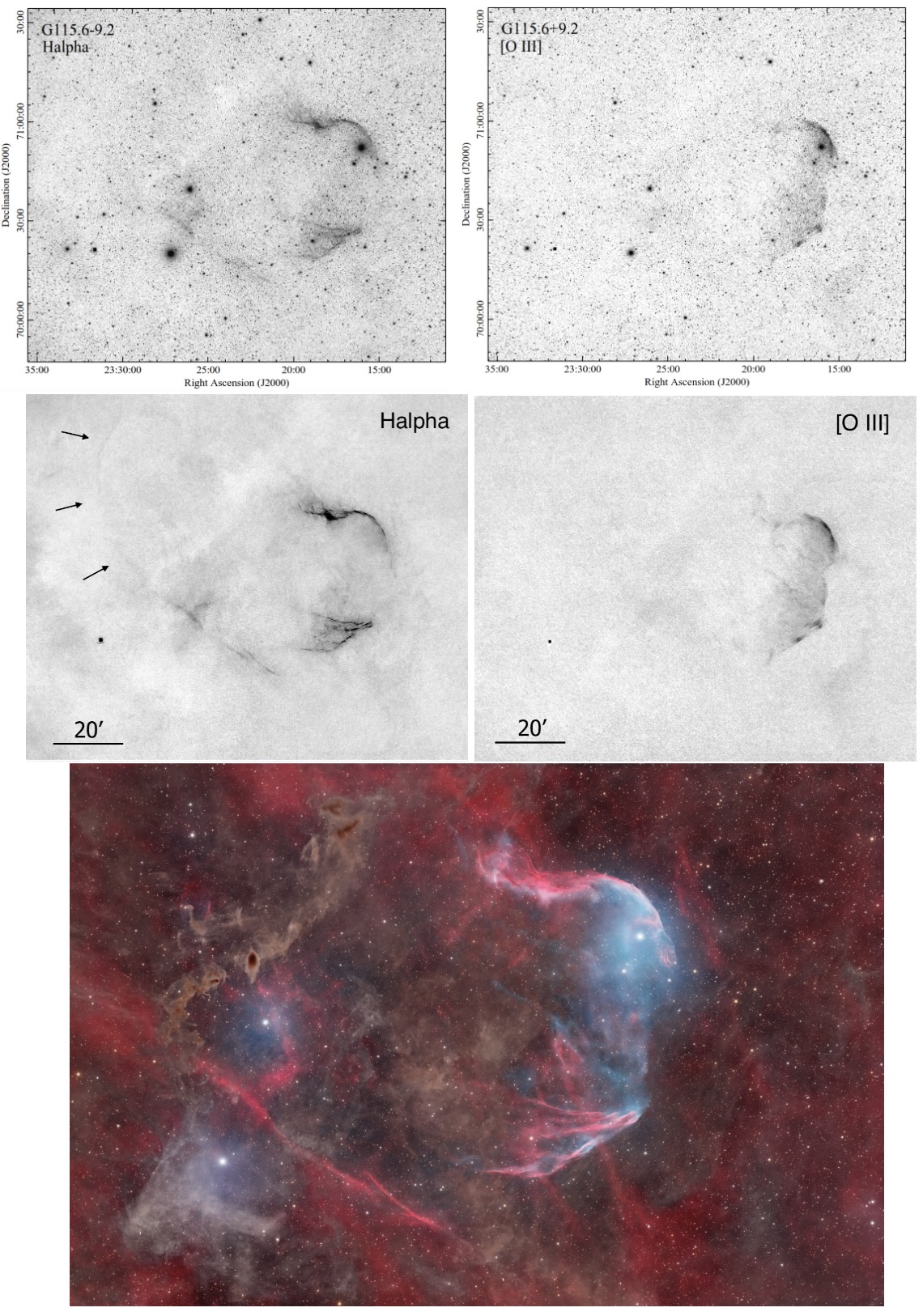} 
\caption{Top: H$\alpha$ and [\ion{O}{3}] images of the G115.6+9.2.
Middle: H$\alpha$ and [\ion{O}{3}] images now with stars removed. 
Arrows point to a faint emission extension to the east-northeast in the H$\alpha$ image.
Bottom: Color composite 
of G115.6+9.2 revealing faint diffuse H$\alpha$ emission surrounding the remnant 
to its north, west, and south but whose nature and association to the remnant is unclear. 
Note also the presence of numerous dusty ISM clouds seen toward the remnant's eastern areas
\label{G115_WCS} 
} 
\end{center}
\end{figure*}

\begin{figure*}[ht!]
\begin{center}
\includegraphics[angle=0,width=16.8cm]{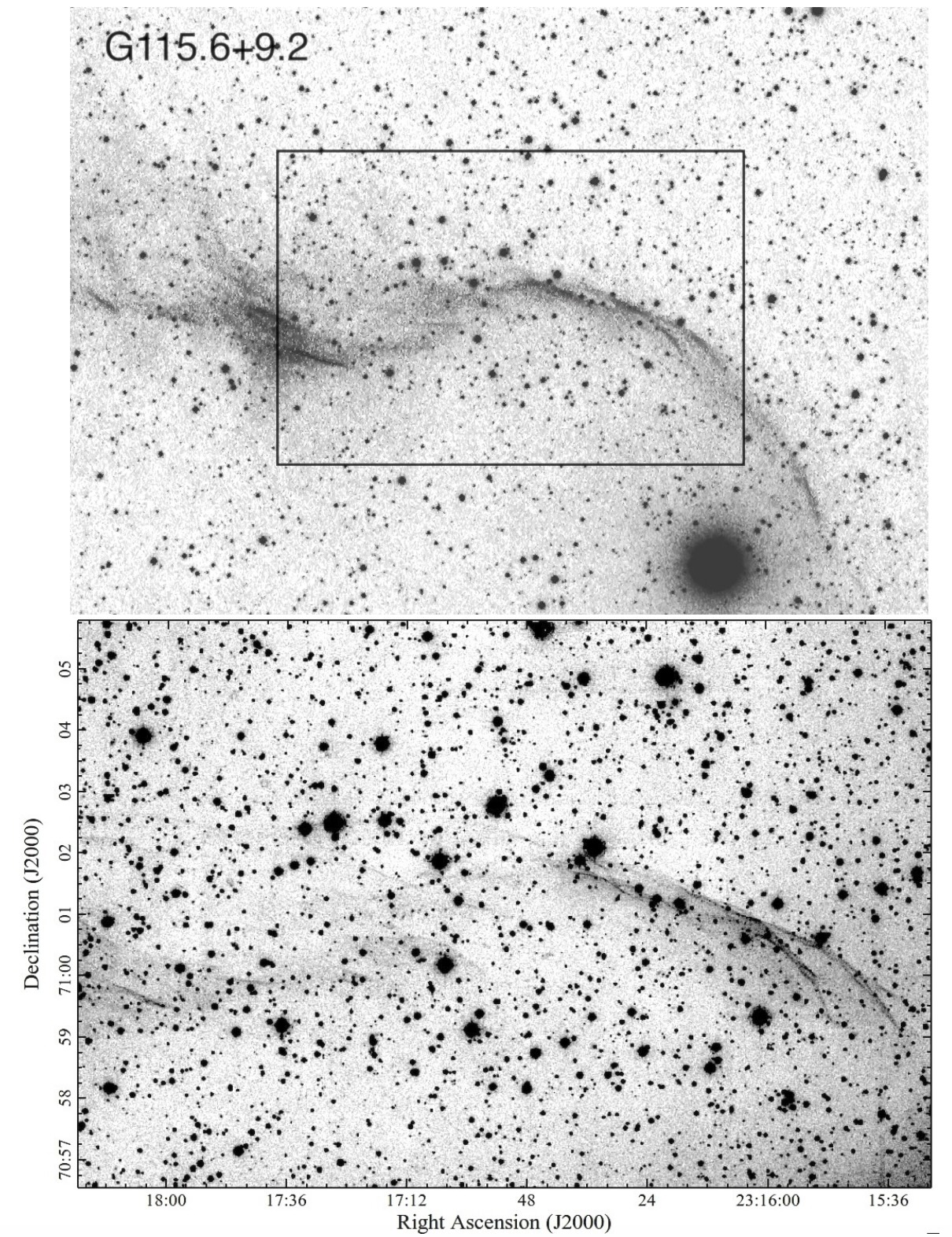}
\caption{Higher resolution of G115.6+9.2's northwestern H$\alpha$ emission.
\label{G115_North} 
} 
\end{center}
\end{figure*}

Finally, while H$\beta$ emission was visible just above the noise level in both spectra, as shown in Fig~\ref{G27_2}, 
a robust H$\alpha$ flux measurement for either slit position was not possible. Part of the cause for this weak detection is that the spectrograph setup for those spectra was relatively 
insensitive to features below 5000 \AA. Hence, we are left with only being able to state that the reddening for both regions appears not to be insignificant, a surprise given the remnant's high Galactic latitude. 

\begin{figure*}[ht!]
\begin{center}
\includegraphics[angle=0,width=16.8cm]{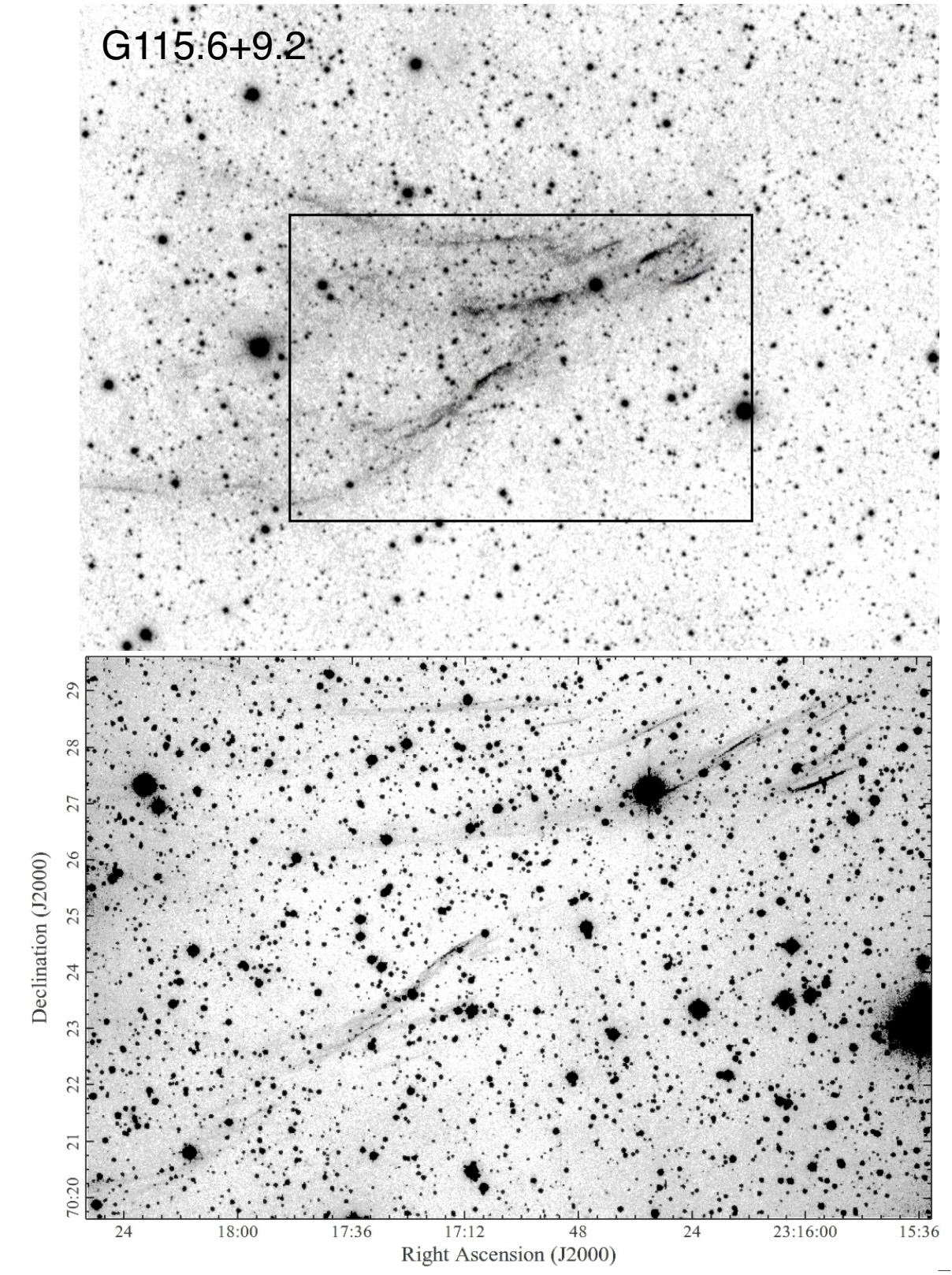}
\caption{Higher resolution of G115.6+9.2's southwestern H$\alpha$ emission.
\label{G115_South} 
} 
\end{center}
\end{figure*}

\begin{figure*}[ht!]
\begin{center}
\includegraphics[angle=0,width=14.5cm]{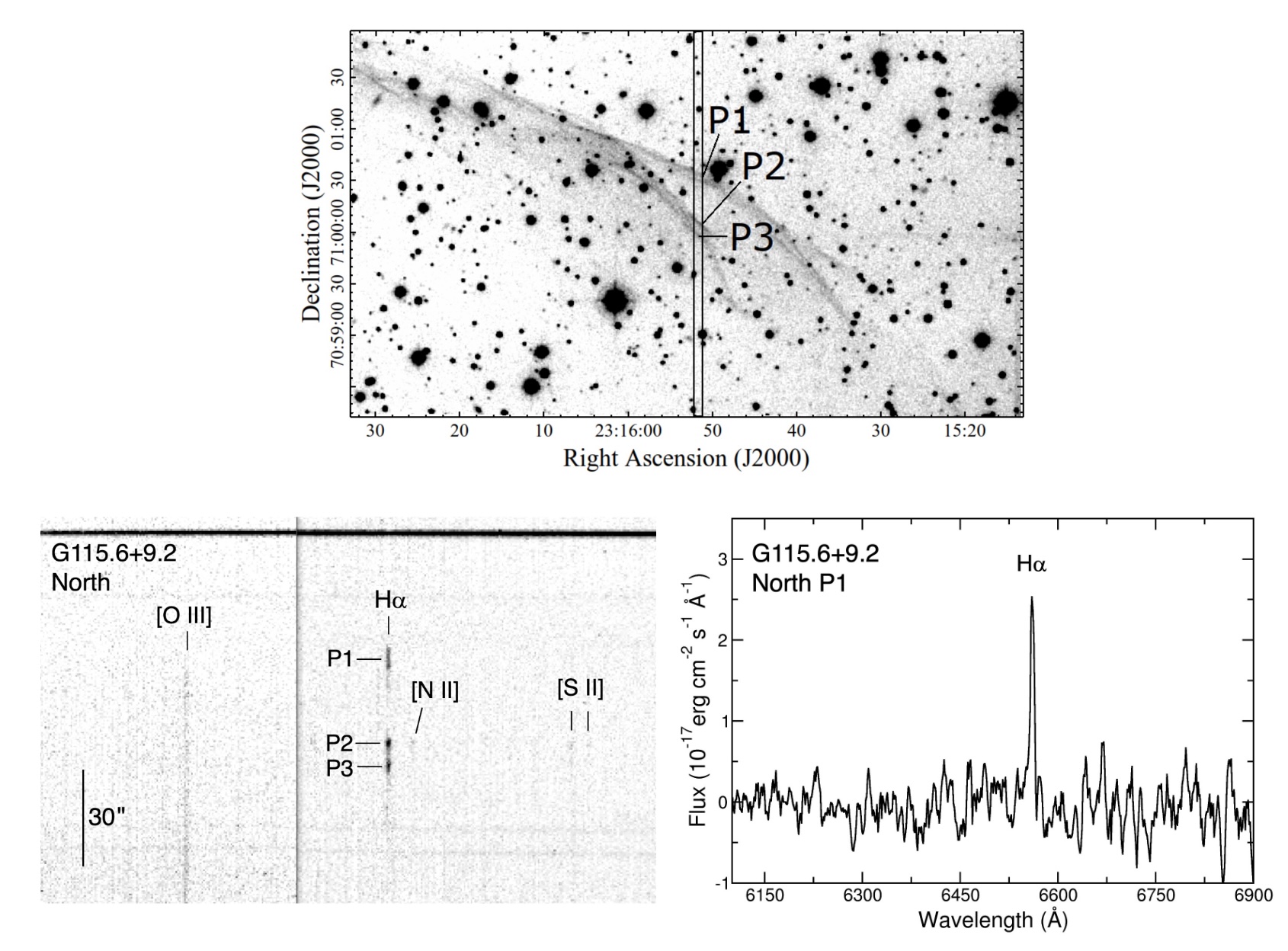}
\caption{Slit location and resulting spectra for G115.6+9.2's northwestern emission.
\label{G115_N_spec} 
} 
\end{center}
\end{figure*}

\subsection{G115.6+9.2 (Scylla)} 

A few years ago, M.\ Drechsler and colleagues noticed some faint
H$\alpha$ filaments located far north in Cepheus, not too far from the well-known supernova remnant CTA~1 (G119.5+10.2). These filaments seem to form a nearly
closed shell structure with a diameter of about one degree and  centered at approximately 
$\alpha$ = 23:21, $\delta$ = +70:40 (J2000).

\begin{figure*}[ht!]
\begin{center}
\includegraphics[angle=0,width=14.5cm]{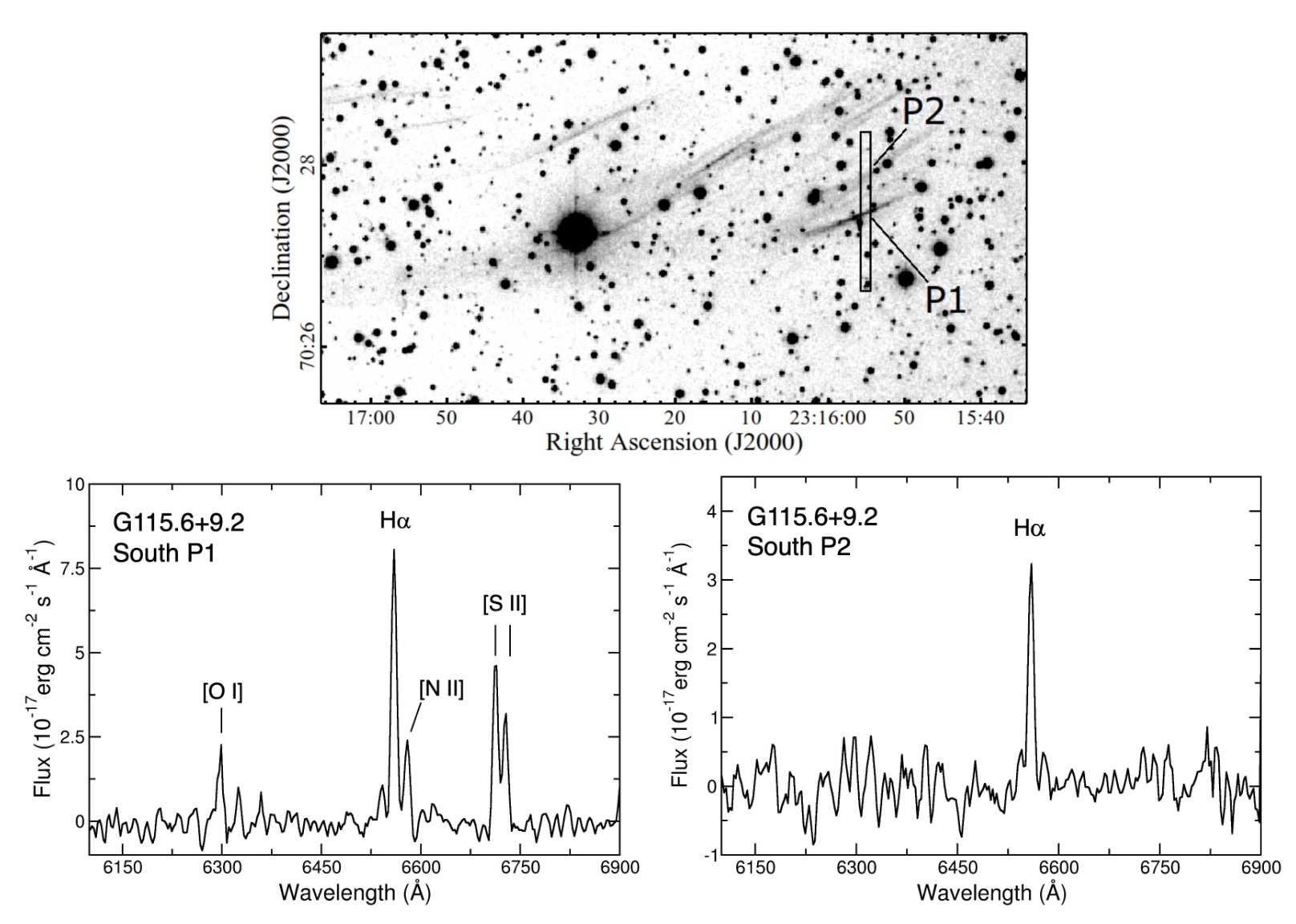}
\caption{Slit location and resulting spectra for G115.6+9.2's southwestern emission.
\label{G115_S_spec} 
} 
\end{center}
\end{figure*}

Not surprising given its faint H$\alpha$ surface brightness, there is no known or suspected concident SNR listed in X-ray or radio catalogs 
(\citep{Safi-Harb2019, Green2025}. 
However, this 
nebulosity, which we will denote as G115.6+9.2, ranks among the northernmost known Galactic supernova remnants\footnote{The remnant discoverers dubbed the nebula “Scylla” inspired by its appearance of multiple thin H$\alpha$ filaments like the tentacles of the mythological man-eating sea monster, Scylla.}.

Recent wide-field, but low resolution 
H$\alpha$ images taken as part of the
Northern Sky 
Narrowband Survey\footnote{https://www.simg.de/index.html} shows G115.6+9.2 to be situated in region relatively free of confusing line-of-sight H$\alpha$ emission. This is shown in Figure~\ref{G115_NSNS} where 
the left image shows an especially wide view of the area around SNR G115.6+9.2 along with SNR G119.5+10.2 (CTA 1), while the right image shows a closer view of the G115.6+9.2 nebula. These images also show faint diffuse H$\alpha$
emission within the nebula's filamentary boundaries, along with some additional faint emission extending farther to the northeast.

Higher resolution H$\alpha$ and \O3 images of G115.6+9.2 taken with the 
equipment listed in Appendix Table A  are presented in Figure~\ref{G115_WCS}.
The remnant's \O3 emission is seen to be mainly limited to just its westernmost areas,  quite different from that seen for its H$\alpha$ emission. This difference is clear in 
the starless H$\alpha$
and \O3 images, along with a color composite image
composted of H$\alpha$, \O3, and RGB images
This color image suggest that there is extremely faint emission off to the east-northeast marked by arrows in the starless H$\alpha$ image, as well as
faint, largely diffuse emission  off to the remnant's west, and faint diffuse emission south of the remnant's southeastern filament.

In general, the remnant's emission consists of both thin filaments and regions of mainly diffuse \O3 emission. The combination of both type of emission can be seen in Figures~\ref{G115_North} and \ref{G115_South}. Whereas wide view images detect  faint, diffuse emission, higher resolution 2.4m MDM images reveal a broken shell of thin H$\alpha$ filaments plus fainter diffuse emission along both the remnant's northwestern and southwestern limbs. 

The presence of thin filaments is a hallmark of ISM shocks seen in SNRs, but a
firm confirmation of the shock nature of 
G115.6+9.2's emission filaments comes from low-resolution spectral observations. Figures~\ref{G115_N_spec} and
\ref{G115_S_spec} show the spectra taken of the nebula's northwestern and southwestern limbs.
 
In Figure~\ref{G115_N_spec}'s upper panel, shows the location of our N-S slit which sampled three emission filaments, P1 - P3 along the remnant's northwestern band of filaments. A 2D image of detected emission around H$\alpha$ and [\ion{O}{3}]
$\lambda$5007 is shown in the lower left panel. Only H$\alpha$ emission was detected in a 3000~s exposure at filament P1 (shown in the lower right panel), whereas  only faint [\ion{N}{2}] $\lambda$6583 and 
[\ion{S}{2}] $\lambda\lambda$6716,6731 emission was weakly detected.

Figure~\ref{G115_S_spec} shows our spectra of the remnant's southwestern filaments.
Spectra for P1 exhibited strong [\ion{S}{2}] emission relative to that of H$\alpha$, a strong indicator of shock emission, whereas spectra of P2 only showed H$\alpha$ and no detectable [\ion{S}{2}] or  [\ion{N}{2}] emission.

For both the regions where we obtained spectra, a nonradiative or Balmer dominated type of emission was seen for some filaments. In addition, we measured heliocentric radial velocities of $-150$ to 
$-175$ km s$^{-1}$ for both northwest and southwest filaments. Although  such high radial velocities are unexpected given the filaments' locations along the remnant's edges, shock velocities $\gtrsim$ 150 km s$^{-1}$ is consistent with Balmer dominated, nonradiative shock filaments
\citep{Chevalier1978,Heng2007,Heng2010}
like we observed in these regions.

With its highly non-spherical structure, the G115.6+9.2's precise center and angular size are uncertain. In Table 1, we list its dimensions as $1.1\degr \times 0.9\degr$ 
but this is only the remnant's minimum size.
For example, its east-west dimension is at least one degree and could be as large as 2 degrees as indicated by faint H$\alpha$ filaments farther to the east as shown in Figure~\ref{G115_WCS}. Moreover, if some of the surrounding diffuse H$\alpha$
emission is associated with the remnant and not faint general Galactic emission,
then the remnant is considerably larger.
Given its high Galactic latitude, its distance is unlikely to be greater than $\sim$3 kpc (see $\S7.3.2$).

The relatively high velocities seen in the western filaments was a bit unexpected.
One the other hand, much like that seen for G27.9-17.6, high shock velocities may be simply a result of the remnant's location 
roughly nine degrees off the Galactic plane in a seemingly  empty and possibly low density region (see Fig.~\ref{G115_NSNS}).
Moreover, such high shock velocities may indicate the remnant is not too old.

\subsection{G190.5-25.3: A Huge Remnant Inside the Orion-Eridanus Superbubble }


A small and nearly circular grouping of about a half dozen small emission clumps, each $1'$ to $2'$ in size covering an area about 20$'$ in diameter and centered at   
$\alpha$ = 04:52:45, $\delta$= +07:41:10 (J2000)
were discovered in 2021 by X.\ Strottner and 
M. Drechsler\footnote{Astrobin IOTD: 2022-02-19}.
These nebulae lie within the northeastern section of the very large 20 by 45 degree H$\alpha$ and 21 cm radio emission shell known as the Orion-Eridanus Superbubble
\citep{Heiles1976,Heiles1979,Reynolds1979,Burrows1993, Brown1995,Bally2008,Ochsendorf2015}.

\begin{figure*}[ht]
\begin{center}
\includegraphics[angle=0,width=13.3cm]{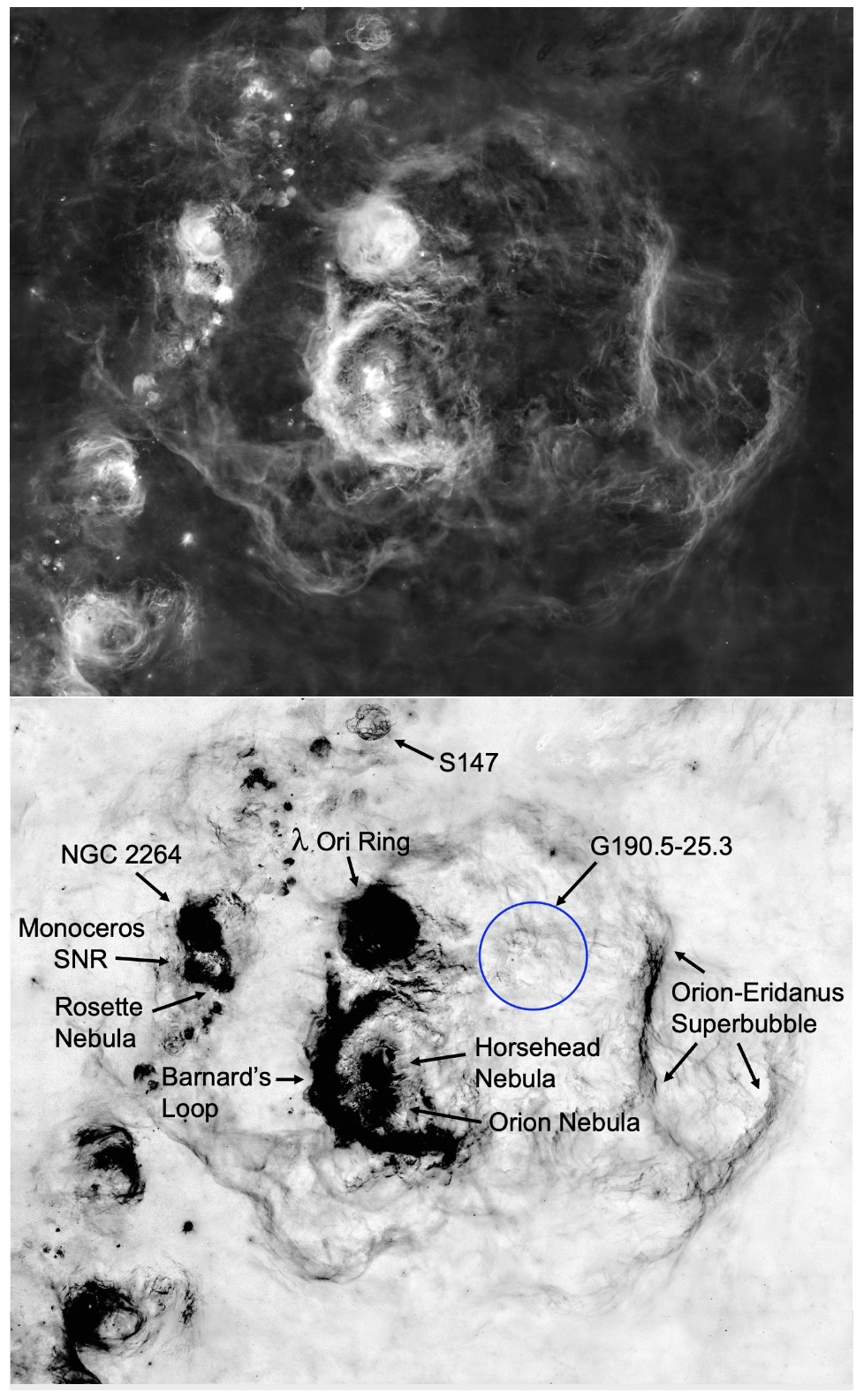}
\caption{Top: A $70\degr \times 60\degr$ positive MDW H$\alpha$ image of the Orion Nebula, Barnard's Loop, and the Orion-Eridanus Superbubble and other nearby Galactic nebulae.
Bottom: Annotated negative version showing the location of the G190.5-25.3 nebula along with the S147 and Monoceros Loop SNRs. The Orion-Eridanus Superbubble mainly consists of 
of Barnard's Loop, curved emissions to the north and west of $\lambda$ Ori, plus western filamentary emissions in the form of a near vertical emission feature (Arc A), a shorter vertical emission farther to the west (Arc B), plus a small, curved southern emission tail (Arc C).
\label{Walker} 
} 
\end{center}
\end{figure*}

\begin{figure*}
\begin{center}
\includegraphics[angle=0,width=17.9cm]{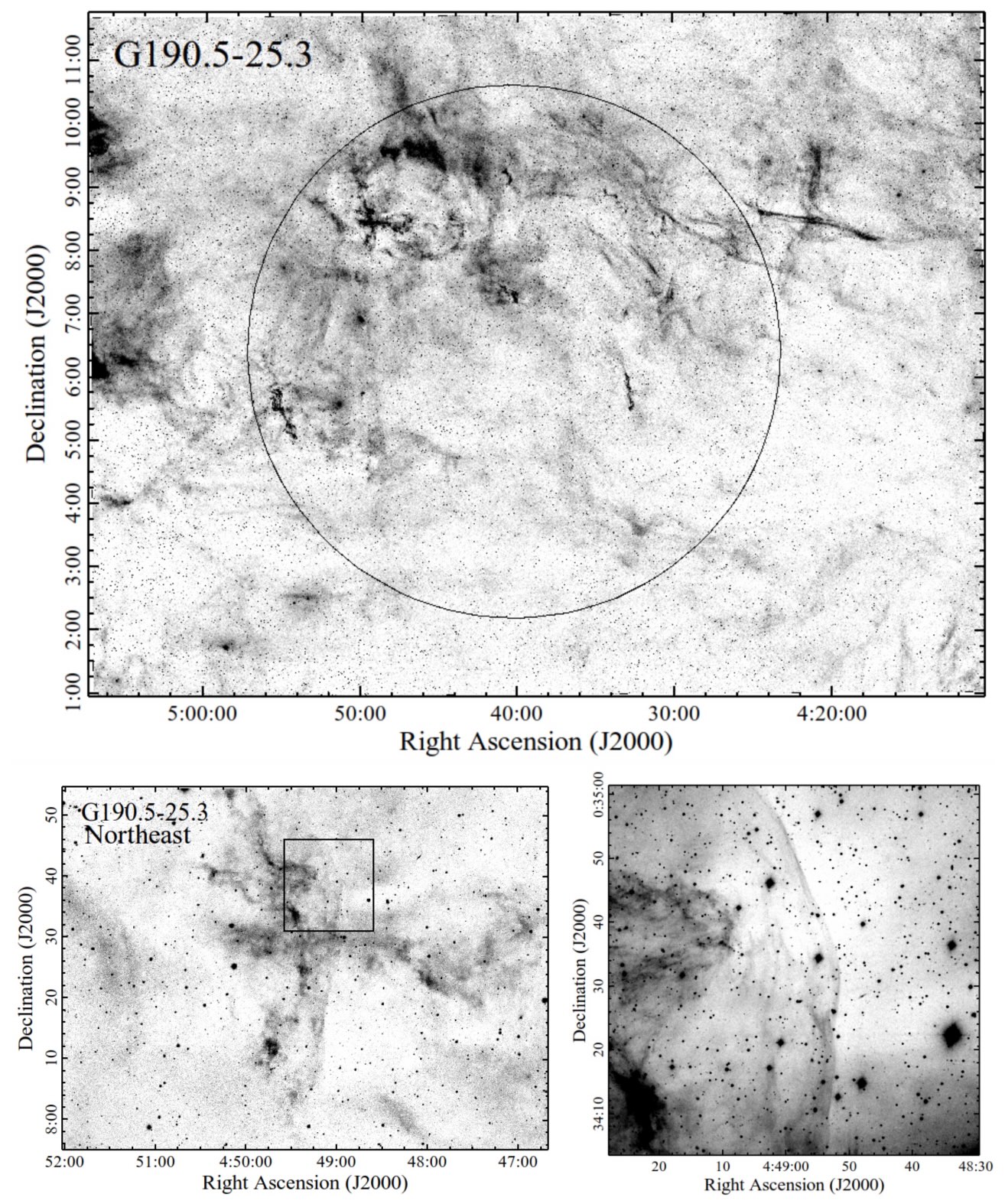}
\caption{Top: A $14.1 \times 10.8$ degree H$\alpha$ view of the G190.5-25.3 nebulosity and surroundings. Circle is 8.5 degrees in diameter. Bottom: Left image shows a portion of G190.5-25.3's NE which suggest the presence of a $40'$ long shock front, with the right image showing a deeper and higher resolution of the northern portion of this apparent interstellar shock. 
\label{G190_MDW} 
} 
\end{center}
\end{figure*}

\begin{figure*}
\begin{center}
\includegraphics[angle=0,width=15.7cm]{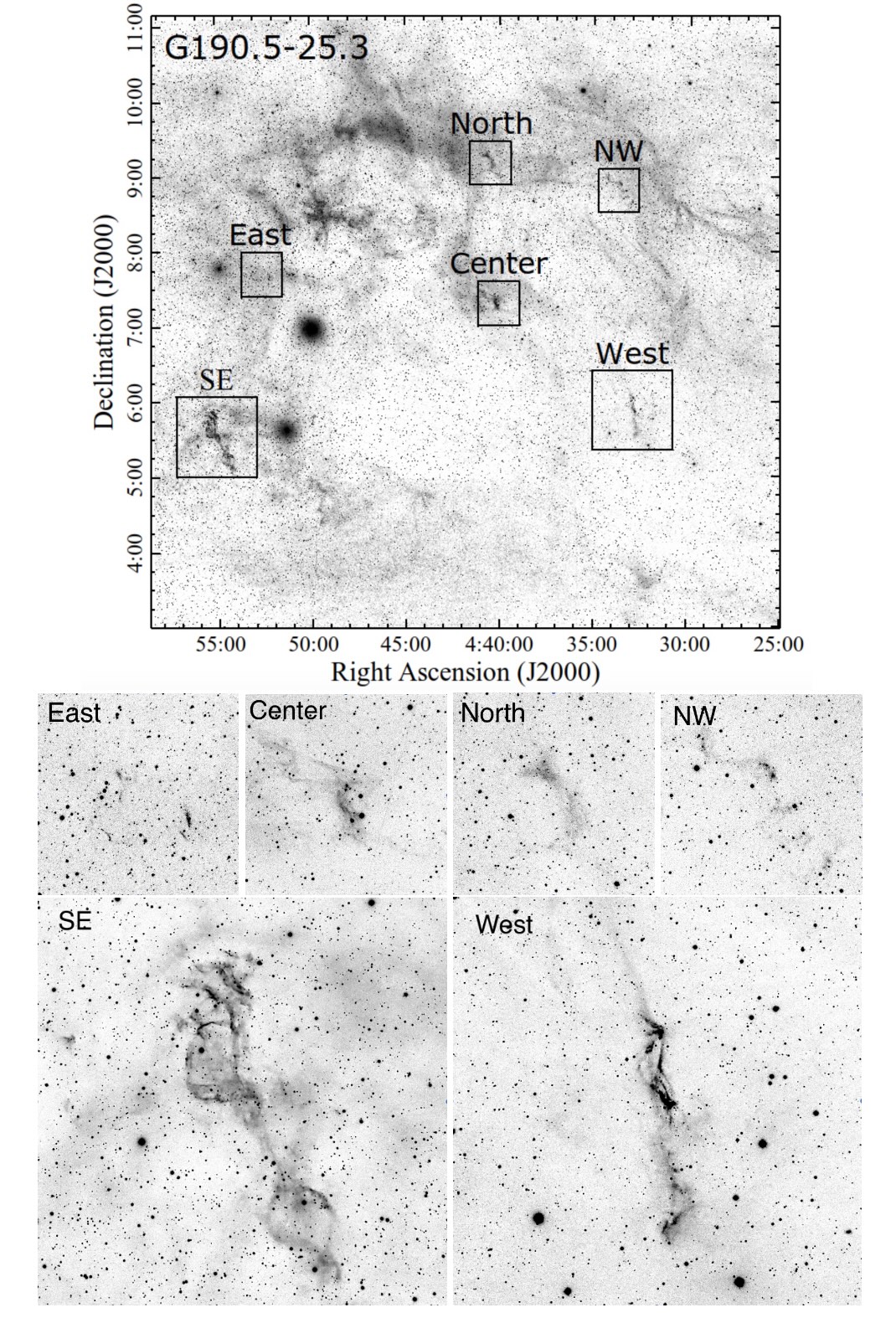}
\caption{A view of the G190.5-25.3 nebulosity with six sections marked that are showed below at higher resolution. Sizes of the small and larger close-up images are $30' \times 30'$ and 
$60' \times 60'$, respectively.
\label{G190_boxes}} 
\end{center}
\end{figure*}

\begin{figure*}
\begin{center}
\includegraphics[angle=0,width=17.1cm]{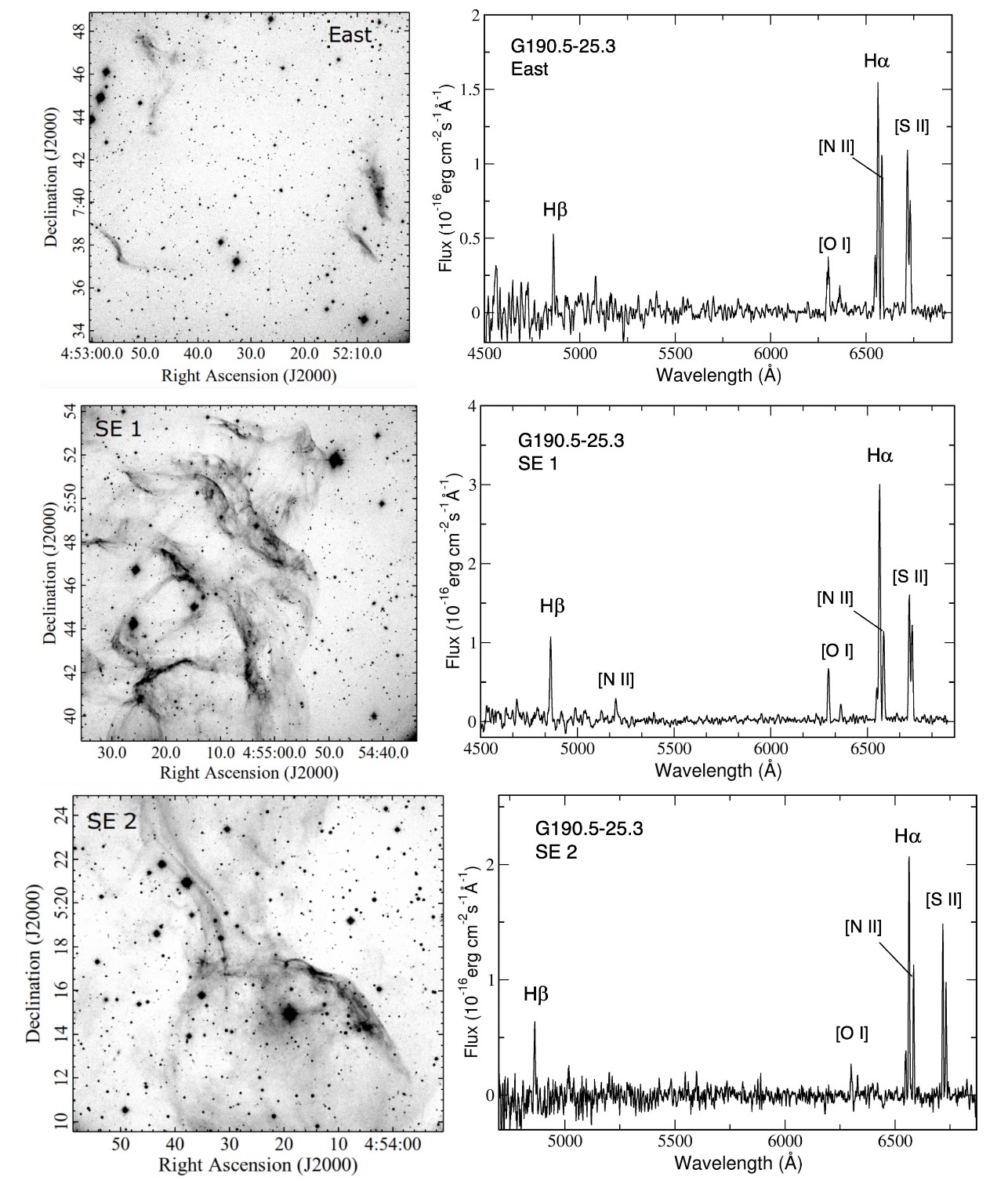} 
\caption{Top: Image of the cluster of small shocked eastern filaments which lead to the discovery of the G190.5-25.3 remnant. Shown is the spectrum of the bright western one.
Middle: The northern portion of  southeastern filaments (SE 1) with the spectrum of the bright filament located east of center shown. Spectrum of the southeastern filament (SE 1B) was also taken. Bottom: Image of the more southern section of these filaments along with spectrum of its brighter western curved filament. }
\label{G190_Set1} 
\end{center}
\end{figure*}

\begin{figure*}
\begin{center}
\includegraphics[angle=0,width=17.90cm]{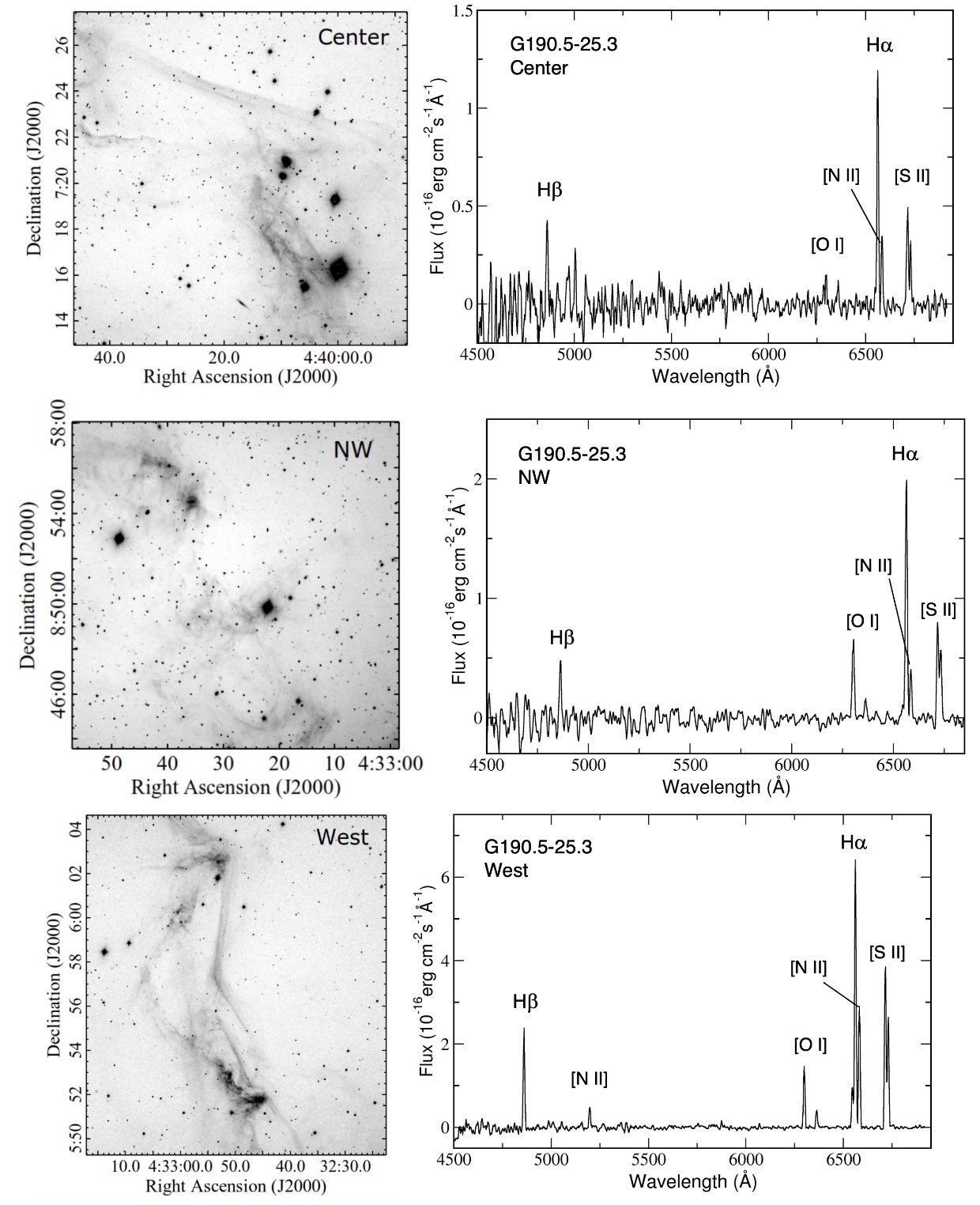} 
\caption{Top: H$\alpha$ image of filaments
located near the center of the G190.5-25.3 remnant along with spectra of the brighter,
lower right filaments. Middle: Group of northwestern emission knots with the spectrum shown of the brighter more northern one. Bottom: H$\alpha$ image of the northern portion of a long eastern filament with the spectrum of its bright southern knot.
}
\label{G190_Set2} 
\end{center}
\end{figure*}

\begin{figure*}[ht!]
\begin{center}
\includegraphics[angle=0,width=17.12cm]{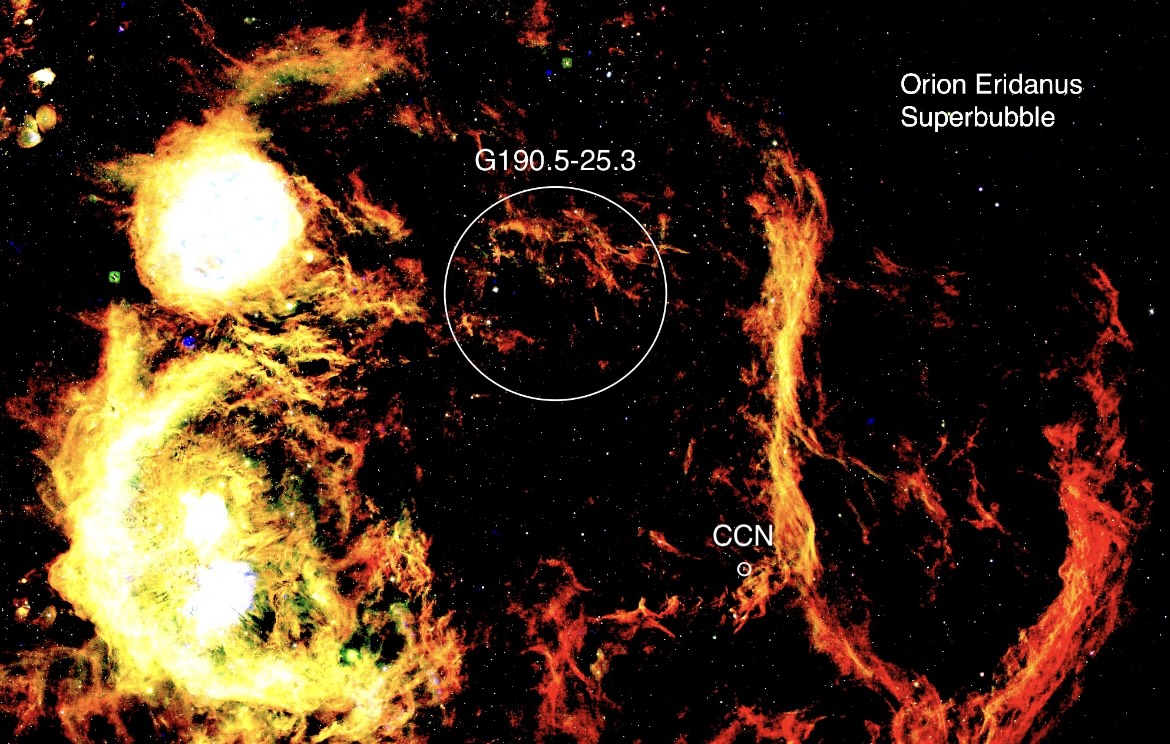} 
\caption{A false-color composite of the Orion Nebula, Barnard's Loop and the  Orion-Eridanus Superbubble generated from NSNS images.
Here [\ion{S}{2}] emission is mapped to red, 
H$\alpha$ emission to green, and [\ion{O}{3}] emission to blue.
This image highlights emission regions strong in [\ion{S}{2}] emissions relative to H$\alpha$.  Notice how the collective features of G190.5-25.3 stand out inside the superbubble shell. 
The circle around G190.5-25.3 is 8.5 degrees in diameter.
The location of the shocked Criss-Cross Nebula (CCN) discovered by \citet{ZW1997} lies near other strong red
[\ion{S}{2}] filaments farther south.
North is up, east to the left.
\label{NSNS_color} 
} 
\end{center}
\end{figure*}

Because these emission nebulae showed relatively strong 
[\ion{S}{2}] 
$\lambda\lambda$6716,6731 emission in exploratory spectral observations by L.\ Mulato
compared to 
their H$\alpha$ emission, they seemed to be shocked clouds of interstellar gas
and thus possibly part of an as  yet unidentified SNR. However, 
no known or suspected SNR lies coincident or near with their location \citep{Green2025}. 

The finding of shocked interstellar clouds in the direction to and possibly
inside the Orion-Eridanus Superbubble is reminiscent of a 1997 discovery by \citet{ZW1997}
of a small nebulosity dubbed the Criss-Cross Nebula (CCN) also located toward the Orion-Eridanus shell
which too showed a high 
[\ion{S}{2}]/H$\alpha$ emission line ratio of 1.15 indicative of shocks. 

The origin of this more recently discovered shocked clouds in the
Orion-Eridanus Superbubble
remained a puzzle until we made an examination of a deep 
H$\alpha$ mosaic image of the region taken as part of the MDW Sky Survey that revealed these clouds were part a $\sim$ 8$\degr$ emission shell structure projected to lie inside the even larger
Orion-Eridanus emission Superbubble. 
The cluster of shocked emission knots were found 
along this partial H$\alpha$ shell along its northeastern shell. Labeled G190.5-25.3 based on its Galactic center coordinates, 
it can be seen in Figure~\ref{Walker} which is wide 70 by 60 degree view of the Galactic plane covering the Orion-Eridanus Superbubble and neighboring nebulae.

The Orion-Eridanus Superbubble is believed to be one of 
the closest sites of high-mass star formation and was formed by the stellar winds of dozens of high mass stars plus a series of supernova explosions over the last few million years
which has swept-up and compressed local gas into the present day shell of ionized and neutral gas detected in 21 cm maps and deep H$\alpha$ images. The Orion-Eridanus Superbubble lies at an minimum estimated distance of 180 pc but may extend out to $\sim$500 pc. It has an often quoted expansion velocity of between 15 and 20 km s$^{-1}$ \citep{Reynolds1979} but could be much higher, around 40 km s$^{-1}$ \citep{Cowie1978, Brown1995, Welty2002, Ochsendorf2015}.

Although in principle these small shocked clouds could simply be part of the overall expansion of the Orion-Eridanus Superbubble, but in order for them to display such strong [\ion{S}{2}] emission, they must have recently experienced shocks above
50 - 60 km s$^{-1}$ \citep{Raymond79,Shull1979,Cox85,Hartigan1987}, 
and thus well above the estimated expansion velocity of the Orion-Eridanus bubble. Moreover, a handful of these clouds displayed weak [\ion{O}{3}] emission around their presumably low density edges indicating the presence of shocks just above 80 km s$^{-1}$. This realization lead us to conduct a  spectroscopic survey a few of the brighter emission features located within the $\sim$8 degree broken emission shell.

Figure~\ref{G190_MDW} shows a deep H$\alpha$ image of G190.5-25.3. The center of this large emission  feature is not well determined, so its coordinate name is only approximate. Nonetheless, this partial shell consists of a few bright emission features concentrated in the east and north where there is considerable diffuse emission, with little emission seen to the south and west. 

While there is no known nonthermal radio emission reported for this emission inside the Orion-Eridanus Superbubble that might suggest the presence of a SNR, there is
considerable
morphological evidence for the presence of shocks in the
G190.5-25.3 shell. Several features in this nebula
show sharp filaments suggestive for the presence of wide spread shocks. One example is
shown in the lower panels of Figure~\ref{G190_MDW} where we find a $40'$
long and curved emission front in the G190.5-25.3's northeastern region. The sharp edge 
of the H$\alpha$ emission seen in the higher resolution
image (right-hand lower panel) is similar to that commonly seen in SNRs.

The presence of such shock-like emission features lead us to obtain optical spectra of several of its
brighter features. 
Spectra were taken at six regions is shown in Figure~\ref{G190_boxes}. The lower portion of this figure shows enlargements 
of these regions.  These seemingly small emission features are, in fact, relatively large in angular size, with both the SE and West emission features extend nearly a degree in length. 

Figures~\ref{G190_Set1} and \ref{G190_Set2} present the resulting spectra of six regions of G190.5-25.3 along with seeing limited images of these regions.
Relative flux measurements are listed in Table 2.
Although the strengths of the emission lines vary
from region to region, all spectra show classic interstellar shock spectra with
relatively strong [\ion{S}{2}] $\lambda\lambda$6716,6731 line emission relative to that of H$\alpha$
(i.e., [S~II]/H$\alpha$ = 0.63 to 1.19) plus strong 
[\ion{O}{1}] $\lambda$6300 line emission. 
Importantly, many of these images reveal sharp filamentary morphology strongly suggestive of interstellar shocks; namely, the SE1, SE2, Center, and West regions.

No significant [\ion{O}{3}] $\lambda$5007 line emission was detected
in any of the six regions indicating relatively low velocity shocks, i.e., below 80 km s$^{-1}$ across much of G190.5-25.3. The [\ion{S}{2}] 6716/6731 ratio was always seen near the lower density limit of 1.43 \citep{Osterbrock2006} indicating densities
of $<$ 200 cm$^{-3}$.
In addition, extinction estimates were all quite small, with $E(B - V)$ between 0 and 0.15, the lowest of which was for the West filament. For these estimates we assumed an intrinsic 
H$\alpha$/H$\beta$ value of 2.86 corresponding to an
electron temperature of $10^{4}$ K and an electron density of $10^2$ cm$^{-3}$ for Case B recombination
\citep{Osterbrock2006}.  Our values are similar to those found for stars in the Orion-Eridanus bubble \citep{Cowie1979}.

We note that the identification of a SNR inside a OB association supperbubble is unusual, with few if any Galactic cases known and only a handful examples seen in the Magellanic clouds. 
Despite the clear evidence for shocks in the G190.5-25.3 nebula, both the superbubble's 40 km s$^{-1}$  plus detection of the shocked Criss-Cross Nebula (CCN) elsewhere in the superbubble, it raises the question whether what we observe as the G190.5-25.3 nebula is truly a SNR separate from the large and complex expansion of the enormous Orion-Eridanus superbubble which itself is believed generated by multiple SNe.  

If G190.5-25.3 is really a separate and identifiable SNR due to its strong [\ion{S}{2}] emission and shock-like optical spectra, might it be identifiable in 
[\ion{S}{2}] emission line images of the Orion-Eridanus superbubble?
Figure~\ref{NSNS_color} addresses this question.  It shows a false color image of the Orion-Eridanus Superbubble using images from the NSNS data collection. G190.5-25.3's
[\ion{S}{2}] line emission is emphasized in this image and mapped as red whereas
H$\alpha$ emission is diminished in strength and colored green.
This is a stretched version of the original NSNS image where H$\alpha$ emission is seen to largely disappear leaving regions with large [S~II]/H$\alpha$ values appear more clearly. 

In this figure  
G190.5-25.3 stands out as an organized group of filaments in an otherwise empty region of the superbubble. Similarly, the
Criss-Cross Nebula (CCN) is also seen lying far to the southwest, lying at the northern edge of a large, more southern shell of emission which exhibits unusually strong [\ion{S}{2}] emission. The location of the shocked CCN at the northwestern edge of this southern shell might signal additional shocked gas here of which the CCN may be just a piece. This should be investigated further. 

Finally, we note that the bright
red diffuse [\ion{S}{2}] emission in the Eridanus Arcs farther to the west appears bright mainly due to their great [\ion{S}{2}] flux. For example, analysis by \citet{Madsen2006} shows that the Eridanus Arc~A exhibits relatively low  [\ion{S}{2}] $\lambda$6716/H$\alpha$ values between 0.17 and 0.20, hence well below that seen in either G190.5-25.3 filaments or the CCN and hence inconsistent with shocks. 

In summary, shock-like spectra plus the shock-like morphology of major portions of a large, $\sim8\degr$ diameter partial shell seen in the direction of  the Orion-Eridanus Superbubble 
suggests it is Galactic SNR.
Specifically, we find the G190.5-25.3 nebulosity contains
several discrete features showing shock-like morphologies and emission like ratios consistent with low velocity shocks, i.e., below $\sim$80 km s$^{-1}$. We conclude that G190.5-25.3 is likely an old SNR 
and may represent one of the more recent SN events in this region that led to the formation of this superbubble.
Investigations of high-velocity Na I or Ca II absorptions in stars located in G190.5-25.3's direction could test our conclusion.

\begin{figure*}[h]
\begin{center}
\includegraphics[angle=0,width=17.0cm]{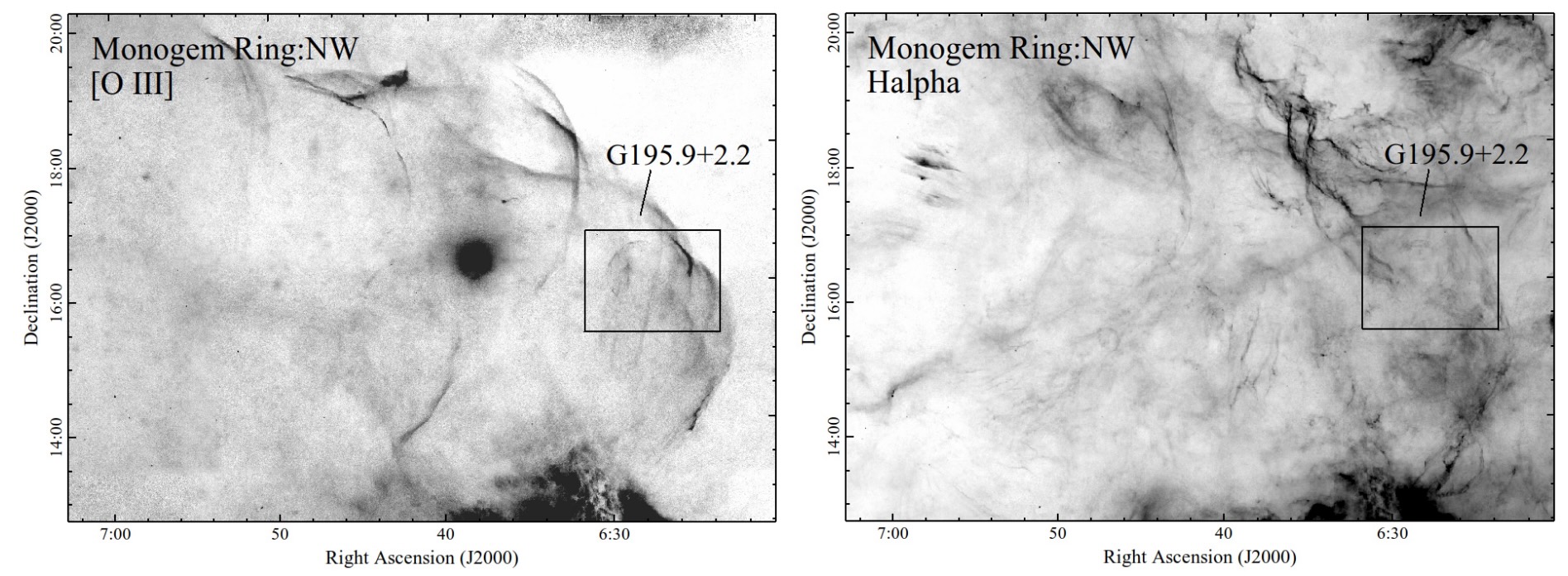} 
\caption{A wide view the northwestern portion of the Monogem Ring SNR showing the G195.9+2.2 remnant nearly lost among the Monogem's bright \O3 and H$\alpha$ emission filaments.
\label{G195_1}
} 
\end{center}
\end{figure*}

\begin{figure*}[h]
\begin{center}
\includegraphics[angle=0,width=17.9cm]{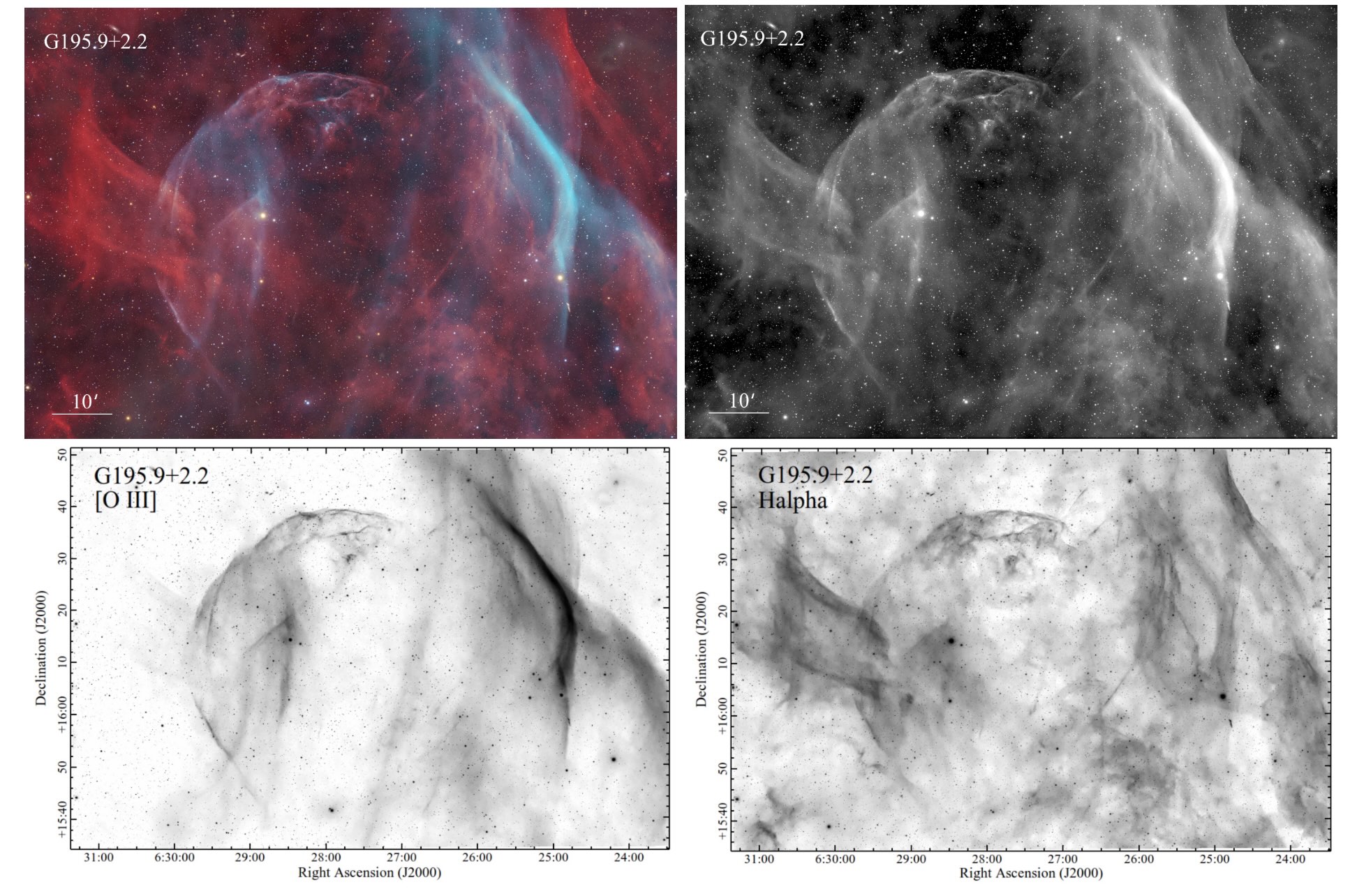}  
\caption{Top Left: False-color image of G195.9+2.2 composed of [\ion{O}{3}] (blue), H$\alpha$ (red).
Top Right: A black and white version at high contrast to enhance the SNR's visibility.
Bottom: Individual [\ion{O}{3}]  and H$\alpha$ images of G239.9+7.0.
\label{G195_fig1} 
} 
\end{center}
\end{figure*}

\begin{figure*}[ht!]
\begin{center}
\includegraphics[angle=0,width=16.0cm]{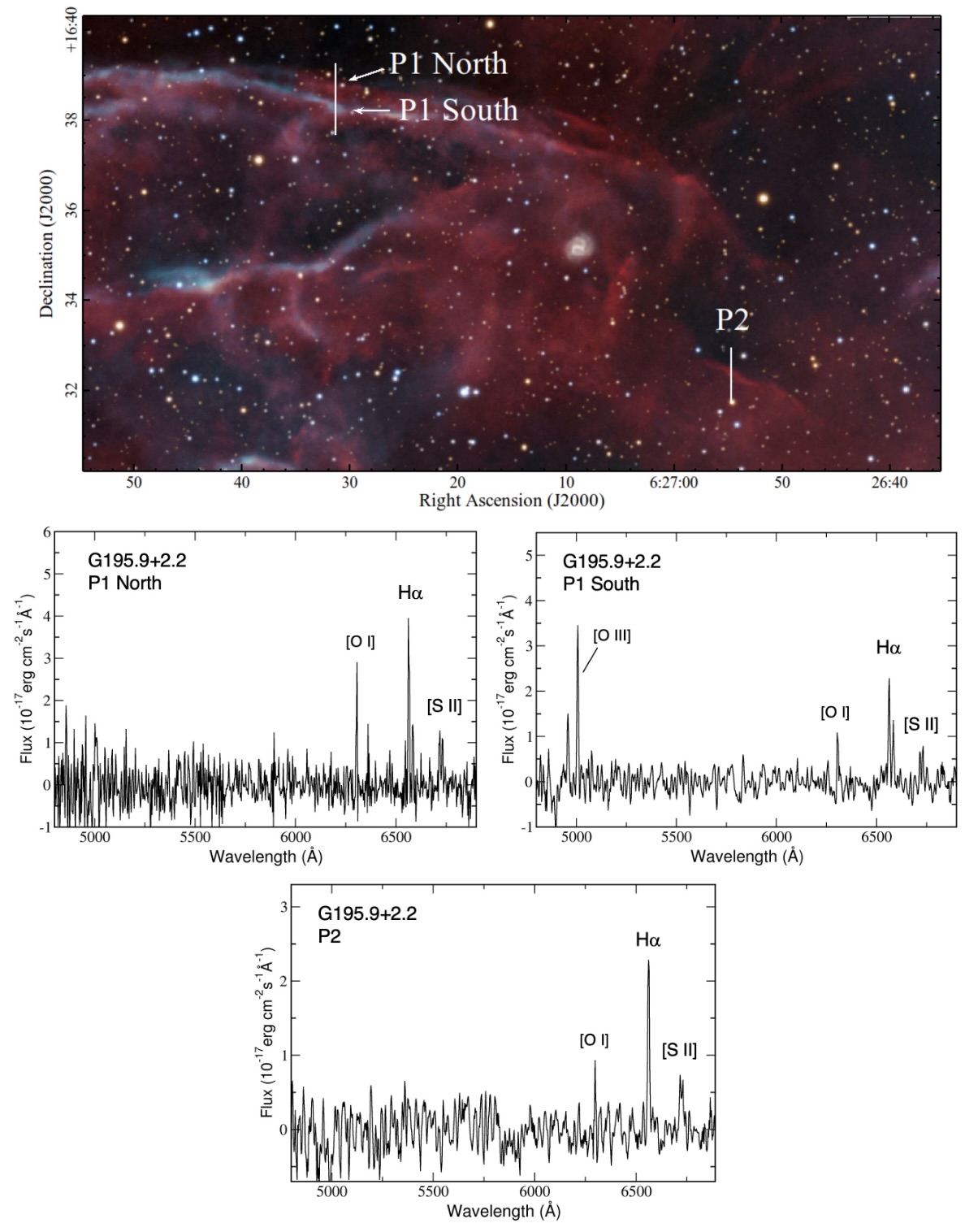}
\caption{Top: Slit locations in G195.9+2.2. Bottom: Resulting low-dispersion spectra.
\label{G195_fig3} 
} 
\end{center}
\end{figure*}

\subsection{G195.9+2.2: The Hidden Supernova Remnant}

During our survey of the Monogem Ring SNR, initial
H$\alpha$ and \O3 images revealed what appeared to be parts of a previously unknown emission shell. A subsequent imaging campaign by us in early 2025 with over 300 hours of exposures confirmed the presence a partial one degree diameter ring of emission located near the Monogem's northwest rim 
centered at $\alpha$ = 06:27:50, $\delta$ = +16:10:30 (J2000).

This nebula
is composed of several sharp \O3 and H$\alpha$ filaments arranged in a partial spherical shape. It is nearly invisible in both the MDW H$\alpha$ survey and in the multi-filter NSNS due to its faintness and confusion with 
brighter and unrelated H$\alpha$ and \O3  emissions, much of which is associated with the Monogem SNR. 

The difficulty in identifying the G195.9+2.2 remnant is demonstrated in  
Figure~\ref{G195_1}. Whereas part of the remnant's  shell can be seen in
the \O3 image, the shell's
H$\alpha$ emission is largely lost in the
overlapping Monogem Ring related 
H$\alpha$ emission (see discussions on the Monogem's emissions below in $\S6$).

\begin{figure*}[ht!]
\begin{center}
\includegraphics[angle=0,width=18.0cm]{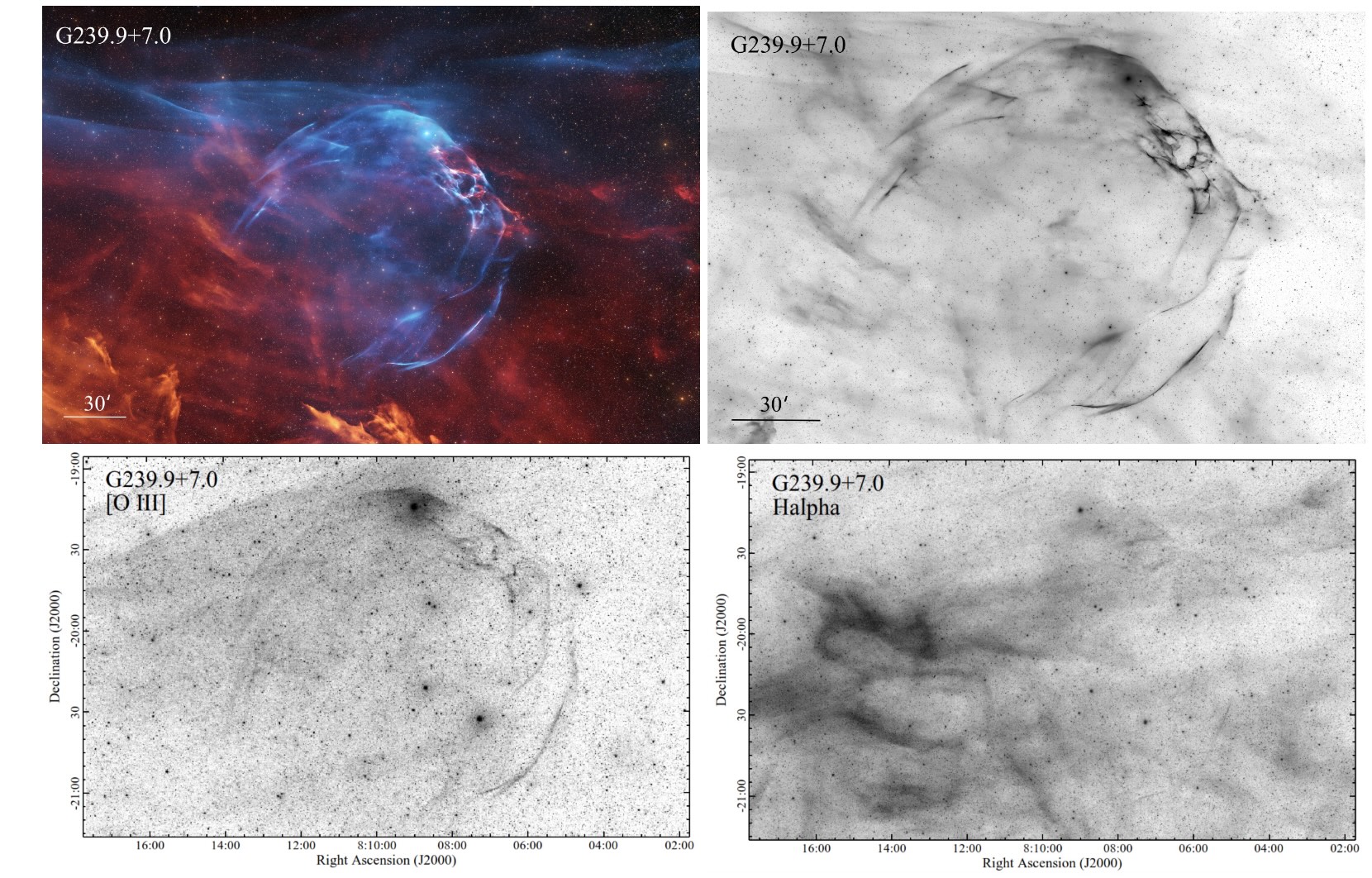} 
\caption{Top: False-color image of G239.9+7.0 composed of [\ion{O}{3}] (blue), H$\alpha$ (red), and a
black and white negative enlarged image. Note the sharp NE and SW filaments. 
Bottom: Separate [\ion{O}{3}]  and H$\alpha$ images of G239.9+7.0.
\label{G239_1} 
} 
\end{center}
\end{figure*}

\begin{figure*}[t!]
\begin{center}
\includegraphics[angle=0,width=17.0cm]{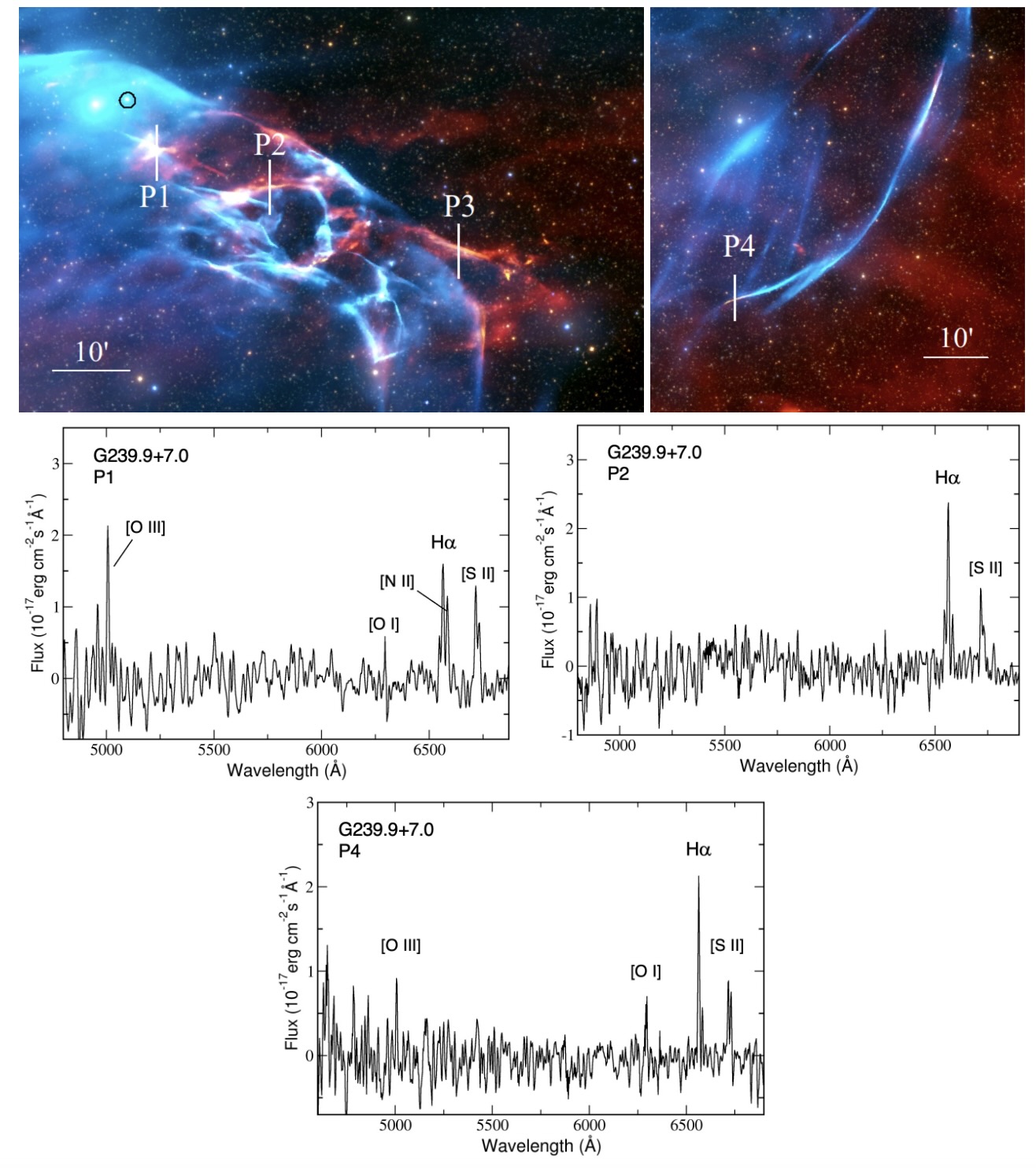} 
\caption{Top: False-color images of G239.9+7.0 showing locations of three slit positions, P1 - P3,  located along the nebula's northern region plus slit position P4 at the southern end of a long G239.9+7.0 filament in the southwest. The small black circle in the upper left marks the location 
of the 40$''$ diameter planetary nebula Sa2-21 \citep{Sanduleak1975} with the bright star, 16 Pup (V = 4.39), to the east of it. 
Bottom: Low-dispersion spectra taken at slit positions P1, P2, and P4. 
\label{G239_slits} 
} 
\end{center}
\end{figure*}

The G195.9+2.2 nebula is much clearer in  
Figure~\ref{G195_fig1} which presents  false color and black and white images
of the remnant. It is best seen in \O3 (blue) which is mainly concentrated along it northern and eastern regions.
The bright \O3 filament west of the remnant is not part of this SNR but rather a filament associated with the Monogem remnant.
The right panel shows a black and while version in which 
G195.9+2.2 can be more easily seen. Separate \O3 and H$\alpha$ images are shown in the lower panels. Whereas a partial shell
is easily seen in the \O3 images, the remnant's many sharp H$\alpha$ filaments almost get lost in unrelated emission features.

Low-dispersion spectra were obtained at two locations in the G195.9+2.2 shell, one along its northern limb and a second off to the northwest. The precise slit positions and the resulting spectra are shown in Fig.~\ref{G195_fig3}. The northern slit (P1) intercepted two different filaments, one showing strong H$\alpha$ emission with no
\O3 line emission, while a second, adjacent southern filament
showed prominent \O3 emission. Slit position P2  sampled a filament off to the NW in which no \O3 emission was present based on images.

In all three cases, strong [\ion{S}{2}] line emission was detected (see Table 2); specifically, [\ion{S}{2}]/H$\alpha$ $\simeq$ 0.60. Such strong [\ion{S}{2}] emission plus significant [\ion{O}{1}] $\lambda$6300 and filamentary emission are all indicators for the presence of shocks. The nebula's clear visibility in \O3  suggests a common shock velocity of $100 - 120$ km s$^{-1}$.

\subsection{G239.9+7.0}

SNR G239.9+7.0 is a faint 2.1 degree diameter nebula in the constellation Puppis. 
It was discovered
by B.\ Falls in late 2024 while conducting an optical emission-line survey in the nearby Gum/Vela nebular complex. Like several other recently discovered Galactic SNRs, it escaped earlier detection in H$\alpha$ surveys since it is predominately seen in [\ion{O}{3}] line emission.

The top panels of Figure~\ref{G239_1} shows a false-color composite image of the remnant along  with an enlarged and
negative version to highlight the nebula's sharper emission features. The bottom panels of Figure~\ref{G239_1} show
separate 
[\ion{O}{3}] and H$\alpha$ images of G239.9+7.0. Without the remnant's \O3 emission, one would not suspect the small handful of H$\alpha$ filaments along its northwestern limb were part of a nearly 2 degree diameter and otherwise unrecognized SNR.

Due to the faintness of features in this nebula, we chose 
$4 \times 4$ pixel on-chip binning (1.1$'' \times 1.1''$), a 3$''$ wide slit strategy together and a 3000 s exposure to improve the S/N of emission line detections. However, the resulting S/N for the spectra was still relatively low. In addition, the spectral observations for G239.9+7.0 were taken under a variety of photometric and non-photometric conditions. Data taken at slit position P3 was especially compromised apparently due to thin clouds, a hazard when observing southern hemisphere objects at high airmasses and close to the horizon from a northern hemisphere observatory.

\begin{figure*}[ht]
\begin{center}
\includegraphics[angle=0,width=15.0cm]{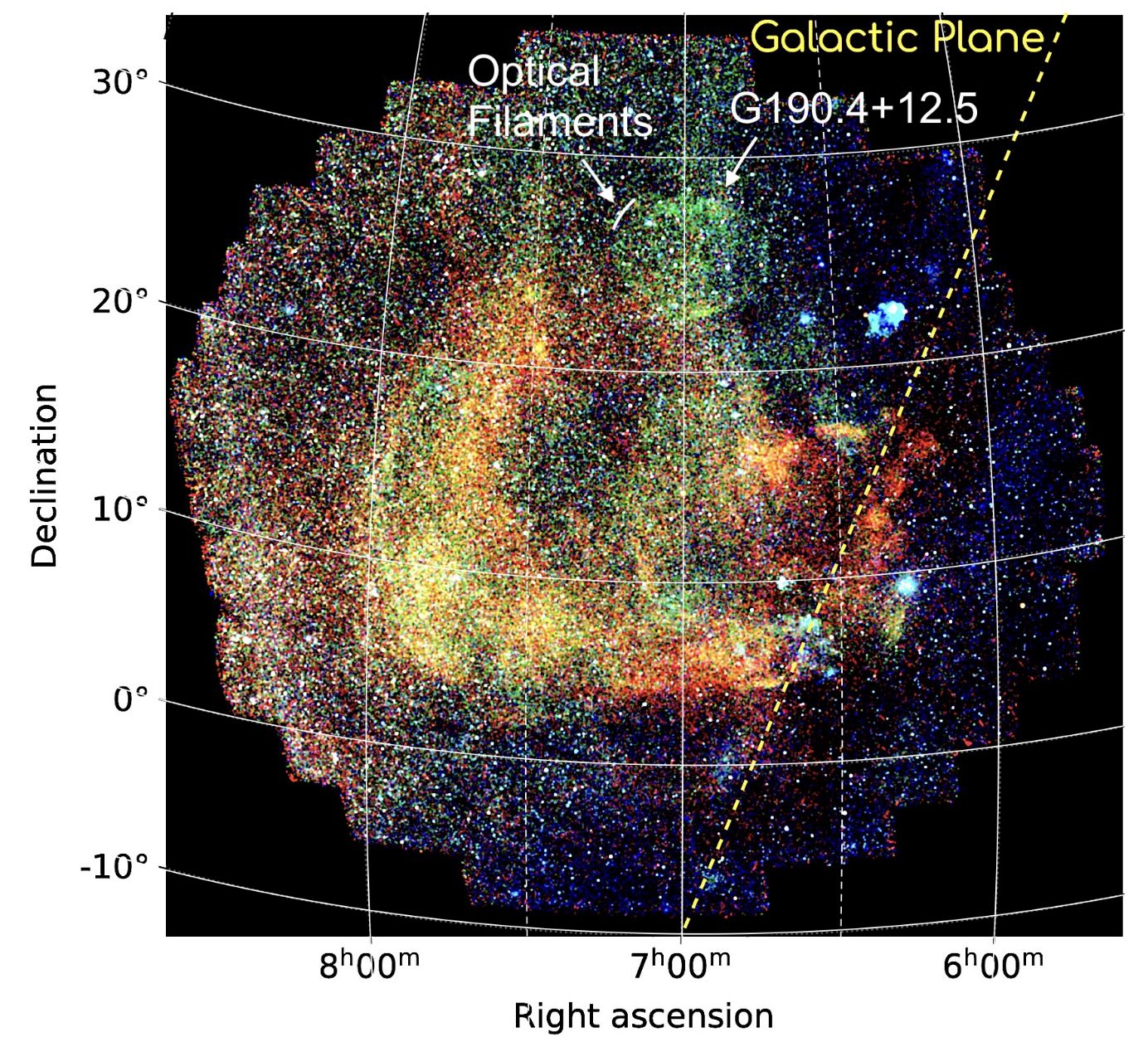} \\
\caption{Three color eROSITA X-ray image of the Monogem Ring taken from \citet{Knies2024} with the candidate SNR G190.4+12.5 indicated along with the location of optical filaments near its northeastern rim.
\label{Xray_1} 
} 
\end{center}
\end{figure*}

Nonetheless, the spectra for three regions of G239.9+7.0 
(Fig.~\ref{G239_slits})
clearly show strong 
[\ion{S}{2}] emission relative to H$\alpha$; specifically we find values of 0.67, 0.75, and 1.42 for [\ion{S}{2}]/H$\alpha$, well above the standard 0.40 ratio criteria for 
the identification of shocked gas. Even our weak  spectrum for slit position P3 taken under partially cloudy conditions indicated stronger 
[\ion{S}{2}] emission than that usually seen in H~II regions. Such strong [\ion{S}{2}] along with significant \O3 emission is further evidence for the presence of interstellar shocks.  The presence of several
thin, sharp filaments  (Fig.~\ref{G239_1})
completes a picture for shock emission from G239.9+7.0.

Consequently, we conclude that G239.9+7.0 is a new Galactic SNR. Its 2.1$\degr$ diameter and 7.0$\degr$ Galactic latitude places it among the largest and highest Galactic latitude SNRs known. Like some other recently discovered Galactic SNRs (G107.7-5.1 and G209.9-8.2; \citealt{Fesen2024}), this remnant's optical emission is dominated by 
\O3 line emission. This and the faintness of its emission helps explain why it had not been detected earlier.

\section{SNR Candidates}

\subsection{G190.4+12.5}

In their eROSITA investigation of the Monogem Ring's X-ray emission, \citet{Knies2024} found a $\sim6\degr$ diameter spherical patch of
soft X-ray emission located on the Monogem's northern limb. Its striking appearance in the eROSITA data led them to propose it as a SNR candidate named G190.4+12.5 with an estimated distance and age in the range of  1 - 2 kpc and $40 - 60 \times 10^{3}$ yr. 

The northern edge of our Mongem imaging survey covered the southern part of this SNR candidate and revealed a faint $\sim$1.5$\degr$ \O3 emission near the northeastern limb of this X-ray emission feature. Due in part to the region's  
complex H$\alpha$ emissions, no obvious H$\alpha$ counterpart filaments were seen. Although these \O3 
filaments may be associated with G190.4+12.5,
the detected filaments appear to lie a bit farther east of candidates SNR's X-ray rim. This is shown in Figure~\ref{Xray_1}. 

\begin{figure*}[ht!]
\begin{center}
\includegraphics[angle=0,width=16.8cm]{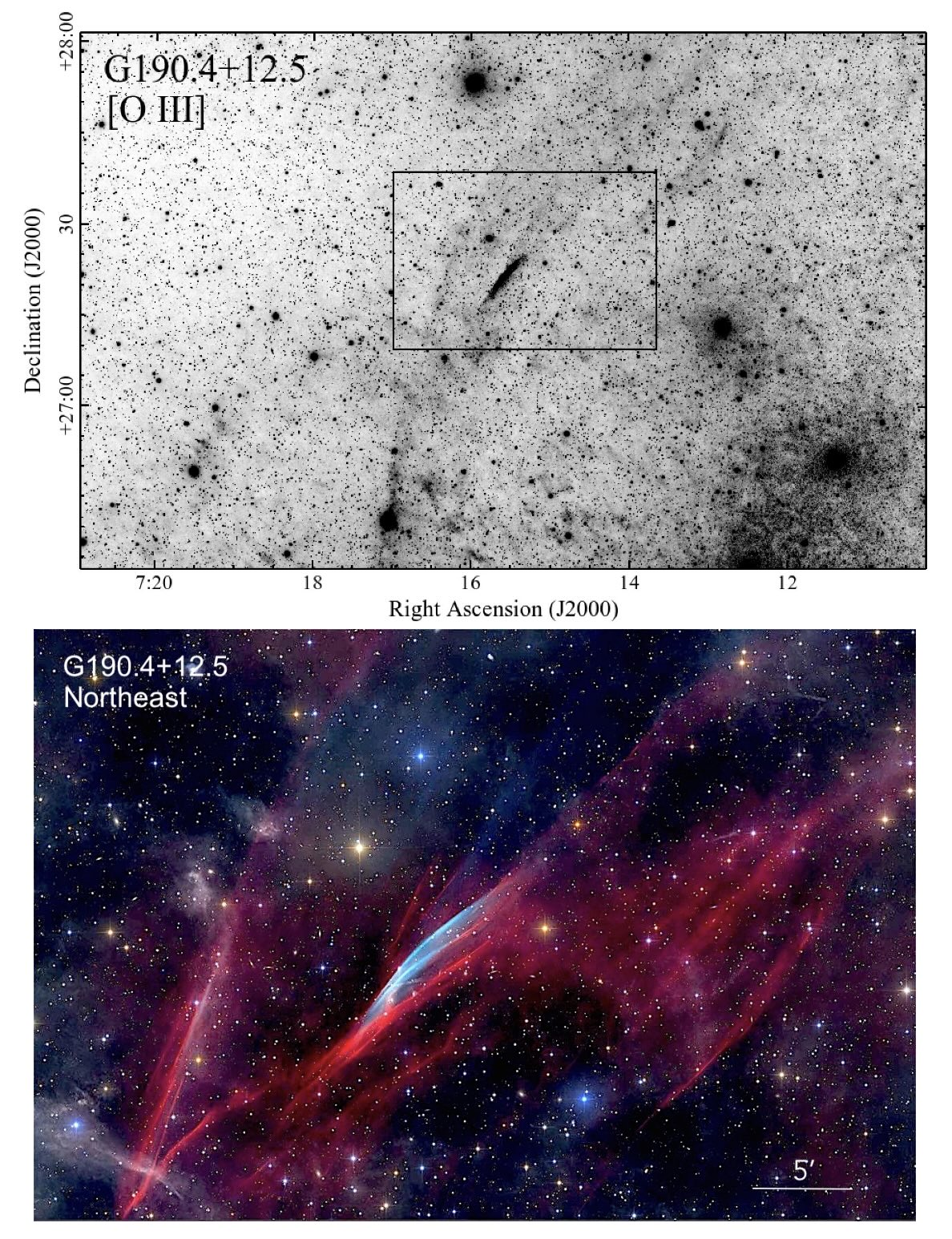}
\caption{Top panel shows our survey \O3 image near the northeastern rim of the eROSITA candidate SNR G109.4+12.5 with the black box indicating the region where deeper and higher resolution
data were taken. Lower panel shows color composite of these higher resolution exposures made from H$\alpha$ (red), \O3 (blue) plus R,G,B images.
\label{Xray_2} 
} 
\end{center}
\end{figure*}

The detected filamentary \O3 emission near G190.4+12.5's northeastern limb is shown in the top panel of Figure~\ref{Xray_2}. One section is especially bright,
while fainter and gently curved \O3 emission is seen to extend in declination from roughly +26.5 to +28.0 degrees.
A color composite image of a section of this area is shown in the figure's lower panel. The sharp curved  blue filaments constitute the brightest part of the upper panel's detected \O3 emission, with the thin H$\alpha$ filament seen along the color image's eastern (left) side can also be seen present in the \O3 survey image.

Such sharply curved optical filaments around and close to the G190.4+12.5's northeastern boundary are suggestive that they are shocked ISM and possibly related to this suspected X-ray emission remnant. Follow-up
optical spectroscopy can help to secure their shocked nature and, in turn, determine if G1904.4+12.5 is a new eROSITA discovered SNR.

\subsection{G191.4+11.1}

\begin{figure*}
\begin{center}
\includegraphics[angle=0,width=16.5cm]{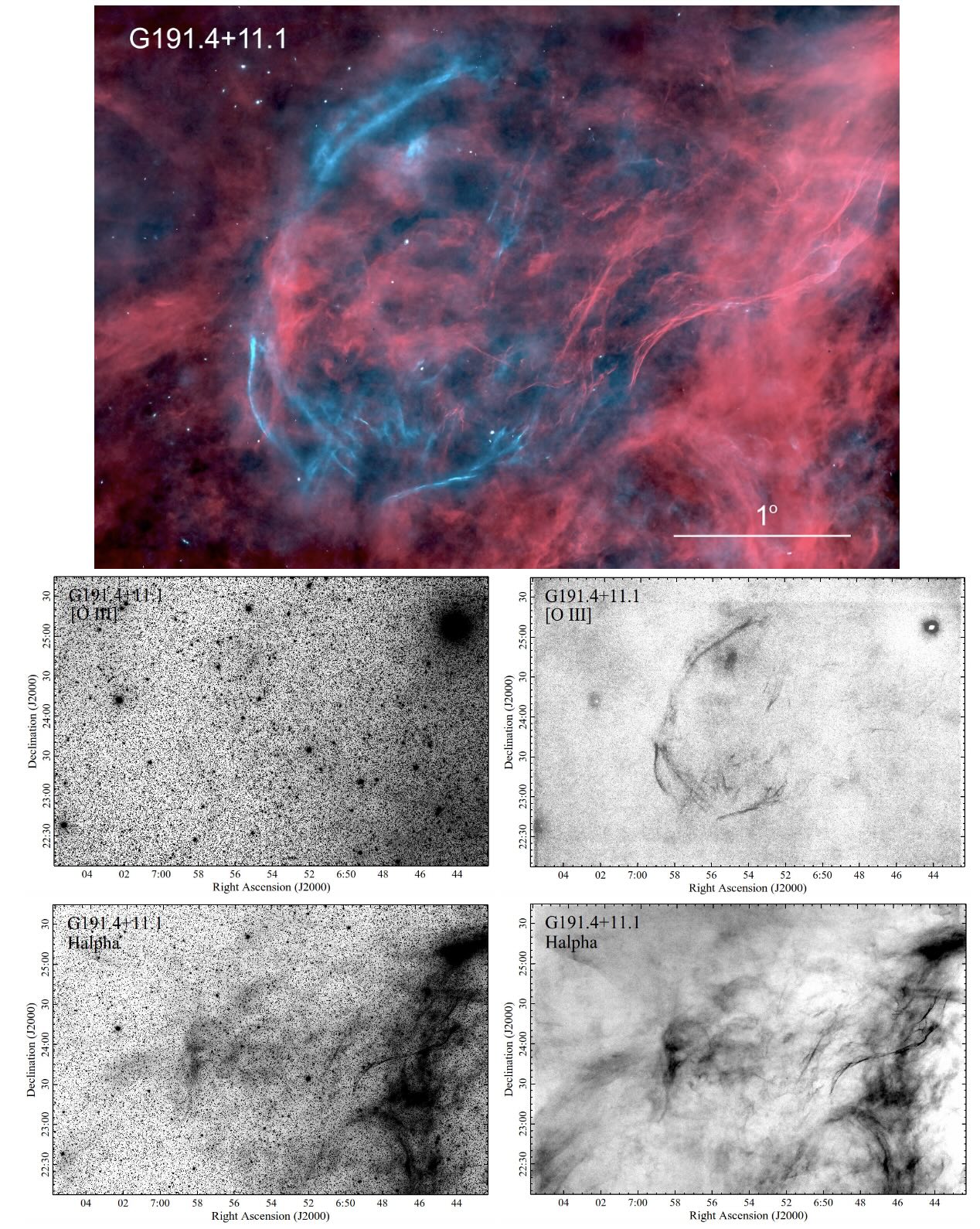} 
\caption{Top: Color composite of our H$\alpha$ (red) and \O3 (blue) images of G191.4+11.1. North is up, east to the left.
Bottom panels: Deep H$\alpha$ images of G191.4+11.1 with and without stars. North is up, east to the left.
\label{G191_1} 
} 
\end{center}
\end{figure*}

Our imaging survey of the Monogem Ring region detected a very faint and partial emission shell seemingly
brightest in \O3 emission located along the very northern edge of the Monogem remnant.  
This discovery led us to obtain much longer and deeper \O3 and H$\alpha$ images of it in order to better determine its size, morphology and possible nature. Even with a strong and focused imaging campaign, our detection is a relatively weak.
The top panel of Figure~\ref{G191_1} shows 
a color composite of our images of this new SNR candidate, G191.4+11.1. This new  nebula is elliptical in shape and relatively large with angular dimensions of $3.2\degr \times 2.2\degr$. 

Figure \ref{G191_1} also shows our individual 
\O3 and H$\alpha$ images, both with and without stars. 
The nebula consists of a partial shell that is entirely missing
its northwestern section. The parts that are visible are \O3 dominated with very few H$\alpha$ filaments. Indeed, most of the brightest H$\alpha$ filaments along the image's western edge are actually parts of the neighboring Gemini H$\alpha$ Ring.

We have not attempted to obtain optical spectra of this nebula to investigate if the nebula emission seen is shocked or photoionized. But given its overall extreme faintness and the lack of bright H$\alpha$ filaments, obtaining confirming optical spectra may be challenging.

\begin{figure*}[h]
\begin{center}
\includegraphics[angle=0,width=17.5cm]{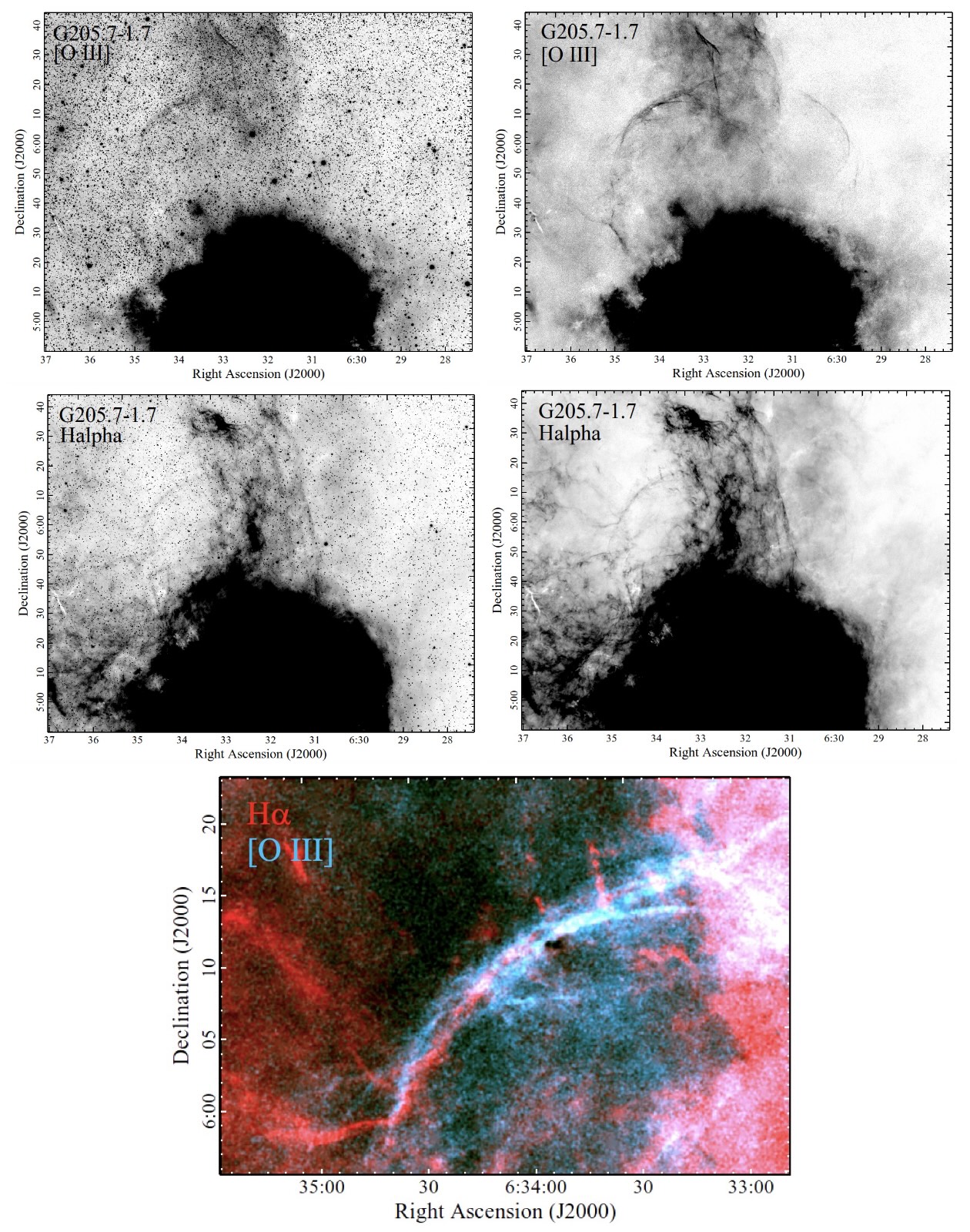}
\caption{Upper panels: Deep \O3 images 
and H$\alpha$ images shown with and without stars of the SNR candidate G205.7-1.7 located above the northern limb of the Rosette Nebula. 
Images as show are tilted 3.0$\degr$ to the east. 
Bottom: Starless color composite
showing the nebula's \O3 emission filaments lying outside of its H$\alpha$ emission.
\label{G205} 
} 
\end{center}
\end{figure*} 

\subsection{G205.7-1.7}

Our Monogem  imaging survey extended down to the bright and well known H~II region, the Rosette Nebula, situated both along the Monogem's southwestern boundary and the southern limb of the Monoceros SNR. Inspection of images of this region revealed an almost perfect half circle of thin \O3 emission filaments extending northward off the Rosette Nebula.
These filaments are shown in Figure~\ref{G205} which presents both  our \O3 and H$\alpha$ images with and without stars.

While this half circle of filaments has been seen before (\citealt{Davies1978}; their Fig.\ 2) and is easily visible on the red DSS2 images of the region between the Rosette Nebula and Monoceros Loop remnant, we know of no previous comment or note about it in the literature.
Not surprisingly given its faintness and thinness, there is also no clear sign of it in radio maps of either the Rosette or Monoceros Loop nebulae 
\citep{Graham1982,Odegard1986,Xiao201}.

Although these 0.7$\degr$ radius filaments appear to overlap with the much brighter and more extensive filamentary emission of the Monoceros Loop SNR
(G205.5+0.5), they are quite singular in appearance and show a distinctively different 
curvature from that of the Monoceros remnant filaments. Moreover, the nearly perfect spherical structure of this arrangements of thin filaments suggests an origin from a point-like source unrelated to the much larger Monoceros remnant.

A SIMBAD search of early type stars within a 10 arc minute radius of the apparent center of this filamentary half-circle (RA: 6:32:06, Dec: +5:41:50) revealed no O or early B type stars
whose winds might have created this nebula.
This is consistent with the findings of
\citet{Villa2025} who lists only one early type star near G205.7-1.7's center, namely GSC 00158-00640, a B5 star lying some 14.2 arc minutes away from our measured shell center. 

As we have discussed above for other SNRs, this SNR candidate which we designate as G205.7-1.7, assuming a center partially hidden by the Rosette Nebula, 
is best seen in the \O3 image, especially in the \O3 image with the stars removed
(Fig.~\ref{G205}). 
It is this nebula's \O3 emission that is most suggestive that this emission structure is a possible SNR.
It retains its filament identity even where it crosses over the Monoceros SNR filaments.
While much of its western half is quite faint, it is still visible in \O3, but is virtually missing in H$\alpha$ except for a small filament along its far southwestern limb.

The sharpness of the nebula's filaments is consistent with shocks and such an interpretation is supported by the filaments'
\O3 emission being located at a  larger radius (i.e., some 10 - 30 arc seconds out ahead) than that of the H$\alpha$ emission.
This is shown in the bottom panel of Figure~\ref{G205}. A similar situation is seen
for the one H$\alpha$ filament seen in the nebula's SW corner, where the \O3 emission is seen again to lie of order 10 - 15 arc seconds farther out relative to the H$\alpha$ emission. 
Such a pattern where \O3 emission lies out ahead of H$\alpha$ is fairly common in well resolved Galactic SNRs like the Cygnus Loop, IC~443, and S147.

Finally, we note that our \O3 image of the ring's eastern part seems to show very faint filaments extending farther to the east off the curved main eastern filament hinting at a possible shock blowout. If confirmed by deeper \O3 images, such a feature could further support the shocked nature of the  
G205.7-1.7 nebula.

\subsection{G305.4-0.7}

\begin{figure}[ht]
\begin{center}
\includegraphics[angle=0,width=8.5cm]{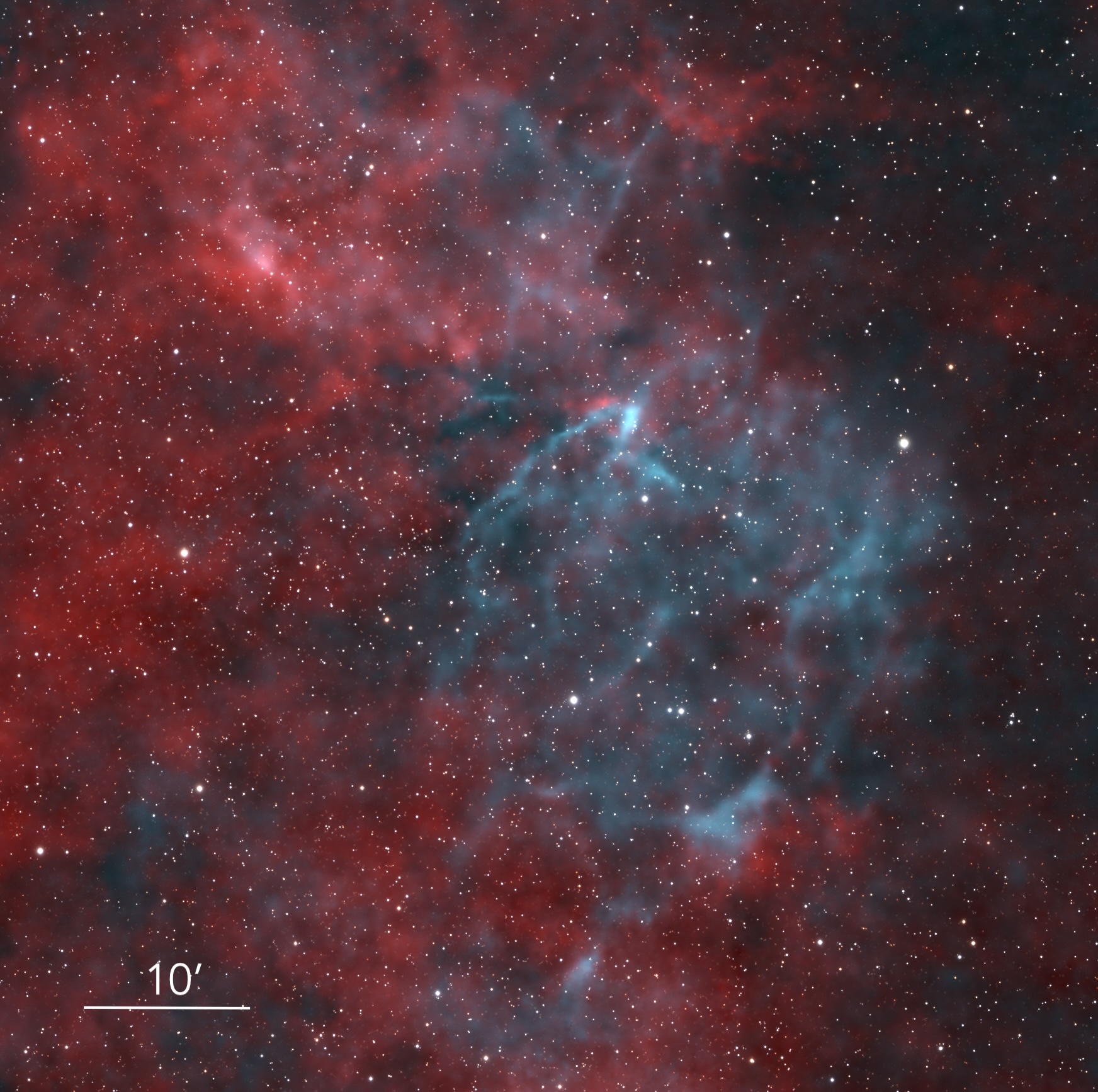} \\
\caption{Color composite of H$\alpha$ (red) and \O3 (blue) images of the SNR candidate
G305.4-0.7 near the open star cluster NGC 5045. North is up, east is to the left.
\label{G305_1} 
} 
\end{center}
\end{figure}

In the course of obtaining $\sim4\degr \times 3\degr$ wide emission line images of the Milky Way's southern hemisphere, M.\ Peitsch and J.\ Rodrigues detected a faint $\sim 30'$ diameter  nebula brightest in \O3 emission situated near the open star cluster NGC 5045 (see Fig.~\ref{G305_1}). 
Higher resolution follow-up images revealed the nebula to have numerous thin and highly curved \O3 filaments organized into a roughly circular shape. A handful of H$\alpha$ filaments that match small sections of the \O3 filamentary nebula were also detected.
However, these H$\alpha$ filaments are nearly lost in a sea of background H$\alpha$ emission, but are more obvious in [\ion{S}{2]} images where the H~II background emission is substantially fainter. 

\begin{figure*}[ht]
\begin{center}
\includegraphics[angle=0,width=18.0cm]{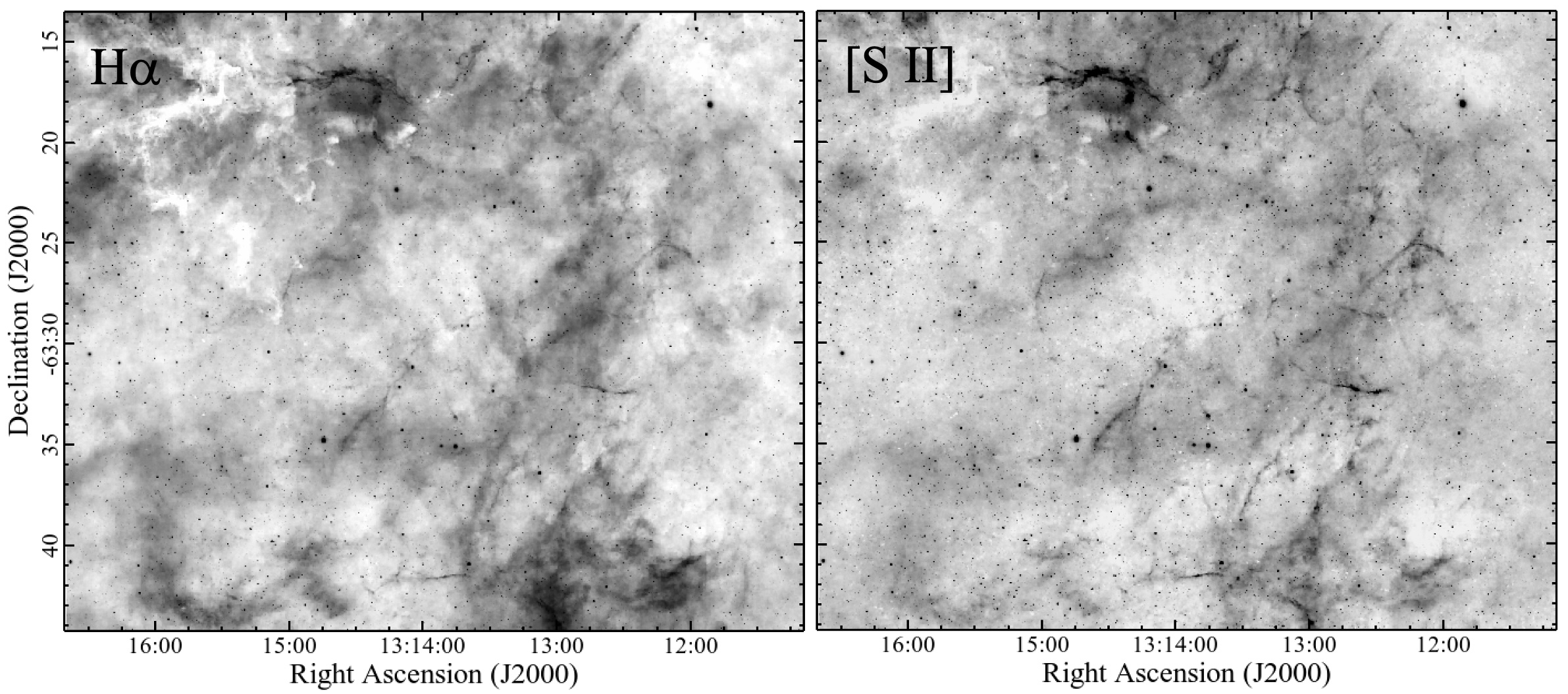} \\
\includegraphics[angle=0,width=17.5cm]{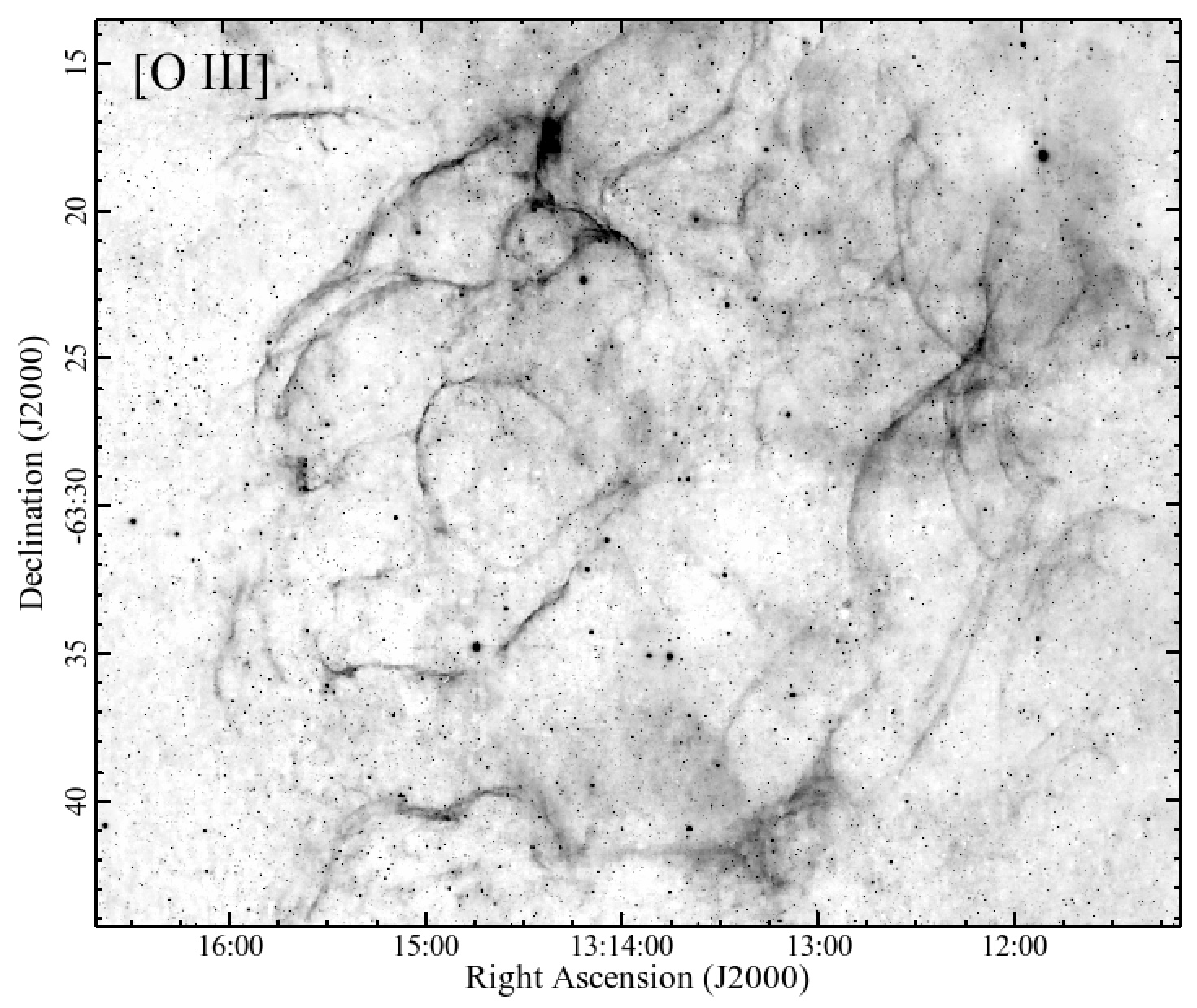} \\
\caption{Deep H$\alpha$, [\ion{S}{2}]
and [\ion{O}{3}] images of SNR candidate G305.4-0.7.
\label{G305_2} 
} 
\end{center}
\end{figure*}

These H$\alpha$, [\ion{S}{2} and \O3 images, 
covering the same exact region,
are shown in Figure~\ref{G305_2}. A few H$\alpha$ filaments are seen, but are much more visible in the [\ion{S}{2}] image. However, an entirely different appearing nebula is seen in the \O3 image with multiple curved and overlapping filaments nearly filling the image. 
Emission is seen even in the nebula's central regions, an arrangement fairly unusual for SNRs.

\begin{figure*}[ht]
\begin{center}
\includegraphics[angle=0,width=18.0cm]{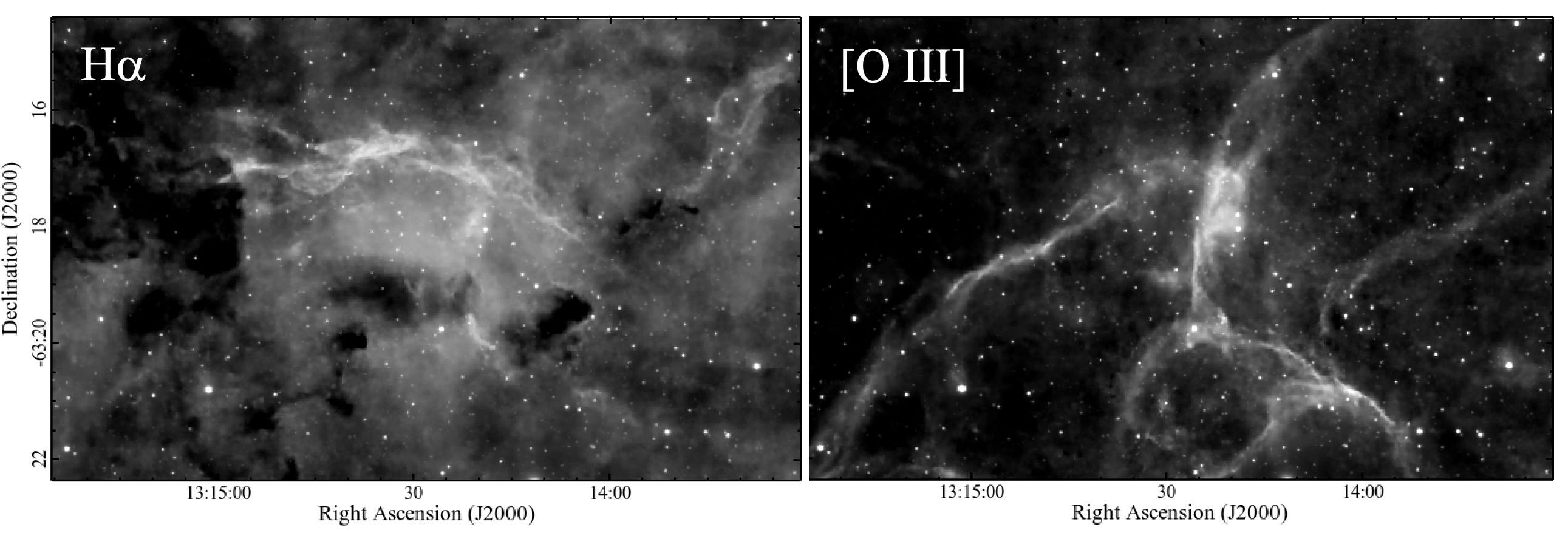} \\
\includegraphics[angle=0,width=18.0cm]{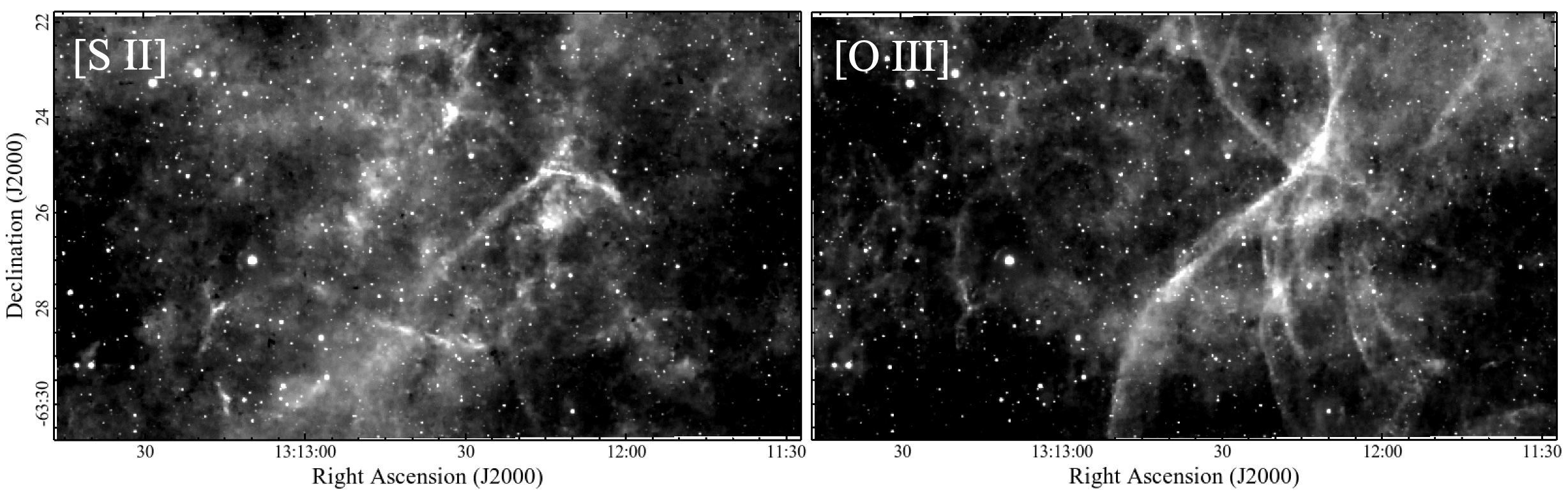} \\
\caption{Comparison of G305.4-0.7's northern (top) and western (bottom) filaments in
H$\alpha$, [\ion{S}{2}] and [\ion{O}{3}] line emissions.  
\label{G305_3} 
} 
\end{center}
\end{figure*}

The nebula's poor correlation between its H$\alpha$ and \O3 emissions is highlighted in the top panel of Figure~\ref{G305_3} where the nebula's brightest H$\alpha$ filament along its northern section shows almost no correspondence in \O3 emission. This raises the question of whether these northern H$\alpha$ are even part of the \O3 emission nebula. However, much better emission agreement is seen in the figure's lower panels where the nebula's western [\ion{S}{2}] and \O3 emissions are compared side by side and show some emission filament correlations.

Without a deep \O3 image of this region, the few short H$\alpha$ and [\ion{S}{2}] filaments seen here would have attracted little attention
in even deep optical SNR surveys. Indeed, this nebula's H$\alpha$ filaments were missed in prior AAO/UKST H$\alpha$ SNR surveys 
\citep{Walker2001,Stupar2008} 
and are virtually invisible in a low resolution published image of this region  (\citealt{Stupar2010}; their Fig.\ 1).

It is this nebula's striking \O3 filamentary structure that
makes this nebulosity such a strong SNR candidate. Although there are a few O,B stars lying in the nebula's direction
(HD 114653: B8 III; HD 114800: B2 III/V; 
HD 114886: O9-B2/V)
the nebula's filaments do not seem to originate from any single point or appear related to any of these stars.
With its multiple and highly curved filaments distributed throughout its structure it closely resembles the well-known remnant S147 (G180.0-1.7) 
\citep{vdb1973,Lozinskaya1976,Furst1986,Ren2018,Greimel2021}. 

Although we view the G305.4-0.7 nebula as a good SNR candidate, it was not detected in the radio in the recent SNR survey using the Australian Square Kilometre Array Pathfinder (ASKAP) 
which covered the G305.4-0.7 region \citep{Ball2025}. 
On one hand, the radio emission structure around
G305.4-0.7 is fairly complex as can be seen in 
in Figure 1 of \citet{Filipovic2025}
which shows an ASKAP 943.5 MHz radio-continuum image of the surrounding environment of G305.4-0.7. But some close neigboring SNRs including two known SNRs (G304.6+0.1 and G306.3-0.9) and three new candidate SNRs 
(G304.2-0.5, G304.4-0.2, and G306.2-0.8)
were detected in this survey.
However, as in the case of G239.9+7.0, some optically confirmed SNRs whose optical emission is dominated by \O3 emission have gone undetected in the radio  \citep{Fesen2024}. Hence, a lack of confirming nonthermal radio emission is not decisive here. In any case, a definitive answer to  G305.4-0.7's nature will require spectral analysis of its optical emission filaments.

\section{Known SNRs}

\subsection{G206.6+6.1 
(aka G206.7+5.9)}

Although not a new SNR, our optical images of G206.6+6.1, especially in \O3,
show it has considerable fine scale details making its overall structure visually striking compared to published radio maps of it.
We have chosen to
include it here for completeness of our survey of 
remnants projected within the boundary of the Monogem SNR. 
The history of its discovery is also a case
of researchers working in different wavelengths 
unaware of each others work on
the same object.

\begin{figure*}[h]
\begin{center}
\includegraphics[angle=0,width=13.5cm]{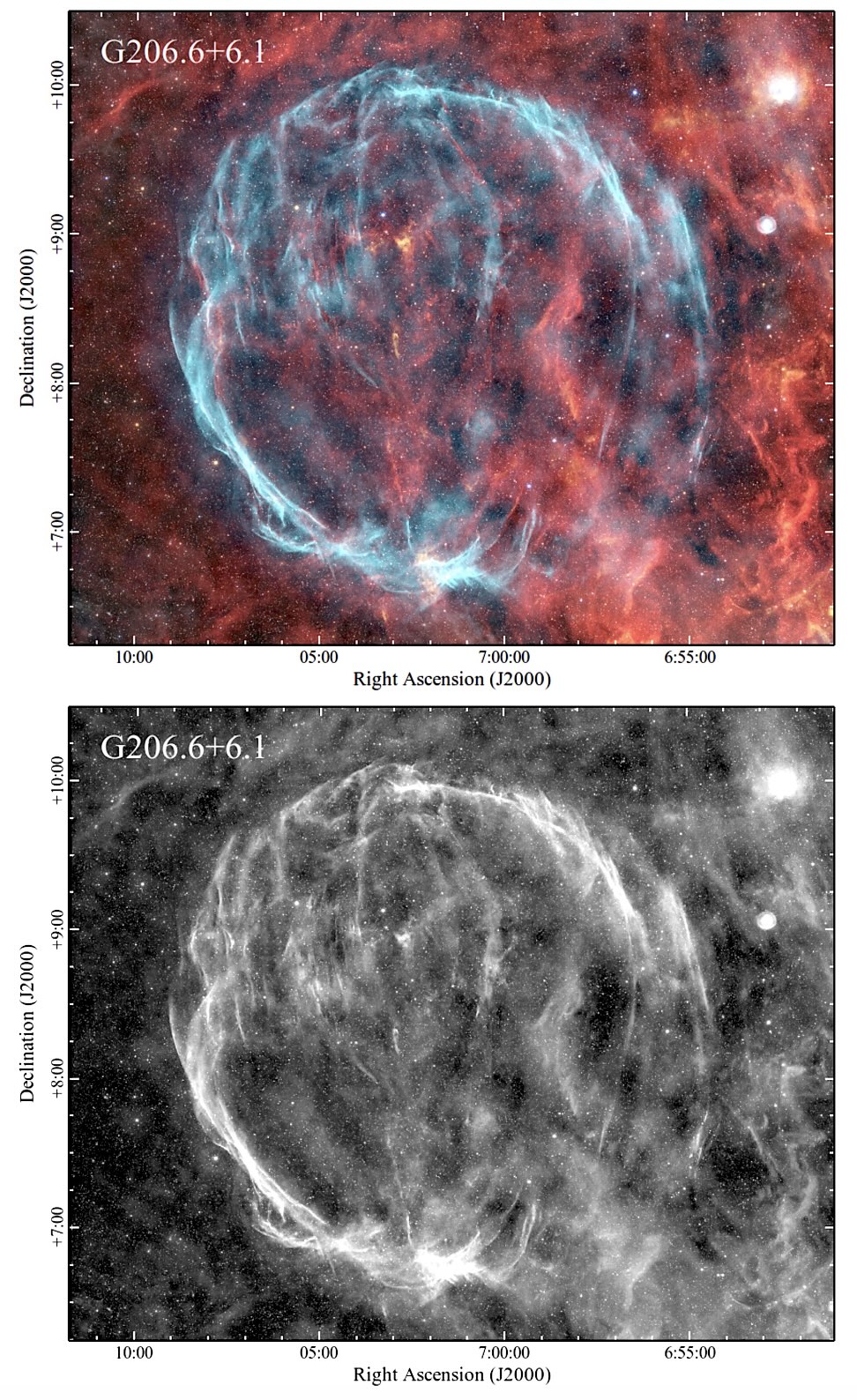} 
\caption{Top: A false-color  composite image of the 3.6$\degr$ diameter G206.6+6.1 SNR composed of [\ion{O}{3}] (blue), H$\alpha$ (red). Bottom: A black and white and higher contrast version. 
\label{G206_1} 
} 
\end{center}
\end{figure*}

\begin{figure*}[ht!]
\begin{center}
\includegraphics[angle=0,width=17.0cm]{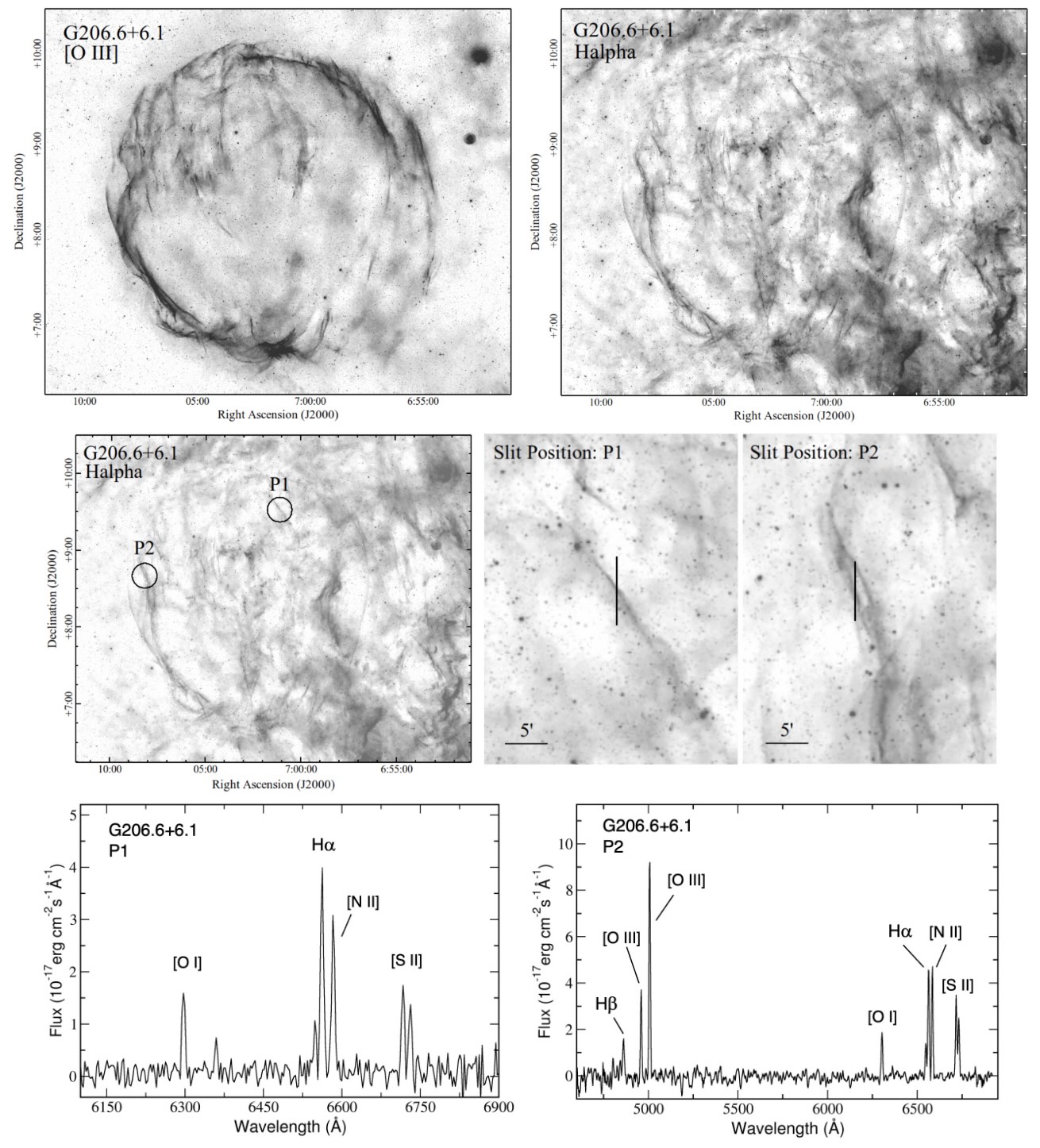} 
\caption{Top Panels: Separate [\ion{O}{3}]  and H$\alpha$ images of G206.6+6.1.
Middle Panels: Marked slit locations in G206.6+6.1 (right) along with   enlargements of slit positions.  
Bottom: Resulting low-dispersion spectra for the two slit positions.
\label{G206_2} 
} 
\end{center}
\end{figure*}

\begin{figure*}[ht!]
\begin{center}
\includegraphics[angle=0,width=17.5cm]{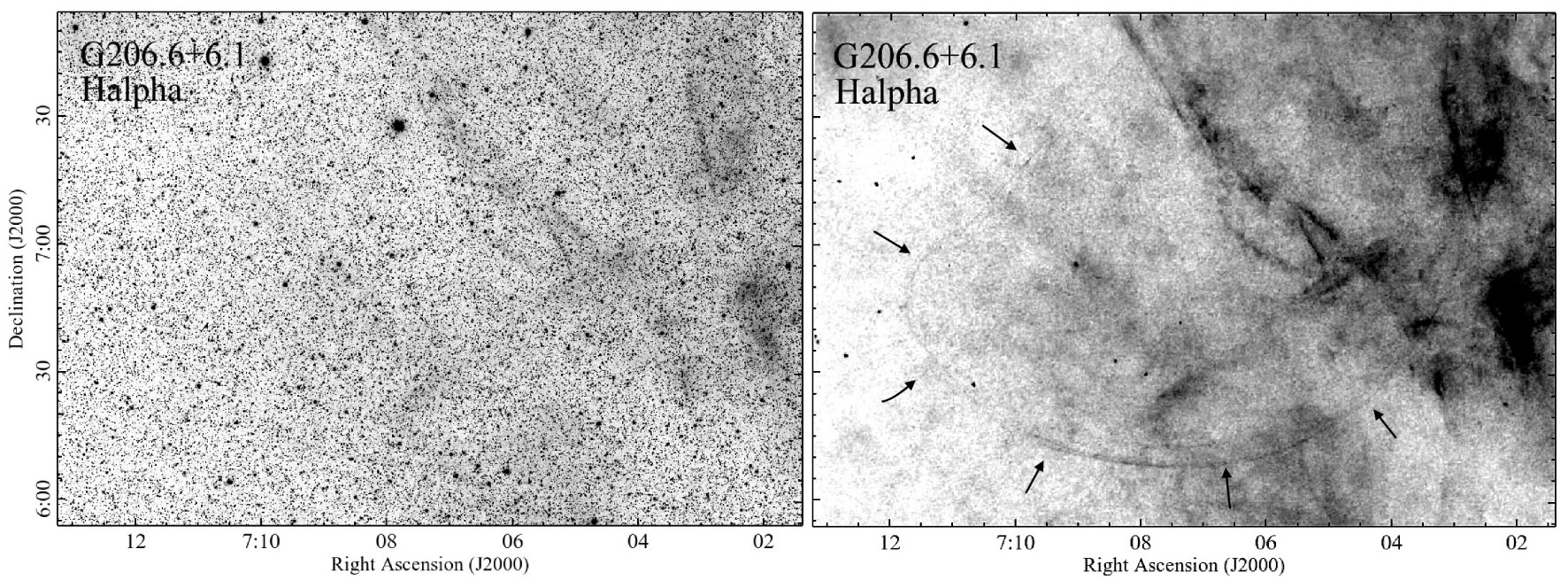} 
\caption{A shock blowout? Star and starless H$\alpha$ images of the southeastern portion of G206.6+6.1. A loop of thin  H$\alpha$ filaments are marked by arrows in the right panel. These filaments are not detected in  our \O3 images.
\label{G206_3} 
} 
\end{center}
\end{figure*}

\begin{figure}[h!]
\begin{center}
\includegraphics[angle=0,width=8.4cm]{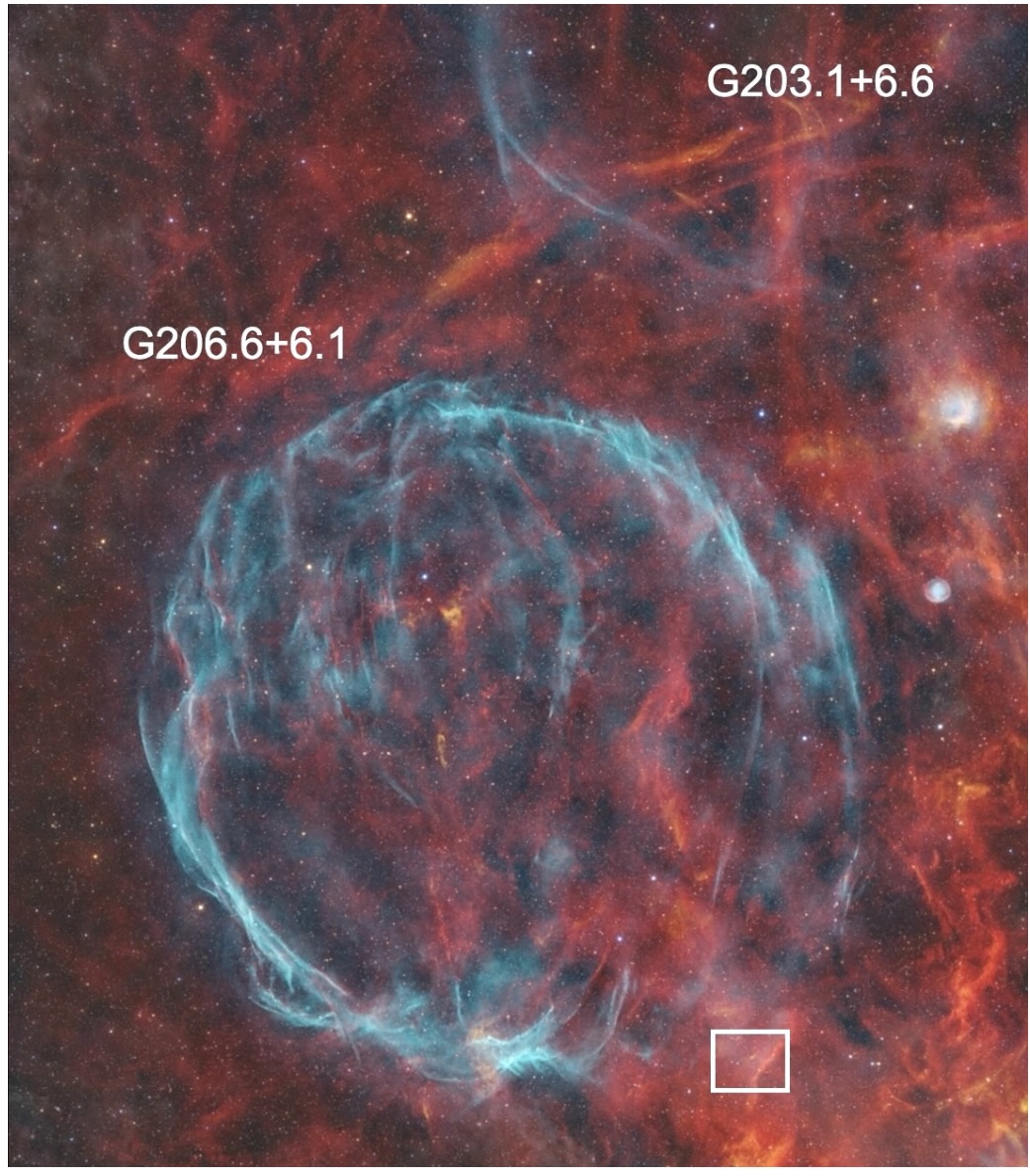}  
\caption{A false-color composite image of the G206.6+6.1 SNR with the small white box indicating the area investigated by 
\citet{Aktekin2025}. Image also shows the southern portion of the neigboring G203.1+6.6 SNR.
\label{G206_4} 
} 
\end{center}
\end{figure}

In a review article on radio emissions from SNRs,  \citet{Reich2002} reported finding 
two low-surface brightness SNR candidates at 1.4 GHz located toward the direction of the Galactic center. These objects were relatively large with diameters of 1.6 and 3 degrees and estimated surface brightnesses of $\Sigma_{\rm 1Ghz}$ = 2.5 and 3.8 $\times 10^{-23}$ 
W m$^{-2}$ Hz$^{-1}$ sr$^{-1}$. 
Although they appeared as limb-brightened shell structures,
the narrow-band total-intensity data plus the lack of any polarization measurements prevented identifying these two objects as SNRs.

These radio emission shells were later observed by
\citet{Gao2022} using the Five-hundred-meter Aperture Spherical radio Telescope (FAST). They found both shells exhibited a non-thermal synchrotron spectra and named them G203.1+6.7 and G206.7+5.9 as confirmed large SNRs. The larger SNR, G206.7+5.9, some 3.5$\degr$ in size showed common SNR bilateral shell morphology, and a double shell on one side. 
Based on morphological correlation between the radio continuum emission and the H~I structures, they estimated a kinematic distance to G206.7+5.9 to be about 440 pc.

Before publication of this 2022 paper, these remnants were 
essentially unknown to both professional and amateur 
astronomers as they were
not listed 
in either the 2014 or 2019 catalog of Galactic SNRs compiled by David Green (\citealt{D_Green2014,D_Green2019}).
Consequently, when a French amateur astrophotographer, L.\ Huet, reported finding a 3 degree diameter \O3 bright 
nebula in 2019
in the anticenter direction (not realizing its coordinates were virtually identical to the radio detected shell G206.7+5.9) he posted it as
a possible nearby planetary nebula or SNR
under the name 
Hu6\footnote{$https://planetarynebulae.net/EN/page_np.php?id=83$}.

In the subsequent years after this optical discovery, several other amateurs attempted to obtain better images of it which could lead to  optical spectra and hence clarification of its nature.
Only when deep H$\alpha$ and \O3 images were taken in 2024 by B.\ Falls 
and a team led by T.\  Schaeffer  was its 
highly filamentary nebula finally seen
which facilitated follow-up spectral observations. These images are the ones we highlight in the following figures and discussion.
Note: Our images indicate the remnant's center of its optical emission
is closer to G206.6+6.1 than the radio determined Galactic coordinates and we will use our revised name for this SNR in the discussion below.

The upper panel of Figure~\ref{G206_1} shows a false color image of G206.6+6.1 composed of our \O3 and H$\alpha$ images colored blue and red, respectively. This image highlights the remnant's dominate \O3 emission relative to that of H$\alpha$ emission. Although there is considerable
H$\alpha$ emission here, it is largely diffuse and quite unlike the remnant's highly filamentary \O3 emission. The remnant's overall
structure is even more apparent in the black and white version seen in the lower panel.

The two upper panels of Figure~\ref{G206_2} shows a comparison of the G206.6+6.1 remnant's \O3 and H$\alpha$ emissions. While there are a few long, curved
H$\alpha$ filaments visible, some of these get lost in the extensive non-filamentary H$\alpha$ emission features seen here. It was this lack of many bright and clearly associated remnant filaments  from low resolution and early discovery images that delayed us from obtaining follow-up spectra. 

The middle left hand panel of this figure shows the location of two H$\alpha$  filaments where spectra were taken; the two other middle panels are blow-ups showing precise slit locations, while the two lower panels present the resulting spectra. Both filaments observed exhibited classic shock-like spectra with strong [\ion{S}{2}] and [\ion{O}{1}] line emissions.  We did not detect H$\beta$ for Filament 1 but did so for Filament 2 where we estimate an $E(B-V)$ reddening value of 0.12 mag.

Interestingly, our H$\alpha$ images also detected 
a very faint loop of thin filaments located off the
southeastern limb of G206.2+6.1. These are shown in the right panel of Figure~\ref{G206_3}. These filaments are not detected in our \O3 images.
This fact and their very thin morphology are suggestive of a
nonradiative shock. Although they could possibly represent a separate SNR, we view it  more likely that these filaments mark a shock  blowout region off G206.6+6.1.

Finally, \citet{Bakis2025a} recently presented low-resolution SHASSA and VTSS 
H$\alpha$ and [\ion{S}{2}] images of a
region off to the southwest of the remnant's \O3 filaments. They report finding [\ion{S}{2}]/H$\alpha$ flux ratios
from 0.51 to 1.15 with an average of 0.74, and clearly
above the 0.4 value indicating the presence of shocks.  They also presented
LAMOST spectra for four regions which also support the shock nature of the observed emission.

However, the region where they took those data is a small area located along  G206.6+6.1's far southwest edge containing only diffuse H$\alpha$ emission.
This is shown as a small white box in Figure~\ref{G206_4}.
Since no filaments were observed here unlike the majority of the remnant's optical emission, there is some  uncertainty if these data represent G206.6+6.1's properties or if the emission they observed is some other shocked material in the direction of the Monogem Ring.


\begin{figure*}[ht]
\begin{center}
\includegraphics[angle=0,width=17.0cm]{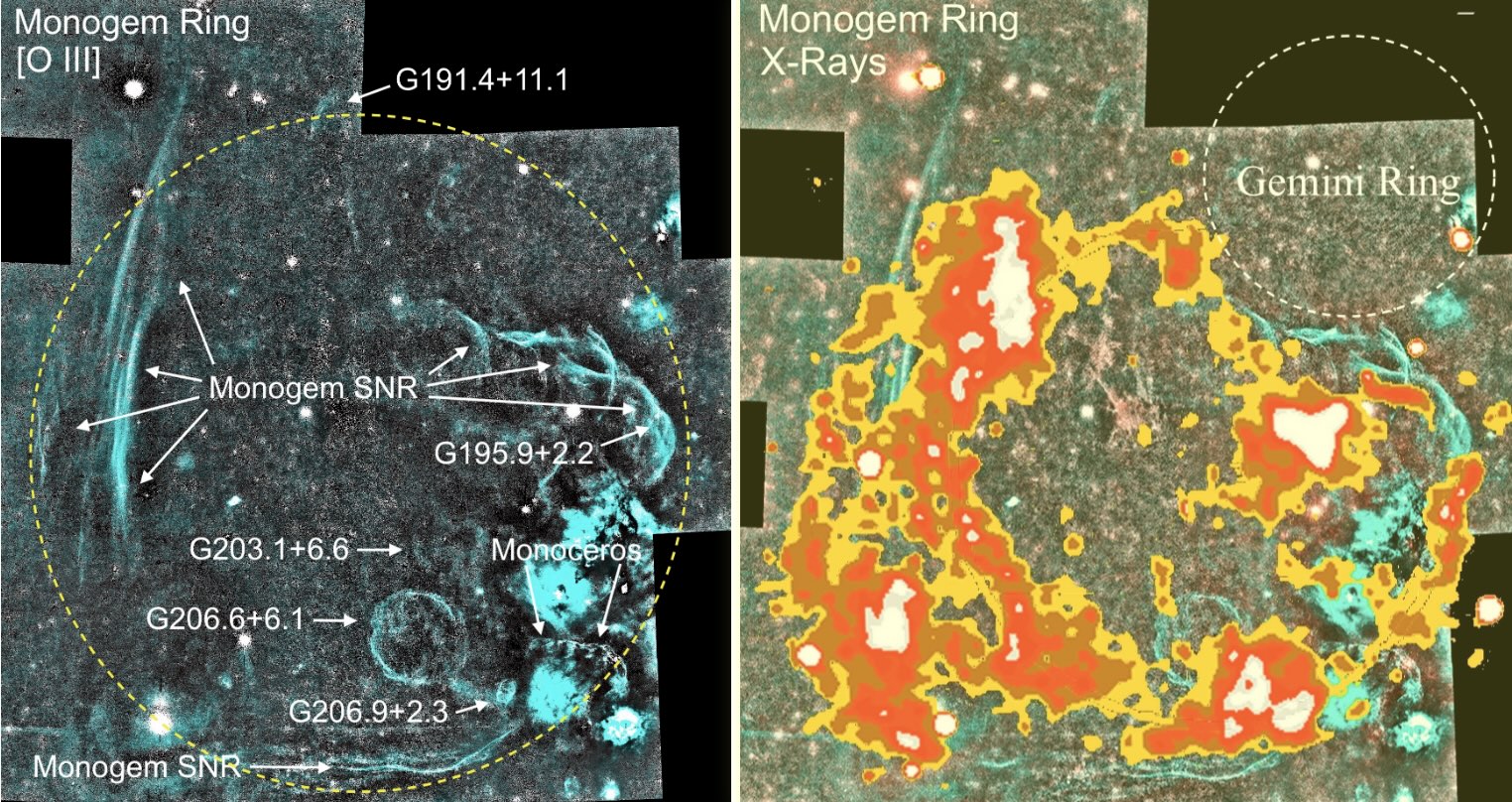} 
\caption{Left: Mosaic of \O3 images of the Monogem Ring. 
 North is up, east to the left.
The dashed yellow circle represents the remnant's eROSITA's  detected 24$\degr$ diameter shell centered at
RA = 107.08$\degr$, Dec = 13.5$\degr$ \citep{Knies2024}. Emission from three known plus two new SNR (G191.4+11.1 and G195.9+2.2) are seen projected toward the Monogem SNR. 
Optical \O3 filaments believed associated with the Monogem SNR's are indicated.
Right: Composite of ROSAT detected Monogem SNR's X-ray emission \citep{Plucinsky1996} overlayed onto our \O3 mosaic.
\label{MG1} 
} 
\end{center}
\end{figure*}

\begin{figure*}[ht!]
\begin{center}
\includegraphics[angle=0,width=17.0cm]{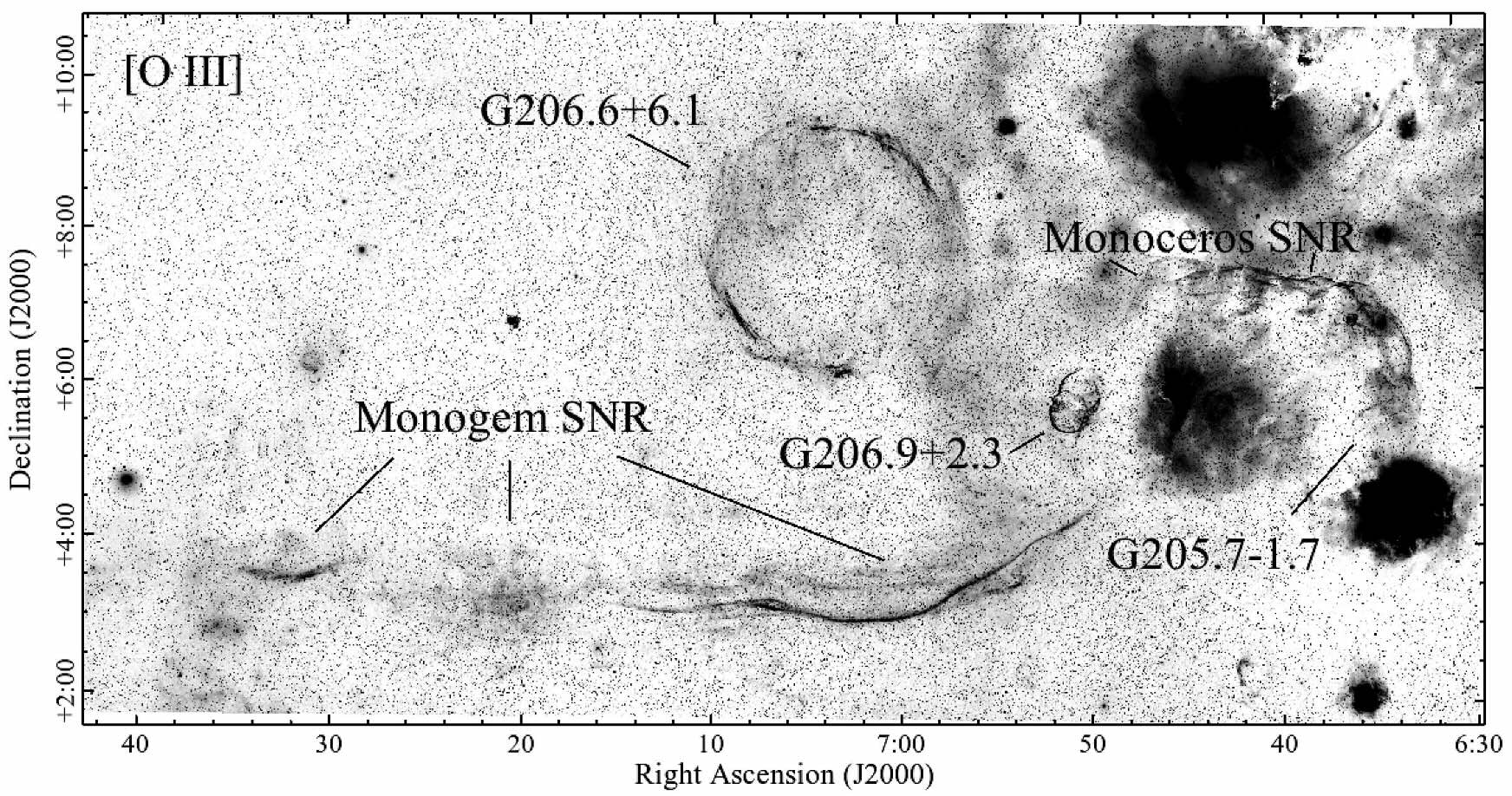} 
\caption{A $9.2\degr \times 18.5\degr$ image mosaic of the southern limb of the Monogem SNR showing a line of \O3 filaments along its southern boundary. Emission from several other SNRs are labeled. The star at the far  lower left is Procyon ($\alpha$ Canis Minor).  
~~~~ 
\label{MG_SW} 
} 
\end{center}
\end{figure*}

\subsection{The Monogem Ring SNR}

A nearly 25 degree diameter soft X-ray emission feature called the Monogem Ring (G203+12)
or the Gemini-Monoceros X-ray enhancement is one of the most striking objects along the Galactic plane in X-rays.  First investigated as a strong Galactic source of soft X-rays during the 1970's \citep{Bunner1971, Bunner1973, Long1977}, its ring like structure was only realized through $HEAO-1$ data
\citep{Nousek1981}. Estimated to lie at a distance of around 300 pc with a diameter of 130 pc, it is thought to be a very old (80,000 - 120,000 yr) single SNR called the Monogem Ring SNR or simply the Monogem SNR. The remnant is located in a region of unusually low interstellar density  $\sim 5 \times 10^{-3}$ 
cm$^{-3}$  \citep{Cordova1989, Plucinsky1996, Knies2024}
and may still be in the adiabatic expansion phase with an estimated initial explosion energy of $0.2 - 1.0 \times 10^{50}$ erg \citep{Plucinsky1996, Knies2024}. 

A recent study of the Monogem SNR using eROSITA X-ray data has, however, found evidence for a significant temperature enhancement in the ring's southeast section suggestive of a second SNR with a younger age $\sim$50,000 yr \citep{Knies2024}. In addition, 
a $\sim6\degr$ diameter SNR candidate, G190.4+12.5, was identified along the ring's northern limb, which we have already discussed above in $\S5.1$.

Despite its enormous angular size and and its relatively close proximity, very little in the way of optical emission associated with the Monogem remnant has been previously reported. These consist of two isolated filaments located around the Monogem's western and southern periphery \citep{Reimers1984, Weinberger2006}, and a $4.5\degr$ long H$\alpha$ seen toward the Monogem's center that was found via depolarized 11 cm radio maps and is suspected to be physically associated with the remnant \citep{Reich2020}.

Reasons for so few previously detected optical emission features for the Monogem SNR include potential confusion with bright neighboring emission nebulae like NGC 2264 and the Monoceros SNR, plus
the Monogem's enormous angular size preventing a comprehensive optical emission-line survey using imaging systems with fields of view less than one degree. 
The remnant's expansion into an
interstellar region of 
exceptionally low density based on analysis of its X-ray properties \citep{Plucinsky1996} also led to
an expectation of little associated optical emission \citep{Weinberger2006}.

However, deep wide field optical emission-line imaging such as the MDW H$\alpha$ survey \citep{Aftab2026}, the Northern Sky Narrowband Survey (NSNS)  \citep{Ziegenbalg2025}, and the wide-field amateur images presented here has now made it possible to more fully investigate the Monogem SNR's optical emissions.
These wide FOV optical images have revealed considerable \O3 and H$\alpha$ remnant related filaments along much of its periphery. 

With the exception of the recently available on-line NSNS images \citep{Ziegenbalg2025}, most of these shock features were not known previously. Below we present wide mosaic images of this SNR in both \O3 and H$\alpha$ emission lines, beginning first with its extensive and impressive \O3 emission filaments.

\subsubsection{The Monogem's \O3 Emission}

The presence of large \O3 filaments around portions of the Monogem SNR's rim can be seen in the left hand panel of Figure~\ref{MG1} which shows a background subtracted $28\degr \times 35\degr$ mosaic made from over 50 \O3 images with most stars digitally removed. The dashed yellow circle marks the approximate size and location of the Monogem SNR as defined eROSITA X-ray data \citep{Knies2024}.
Long \O3 emission filaments are seen along much of the Monogem Ring's eastern and southern limbs,
with shorter and more curved \O3 filaments located along the ring's northwestern rim.

Besides emissions from the Monoceros SNR, NGC~2264 and the Rosette Nebula situated along the image's southwest, this figure shows \O3 emission from 
four known SNRs: G203.1+6.6, G206.6.+6.1, the Monoceros SNR, and G206.9+2.3.
In addition, as we discussed above in 
$\S4.4$ and $\S5.1$, there are two new confirmed or candidate SNRs here as well, namely G191.4+11.1 lying far to the north and G195.9+2.2 largely hidden by Monogem's bright northwestern \O3 filaments.

\begin{figure*}[ht!]
\begin{center}
\includegraphics[angle=0,width=18.0cm]{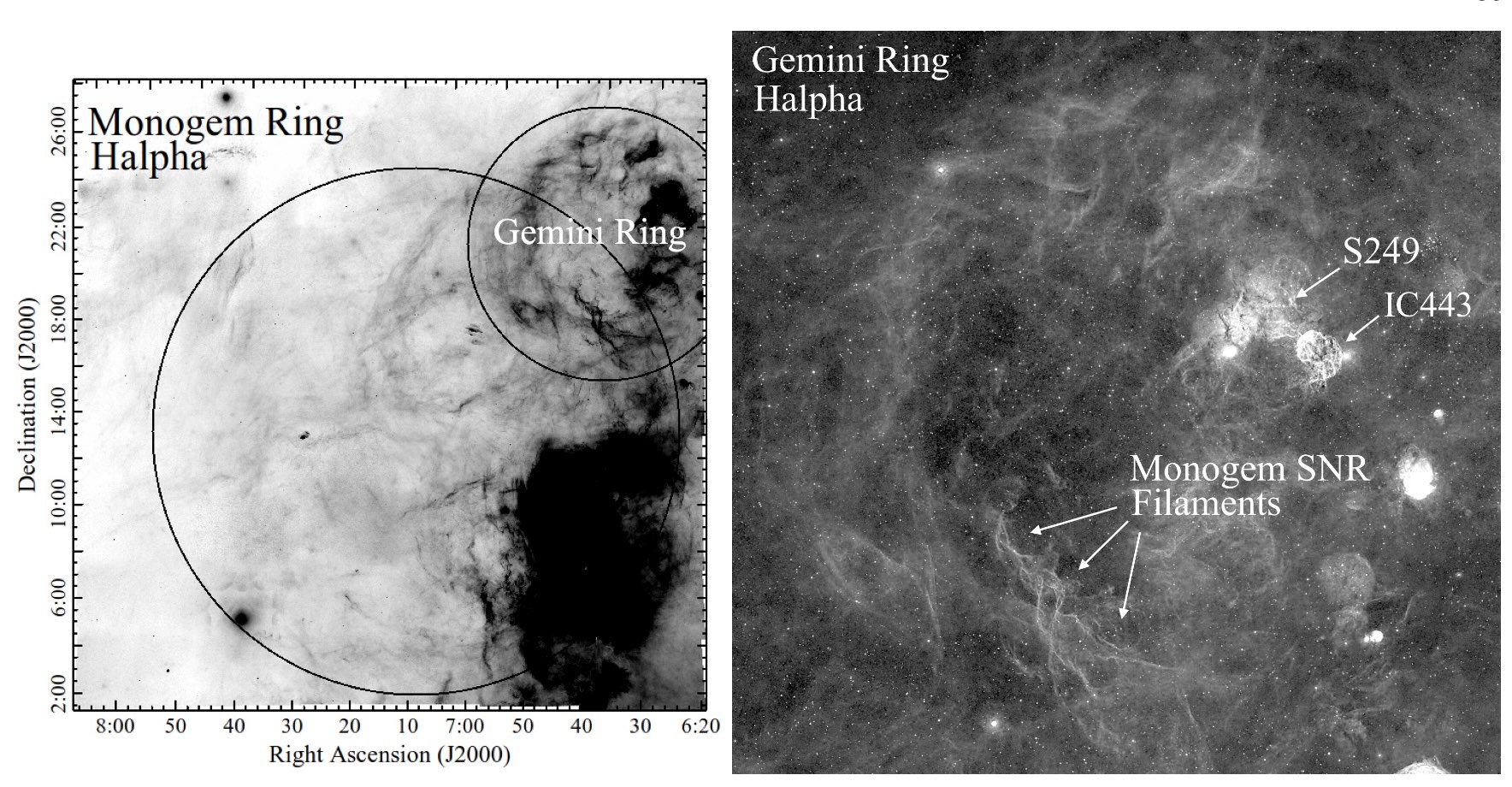} 
\caption{Left: H$\alpha$ mosaic of the Monogem Ring which is indicated by
the large black circle (dia = 23.6$\degr$) centered as seen in eROSITA data \citep{Knies2024}. The smaller black circle indicates the Gemini H$\alpha$ Ring.
Right: NSNS H$\alpha$ image of the Gemini Ring with SNR IC~443, the H~II region S249, and suspected
Monogem SNR shocked filaments marked.
\label{Gemini} 
} 
\end{center}
\end{figure*}

Several large \O3 emission filaments, marked here in this figure by arrows, are arranged along the Monogem's periphery leading us to believe these are related to the remnant. The largest of these are the nearly vertical 15 degree long group of \O3 filaments situated along the Monogem's eastern and northern limbs. A bit farther east of these bright filaments are shorter and much fainter [\ion{O}{3}] filaments. As can be seen in the Figure~\ref{MG1}'s right hand panel, these \O3 filaments lie at the easternmost extent of the Monogem's soft X-ray emission.

Strong \O3 filaments are also seen in the south where we find a line of filaments lying just outside (south) of Monogem's X-ray southernmost emission.
A portion of these filaments can be better seen in 
Figure~\ref{MG_SW} where the \O3 emission here consists of several overlapping filaments. While these \O3 filaments end before reaching the Monoceros SNR, they extend several degrees farther to the east (see Fig.~\ref{MG1}).
Note: The new SNR candidate G205.7-1.7 is marked here which was discussed in 
$\S5.2$.

Optical and X-ray emission coincidences can also be seen in the Monogem's northwest where a group of bright and highly curved \O3 filaments border the remnant's northwestern X-ray emission. This region is where the Monogem SNR is thought to have collided with the 
so-called Gemini H$\alpha$ Ring
(see also Figs.~\ref{Gemini} and \ref{NSNS_NW}).
Located 
at (l, b) $\sim$ ($191.5\degr, +5.0\degr$) this $5.7\degr$ diameter ring of H$\alpha$ emission lies at
the large indentation in 
the Monogem's northwestern X-ray emission.

The Gemini H$\alpha$ Ring was discovered by \citet{Kim2007} in a UV emission-line study of the Monogem Ring  where they detected very strong 
\ion{C}{4} $\lambda\lambda$1548,1550 emission at its southeastern overlap with the 
Monogem Ring.
The Gemini H$\alpha$ Ring is likely a 
stellar-wind bubble due to several
OB-type stars lying in its direction \citep{Knies2018}.
With an estimated distance of $\sim$ 200 - 350 pc similar to that of the Monogem Ring, this suggests its direct contact with the Monogem's hot gas, leading to the observed Monogem's NW indented and distorted X-ray ring morphology.
The curved \O3 filaments seen here are likely a sign of this interaction.

\begin{figure*}
\begin{center}
\includegraphics[angle=0,width=16.0cm]{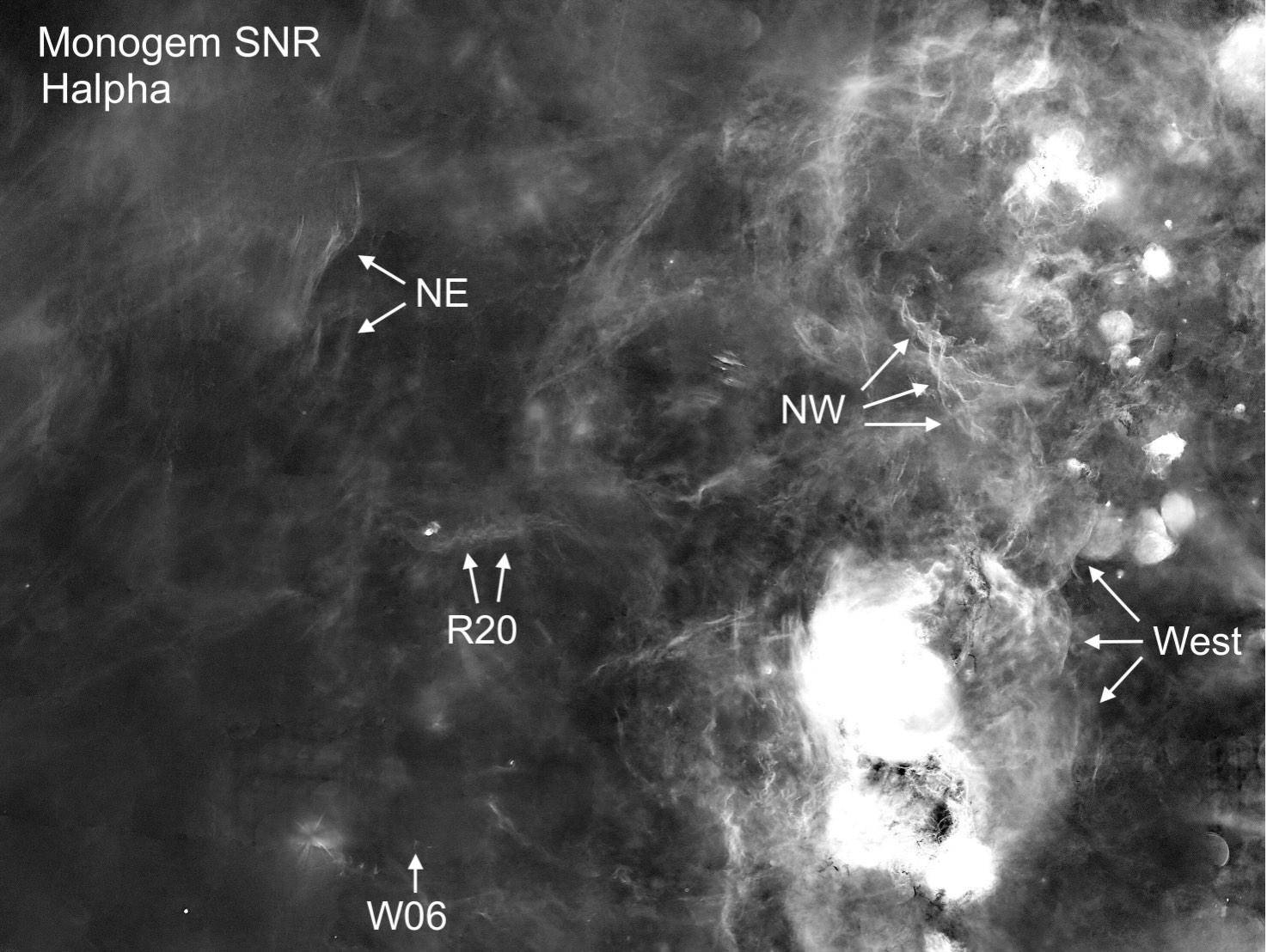}
\caption{A starless MDW H$\alpha$ mosaic image of the Monogem Ring SNR with newly identified NE, NW and west filaments indicated. 
Also indicated are a filament, W06, discovered by \citet{Weinberger2006} and a central filament, R20, discussed by \citet{Reich2020}.
\label{MG_Sean} 
} 
\end{center}
\end{figure*}

\subsubsection{The Monogem's H$\alpha$ Emission}

In addition to its extensive \O3 filaments, the Monogem SNR exhibits considerable H$\alpha$ filaments. Figure~\ref{MG_Sean} shows a very wide MDW H$\alpha$ mosaic of the whole Monogem Ring region with remnant related H$\alpha$ filaments marked, including two previously known 
filaments; W06 \citep{Weinberger2006} and R20 \citep{Reich2020}.

Among the most notable examples are H$\alpha$ filaments in located along the southeastern rim of the Gemini H$\alpha$ Ring and coincident with the bright \O3 filaments just discussed above.
Some of the brighter H$\alpha$ Monogem SNR filaments seen in the Gemini Ring are marked in the right hand panel of 
Figure~\ref{Gemini}\footnote{Although discovered by \citet{Kim2007} and studied by \citet{Knies2018}, neither paper included a deep optical image of it.}.

\begin{figure*}
\begin{center}
\includegraphics[angle=0,width=16.0cm]{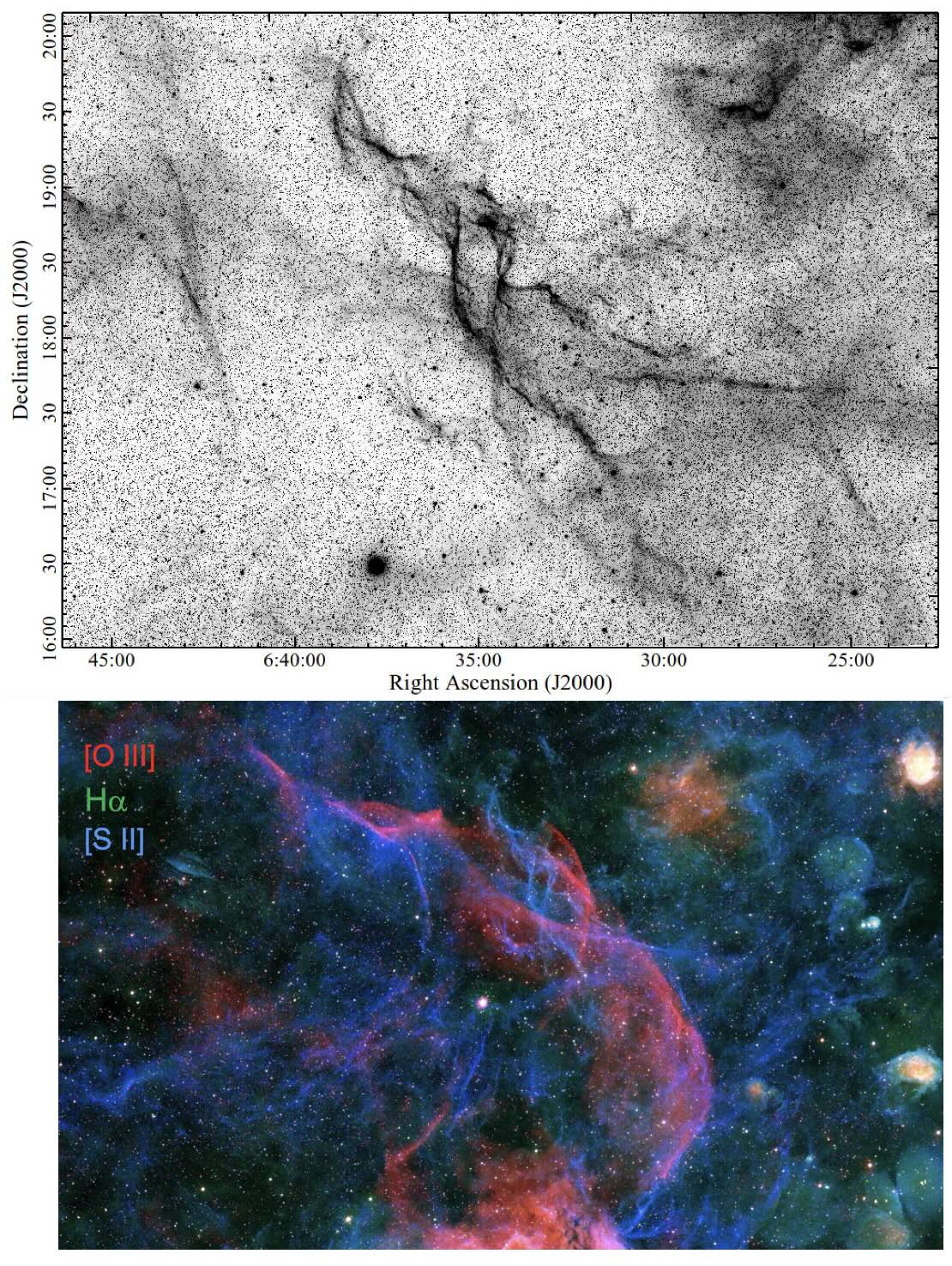} 
\caption{Top: Twisted
H$\alpha$ filaments in the Monogem  Ring - Gemini Ring interaction region.
Bottom: NSNS color image of Monogem's NW filaments. 
Note: The blue twisted [\ion{S}{2}] filaments lie largely behind (to the left) of the red \O3 filament. 
\label{NSNS_NW} 
} 
\end{center}
\end{figure*}
\begin{figure*}
\begin{center}
\includegraphics[angle=0,width=14.0cm]{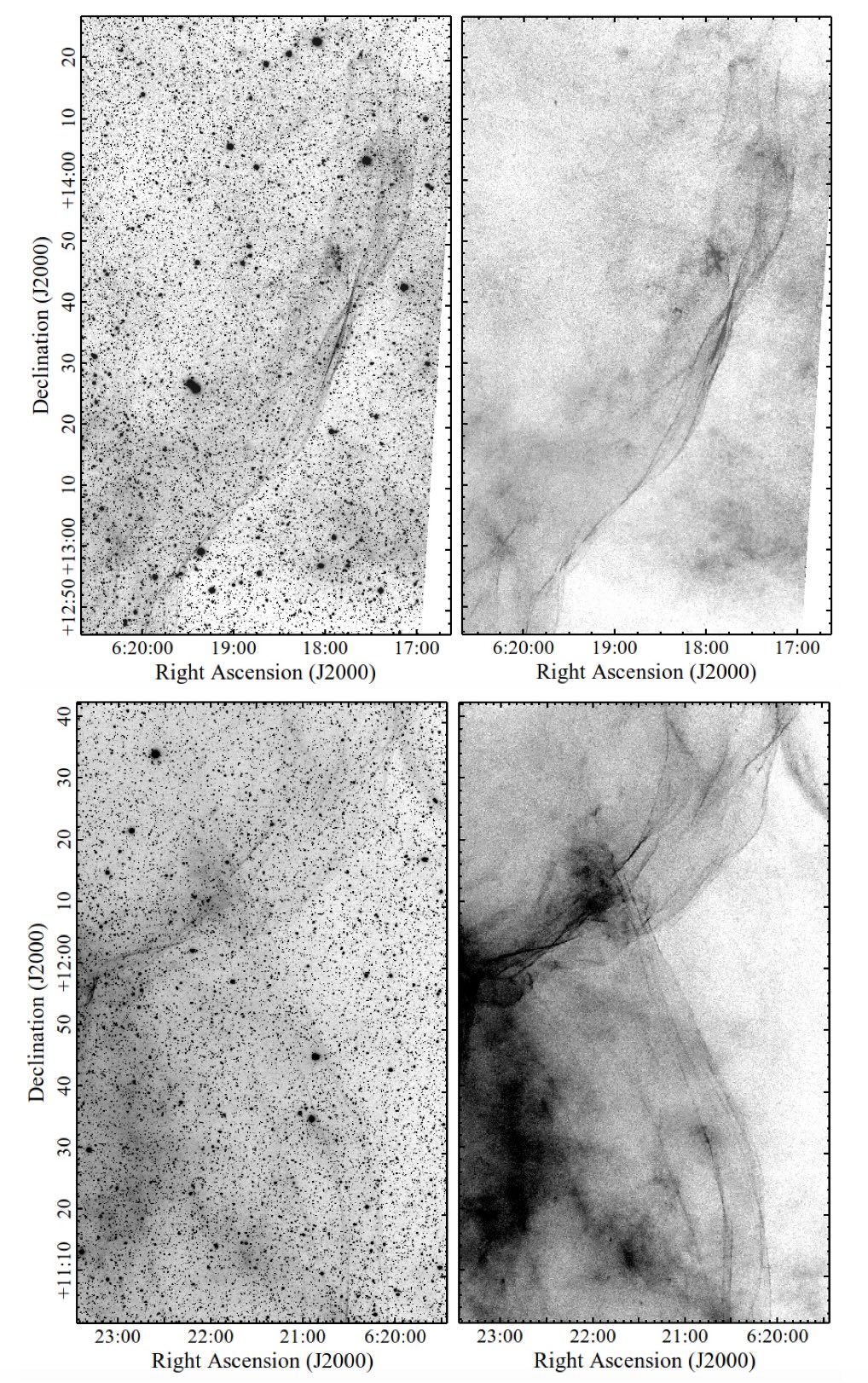} 
\caption{$1\degr \times 1.7\degr$ H$\alpha$ images of Monogem's 
west-central limb with and without stars. 
\label{MG_West_1} 
} 
\end{center}
\end{figure*}

\begin{figure*}
\begin{center}
\includegraphics[angle=0,width=14.0cm]{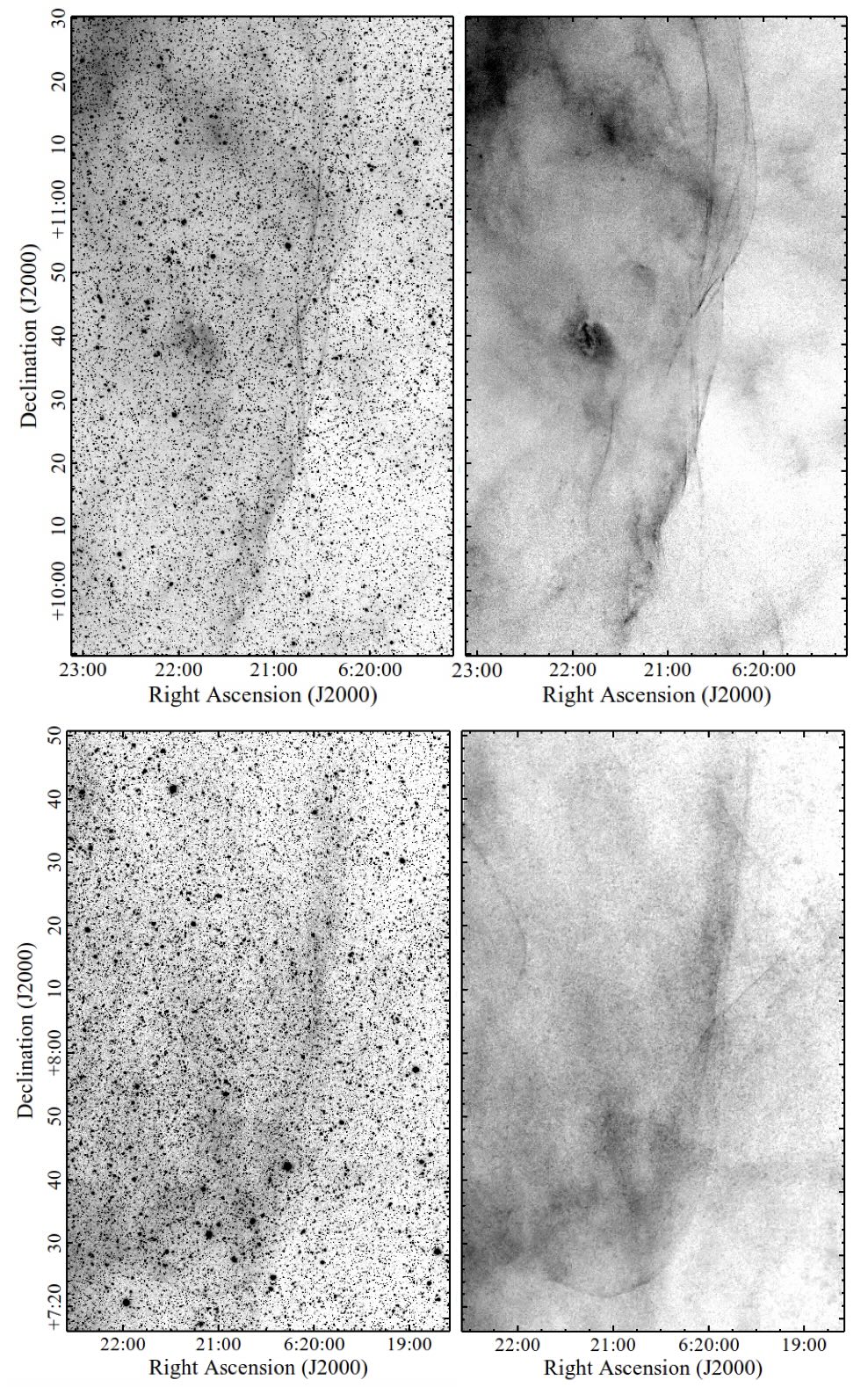} 
\caption{$1\degr \times 1.7\degr$ H$\alpha$ images of Monogem's 
west-central limb with and without stars. 
\label{MG_West_2} 
} 
\end{center}
\end{figure*}
\begin{figure*}
\begin{center}
\includegraphics[angle=0,width=13.5cm]{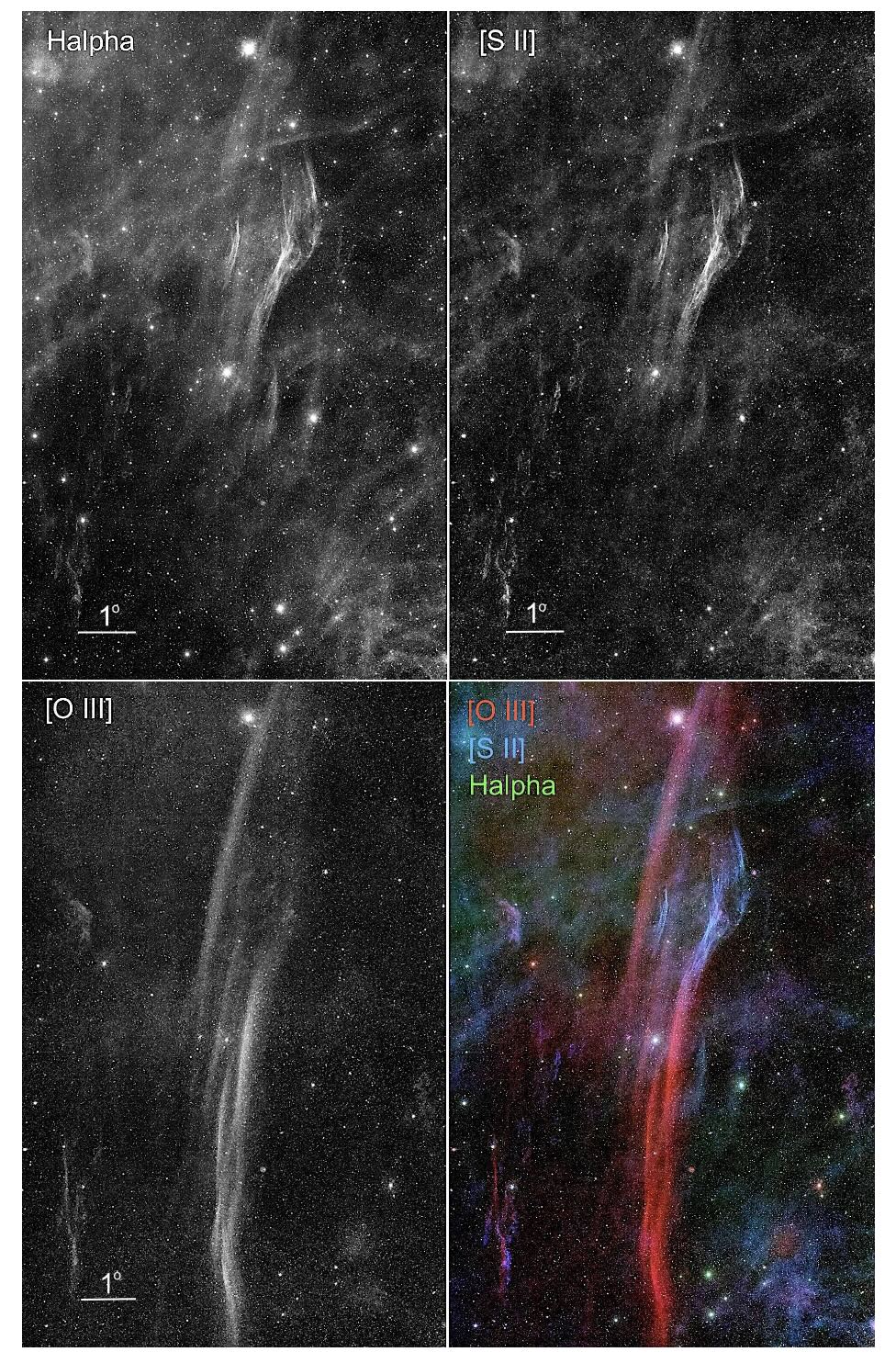} 
\caption{NSNS H$\alpha$, [\ion{S}{2}], and \O3 images of the NE filaments of the Monogem Ring SNR. 
Note how the H$\alpha$ and 
[\ion{S}{2}] filaments intersect with the \O3 bright filaments.
Images are $7.8\degr \times 12.0\degr$ in size. North is up east to the left.
\label{NSNS_NE} 
} 
\end{center}
\end{figure*}
\begin{figure}
\begin{center}
\includegraphics[angle=0,width=8.5cm]{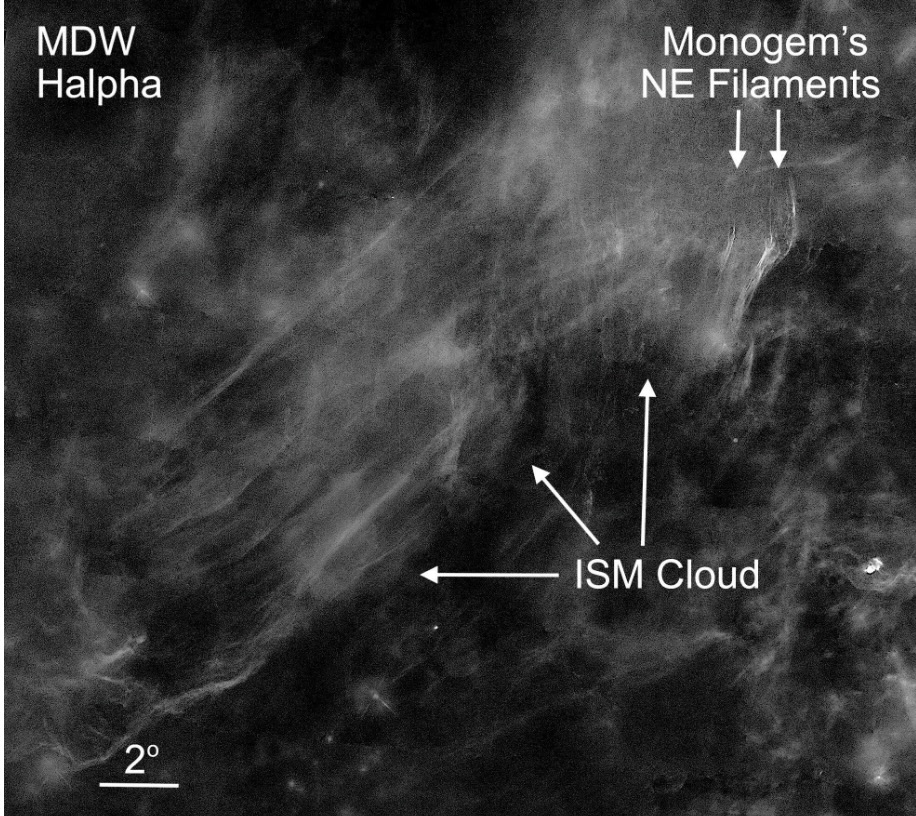}
\caption{A $21.8\degr \times 19.6\degr$ H$\alpha$ image of the region east of the Monogem SNR. North is up, east to the left.
The Monogem's northeastern filaments lie just where its shock has impacted a large ISM cloud that extends over 10 degrees farther to the east.
\label{Sean_FarEast} 
} 
\end{center}
\end{figure}

These bright H$\alpha$ filaments have a tangled or twisted appearance and lie in a region coincident with the bright northeastern \O3 Monogem SNR filaments.
This overlap of \O3 and H$\alpha$ filaments can be seen in Figure~\ref{NSNS_NW}. It is likely that the \O3 filaments are where the Monogem's shock exceeds $\sim$100 km s$^{-1}$, whereas the H$\alpha$ filaments are instances where the shock is slower. The H$\alpha$ filaments here are more clearly seen in Figure~\ref{NSNS_NW}. The \O3 and H$\alpha$ filaments likely
represent a range of shock velocities presumably due to ISM density variations that the Monogem's shock is encountering as it collides
with the Gemini H$\alpha$ Ring.

There are other H$\alpha$ Monogem SNR related  
filaments along the remnant's western limb where our images reveal a long line of very
thin H$\alpha$ filaments.
This is shown in Figures~\ref{MG_West_1} and \ref{MG_West_2} in both star and starless versions
along with WCS coordinates. 
Although also visible in NSNS images, that survey's $10''$ resolution do not reveal their fine scale structure.

These filaments lie several arc minutes west of the Rosette Nebula, the Monoceros SNR, and the H~II region NGC 2264. Although we are not certain they are directly part of the Monogem SNR's shock front and not simply part of the outer fringes of these bright nebulae, their nearly continuous structure over nearly 8 degrees and exquisitely fine filamentary structure suggests a moderate velocity shock over a span greater than the extent of any one of these nebulae. 
These other nebulae also lie 
substantially farther away 
at 1.2 to 2.0 kpc \citep{Zhao2018}
than the Monogem SNR's 0.3 kpc
estimated distance. 
The western position of these filaments also roughly aligns
with the Monogem's western rim.
In addition, their faintness and sharp appearance is also not unlike that seen in shock filaments in other SNRs, such as along the eastern and western boundaries of the Cygnus Loop SNR.

As shown in Figure~\ref{MG_Sean}, there are also a few H$\alpha$ filaments along the Monogem's northeastern (NE) rim. Figure~\ref{NSNS_NE} shows individual NSNS H$\alpha$, [\ion{S}{2}], and \O3 images of these filaments
along with a color composite of these three images. These H$\alpha$ and [\ion{S}{2}] bright filaments line at the northern end of the remnant's \O3 filaments. This suggests that they are different parts of the same shock front, the main difference
being the H$\alpha$ + [\ion{S}{2}] filaments are where the shock has encountered a higher ISM density and hence is slower compared to the higher velocity \O3 filaments which is expanding through a much lower and much larger ISM region.

The relative isolation of these northwestern H$\alpha$ along the whole of the Monogem's eastern limb raises the question of why only there and in just one small region? This maybe answered when we look at H$\alpha$ emission farther to the east. Figure~\ref{Sean_FarEast} shows a H$\alpha$ image 
covering nearly $22\degr \times 20\degr$ both farther east and south. This image shows that the Monogem's NE filaments are located at the western tip of a broad and long H$\alpha$ emitting interstellar
cloud which extends far to the east. The Monogem NE filaments appear to terminate at the south end of this cloud's western tip suggesting the two features are physically linked.

Finally, we note that \citet{Ziegenbalg2025} has estimated emission-line ratios for these NE filaments based on approximate calibrations of his images and his 3.5 nm FWHM wide filters. For the brighter of these 
NE H$\alpha$ filaments he finds a 
I(H$\alpha$)/I([\ion{S}{2}]) ratio around 1.0,  consistent with the presence of shocks.


\section{Discussion}

Over the last two decades, the landscape for deep optical imaging of the night sky has radically changed. The development of affordable large-format detectors and high transmission (T $\geq95$\%) 
filters has led to a revolution in deep and wide field imaging
by both professionals and amateurs.
The ability to combine of literally hundreds of low noise CMOS exposures taken
using small telescopes equipped
with high-throughput filters and sensitive
digital detectors have enabled the detection of many previously unknown Galactic and extragalactic objects and nebulae.

This paper arose out of this new imaging revolution. Deep images obtained by teams of amateurs over the last few years totaling over 200 hours of exposures have revealed five new Galactic supernova remnants. Several of these were found at fairly high Galactic latitudes,
i.e., $|b| < 6\degr$ and hence outside many radio SNR surveys 
\citep{Heywood2022,Dokara2023,Ball2023,Goedhart2024,Ball2025}, although some recent surveys are
changing this \citep{Hurley_Walker2019, Green2025}.

Not all detected new nebulae were followed-up with higher resolution imaging and optical spectra. 
But some that did led to the discovery of  new Galactic remnants and SNR candidates
as described in detail above. In addition, as part of a multi-year campaign of imaging the Monogem Ring remnant, deep wide FOV images have uncovered extensive optical emission filaments associated with this huge remnant which has created a far more complete picture of its widely separated optical emissions and why those emissions are seen where they are. 

\subsection{Optical Imaging for SNR Detection}

For detecting extragalactic SNRs, optical emission-line surveys are much better than either radio or X-rays surveys \citep{Sasaki2020}. But this is not true for Galactic SNRs. Some 95\% of all confirmed Galactic SNRs were first detected at radio wavelengths
\citep{Dubner2015,Dubner2017}. In the few cases where a SNR located far off the Galactic plane is discovered via optical images,  follow-up radio studies are often successful at detecting it; e.g., the recent case of G107.0+9.0 
\citep{Fesen2020,Reich2021}.

However, some optically discovered Galactic SNRs, especially ones whose optical emissions are quite faint and/or dominated by \O3 line emission, apparently can escaped radio detection. An example is G107.7-5.1, 
a large, nearly 3$\degr$ diameter nebula bright in \O3 emission
\citep{Fesen2024}. Although confirmed as a SNR through its optical spectral properties and as a GeV source
\citep{Araya2024,Aktekin2026}, it has not yet been reported as a radio source. 

Several recently discovered SNRs appear much brighter and often more extensive when viewed in \O3 images compared to H$\alpha$ images. This has help make some remnants more  easily visible particularly in complex background/foreground H$\alpha$ emission fields. This is the situation for the remnants G82.2+5.3 and G321.3-3.9 \citep{Fesen2024}. In fact, a few remnants have only been initially detected optically via their \O3 emission. Examples include the remnants G209.9-8.2 and G210.5+1.3 \citep{Fesen2024} and in the new and candidate remnants presented here: G191.4+11.1,  G239.9+7.0, and G305.4-0.7.
Indeed, without deep \O3 images, we would have also missed the new candidate remnant
G205.7-1.7 right next to the bright Rosette Nebula.

This raises the question about just how
rare are such \O3 emission dominate remnants and thus how complete are past SNR surveys that only used H$\alpha$ emission images for remnant detection (e.g., \citealt{Stupar2008, Stupar2011}). Future optical SNR surveys can address this question through the addition of deep \O3 filter images.

\subsection{SNRs with Nonradiative Filaments}

Two of our new SNRs exhibit
nonradiative Balmer dominated line spectra; namely, G27.9-17.6 and G115.6+9.2.
As shown in Table 2, some filaments in these remnants exhibited a spectrum consisting of only H$\alpha$ (see discussion in $\S4.1$ and $\S4.2$).

Nonradiative spectra are relatively rare in SNRs with only a few cases seen; e.g., 
the Cygnus Loop (G74.0-8.5)
\citep{Levenson1998,Sankrit2023,Vuceti2023}
and SN~1006 (G327.6+14.6)
\citep{Ghavamian2002,Heng2007}.
In these objects, there is a high-velocity shock
moving through a relatively low density ISM. From the high Galactic latitude of these examples, it is maybe not surprising that both G27.9-17.6 and G115.6+9.2 show nonradiative spectra since they lie at high latitudes and likely expanding in a low density ambient medium high above the Galactic plane.

\begin{deluxetable*}{cccccccccccc}[ht]
\tablecolumns{12}
\tablecaption{Estimates of the Physical Properties of the New SNRs}
\tablewidth{0pt}
\tablehead{ 
\colhead{SNR} & \colhead{Dia.}  & \colhead{d} & \colhead{R$_{\rm s}$} & \colhead{$|z|$} & \colhead{$N_{\rm H}$} &
\colhead{V$_{\rm cl}$}   & \colhead{$n_{\rm cl}$ }  & \colhead{V$_{\rm s}$}   & \colhead{$n_{\rm 0}$ }
& \colhead{SN Energy} & \colhead{Age} \\
\colhead{ID}  & \colhead{($\degr$)} & \colhead{kpc} & \colhead{(pc)}    & \colhead{(pc)} & \colhead{(cm$^{-2}$)} &
\colhead{(km s$^{-1}$)} & \colhead{(cm$^{-3}$)} & 
\colhead{(km s$^{-1}$)} & \colhead{(cm$^{-3}$)} &
\colhead{(10$^{51}$ erg)} & \colhead{(yr)} 
}
\startdata
G27.8-17.1 & 5 &  $0.5\pm0.25$ &  22d$_{0.5}$  & 150d$_{0.5}$ & $0.8\times 10^{21}$  & 80 - 200  & 1 - 5 &  230 & 0.27 & 0.1 - 0.6 & 37 000 \\
G115.6+9.2 & 1 &  $2.5\pm0.5$  &  22d$_{2.5}$  & 400d$_{2.5}$ & $3.2\times 10^{21}$  & 80 - 200  &$6\pm2$&  290 & 0.46&   0.8     & 30 000 \\
G190.5-25.3& 8 &  $0.45\pm0.1$ &  32d$_{0.45}$ & 210d$_{0.55}$& $9.3 \times 10^{20}$ & 60 - 80   &$\leq7$&  150  &$\sim1$ &1.5 - 3.0& 90 000 \\
G195.9+2.2 & 1 &   $2\pm 1$    &  17d$_{2.0}$  & ~75d$_{2.0}$ & $3.7 \times 10^{21}$ & 100 - 120 & 5 - 10&  320  & $\sim$0.9 & 0.8  & 21,000 \\ 
G239.9+7.0 & 2 &  $1.0\pm0.5$  &  17d$_{1.0}$  & 120d$_{1.0}$ & $1.2 \times 10^{21}$ &  90 - 130 &$\sim6$&  370  & 0.35   & 0.6     &  18,000 \\ 
\enddata6\
\end{deluxetable*}

\subsection{Physical Properties of New SNRs}

For our five new SNR discoveries there are no X-ray or radio data that could help determine some of their physical properties. Instead, what we do have are low-dispersion spectra that can offer estimates of the shock velocities and electron densities of the observed filaments, along with total $N_{\rm H}$ column densities in their directions plus reddening estimates. 

In what follows below, we have estimated filament shock velocities using calculated emission line ratios from a variety of shock models
\citep{Shull1979,Raymond1979,Cox85,Hartigan1987,Jin2025},
H~I column densities toward these remnants from  the \citet{HI4PI2016}, 
and extinction values as a function of Galactic distance using the on-line tool
GALExtin \citep{2021Amores}. However, the remnant properties discussed below and listed in Table 3 are presented as initial rough estimates and hence should be viewed with a good deal of caution.

Despite their large angular sizes, we have assumed these SNRs are still in the adiabatic phase of their evolution and thus follow Sedov-Taylor expressions: 

\begin{equation} {R_{s}} =
12.9 ~ t_{4}^{2/5} ({\epsilon_0}/{n_0})^{1/5} ~ {\rm pc} 
\end{equation} 

\begin{equation}  V_{s} =
0.4 (R_{s}/t)
\end{equation} 

\begin{equation} {E_{o}} =
1.37 \times 10^{42}{n_{o}}{V_{s}^2} {R_{s}^3} ~ {\rm erg}
\end{equation}

\noindent
where $R_{s}$ is the radius of the shock front in pc, 
$V_{s}$ is the shock velocity in km s$^{-1}$,
$n_{o}$ is the preshock ambient interstellar density in cm$^{-3}$,
$\epsilon_{0}$ is the initial remnant energy in units of $0.75 \times 10^{51}$ erg,
$t_{4}$ is the remnant age in $10^{4}$ yr,
and $E_{\rm o}$ is the SN energy in ergs. 

Equation 3 can also be written in terms of optical filament density, $n_{\rm cl}$,  and filament shock velocity
$V_{cl}$ is in units of 100 km s$^{-1}$ and $R_{s}^{3}$ in pc

\begin{equation} {E_{51}} =
2 \times 10^{-5} n_{cl} V_{cl} R_{s}^{3} ~ {\rm erg}
\end{equation}

We can obtain an estimate the electron density of a remnant's emission filaments from the [\ion{S}{2}]
$\lambda$6716/$\lambda$6731 line ratio \citep{Osterbrock2006} using the $temden$ tool in STScI data reduction software assuming T$_{\rm e}$ = 10$^{4}$ K. 
The equation is
 
\begin{equation}
{\rm N_{[S~II]}} \simeq 45~ \left(\frac{\rm V_{\rm cl}}{100~\rm km~s^{-1}}\right)^2 \left(\frac{n_{\rm cl}}{\rm cm^{-3}}\right) 
\end{equation}  

where $n_{\rm cl}$ is the filament cloud's preshock electron density in cm$^{-3}$,  V$_{\rm cl}$ is the shock
velocity in the cloud responsible for the observed [\ion{S}{2}] emission \citep{Russell1990}.  

Below we estimate some of the physical properties of our five new SNRs with the results  tabulated in Table 3.

\bigskip

\noindent
\subsubsection{G27.8-17.1} 

Our data on this remnant is quite limited. Its large angular size, high Galactic latitude and nonradiative optical shock spectra indicating a high interstellar shock velocity all seem to point to it being a fairly young and relatively nearby object. 
Even at distance of just 1 kpc, its 5$\degr$ angular diameter would mean a physical size of almost 90 pc, the opposite of a young or middle aged remnant. Thus, we have estimated its distance at 0.5$\pm 0.25$ kpc.  

We do not have an $E(B - V)$ measurement from the spectrum taken of its bright southwest filament. However, $N_{\rm H}$ for the remnant's central and eastern directions of $0.80 \times 10^{21}$ cm$^{-2}$ \citep{HI4PI2016} can be used with the following expression 
\citep{Jenkins1974}

\begin{equation}
E(B-V) = \left(\frac{N_{\rm H}}{7.5 \times 10^{21} \rm cm^{-2}} \right)
\end{equation}
to estimate an $E(B-V) \simeq$ 0.1 mag.  While such a low value is  not surprising given its high Galactic latitude, it is nearly twice the 0.19 mag
using the extinction model of \citet{Amores2005} for d = 0.5 kpc.
The preshocked ambient density assuming a $N_{\rm H}$ related distance of 1 kpc is $(0.27 \pm 0.13)$ cm$^{-3}$.

The remnant's SW filament is quite bright in \O3 emission indicating a filament shock velocity above 100 km s$^{-1}$.  In contrast, our spectrum of the filament's western end
showed no appreciable \O3 indicating a much slower shock 
around $70 - 80$ km s$^{-1}$. This filament's electron density is near the [\ion{S}{2}] $\lambda$6716/$\lambda$6731 low density limit of 1.43 implies a density $\leq 200$ cm$^{-3}$.
Adopting this and choosing V$_{cl}$ to be 75 km s$^{-1}$ for the filament's shock velocity, we estimate from equation 5 a preshock filament density 
of $n_{\rm cl}$ $\leq$ 7 cm$^{-3}$. If we adopt a filament density of  1  - 5 cm$^{-3}$, we can use equation 2 to estimate
$E_{o}$ $\simeq$ $(0.1 - 0.6) \times 10^{51}$d$^{3}_{0.5}$ erg.

Finally, assuming pressure equilibrium where
\begin{equation}
   n_{\rm cl}V_{\rm cl}^2 = n_{\rm 0} V_{s}^2
\end{equation}

we estimate V$_{s}$ = 230 km s$^{-1}$ and a remnant age of
$\simeq$ 37 000 yr. The ambient interstellar density that the nonradiative filaments along the remnant's eastern rim are encountering has a density of around 0.35 cm$^{-3}$ assuming a filament shock velocity of $\simeq$200 km s$^{-1}$.

If we adopt these values, namely an age of 39,000 yr, an energy $\epsilon_{0}$ of 
$0.3 \times 10^{51}$ erg, and a preshock density of 0.35 cm$^{-3}$, from equation 1
we find $R_{s}$ = 23 pc matching our 
assumed remnant distance and radius (Table 3).

\subsubsection{G115.6+9.2}
We have only slightly more data on this remnant than for G27.8-17.1. With an angular diameter of roughly 1$\degr$, nonradiative spectra for some northern and southern filaments, plus filament radial velocities of over 100 km s$^{-1}$, this remnant appears to be relatively young. Despite its high Galactic latitude, we also suspect this remnant lies at a distance of around 2.5 kpc  which would give it a physical radius around 22 pc consistent with a moderate age remnant and a substantial z distance off the Galactic plane $\simeq$ 325 pc.

We will follow many of the same calculations we did for G27.8-17.1.
From a $N_{\rm H}$ value of $3.2 \times 10^{21}$ cm$^{-2}$ in its direction, we estimate $n_{0}$ $\simeq$0.46 cm$^{-3}$ and we estimate  $E(B-V) = 0.42$ mag. However, higher $E(B-V)$ estimates of 0.55 and 0.71 come from \citet{PlanckCollaboration2014P} and \citet{Schlafly2014} data sets. Such relatively moderate to high values might help 
explain the lack of our detected H$\beta$ emission.

Only the spectrum taken P1 for the bright southern filament showed a radiative type shock spectrum. The density sensitive [\ion{S}{2}] line ratio of 1.24 indicates an electron density of $190\pm60$ cm$^{-3}$ and thus a filament density around 6 cm$^{-3}$. The very weak and sparse [\ion{O}{2}] $\lambda$5007 emission
and seen in our images in the south indicates a shock velocity just above 80 km s$^{-1}$. From equation 5,
we estimate $n_{\rm cl}$ = $6.1\pm2.1$ cm$^{-3}$. Then adopting a radius, R$_{\rm s}$, of 22 pc assuming a distance of 2.5 kpc
and using equation 4, we find 
$E_{0}$ = $0.8d^3_{2.5}$ $\times 10^{51}$ erg. 

As can be seen in Table 3, many of these numbers are close to what we estimated for G27.8-17.1, This is not so unexpected since both remnants exhibit radiative and nonradiative indicating interstellar shocks ranging from 80 to around 200 km s$^{-1}$. Because our estimates of their distances are highly uncertain we are not in a position to estimate meaningful errors for either remnant energy or age.

\subsubsection{G190.5-25.3}
This huge remnant, some 8+ degrees in size, lies in the direction to the even larger Gemini-Eridanus Superbubble
whose distance has been estimated by various means to lie between 200 and 400 pc \citep{Reynolds1979,Brown1995,Lee2009,Joubaud2019}.
If we assume that this remnant lies physically inside the superbubble at a distance of 300 pc as it at first might appear, the remnant would then have a fairly modest physical radius $21$ pc  and a z distance of 140 pc. 

However, we doubt this SNR lies inside the superbubble due to two reasons: a) a near total lack of \O3 emissions which would be a bit surprising if it were really expanding inside a very low density bubble, and b) its over abundance of emission features many of which lie close to the remnant's center and extend far to the west almost to the nearly vertical emission Arc A of the superbubble (see Fig.~\ref{Walker}).
Instead, we suspect the remnant lies either on the near or far side of superbubble and is interacting with swept-up material in the superbubble's thick shell. 

Such scenarios resolves the question of why there is so much emission for a remnant that gives the impression of lying inside a Galactic superbubble, and why there is so much diffuse emission within the remnant that appears to be not shock emission.  Although the SNR could lie at either 200 - 300 or 400 - 500 pc, based on the total lack of \O3 emission suggesting a large and low velocity shock, a larger distance and hence larger remnant would seem more likely.

Assuming a distance of around 450 pc, the remnant has a
physical radius of 28 to 35 pc. The mean of HI4PI $N_{\rm H}$ measurements at and near remnant center is $9.2 \times 10^{20}$ cm$^{-2}$ suggesting a preshock density for the shell of $\simeq1$ cm$^{-3}$, consistent with our [\ion{S}{2}] line ratio measurements near the low density limit.
We estimate a remnant energy around $1 - 3 \times 10^{51}$ erg and an age approaching 100,000 yr, again consistent with an old remnant with a low velocity filament shock and a large physical size. Lastly, we note more 
that with this size, the remnant might have  already transitioned into the snowplow phase.

\subsubsection{G195.9+2.2}
This one degree diameter remnant has probably been missed in previous SNR surveys due to the fact that it is fairly hidden by Monogem's brighter filaments along its northwestern rim  (see Fig.~\ref{G195_fig1}).  The remnant's western half is missing in both H$\alpha$ and \O3, while its eastern and northern parts define an almost perfect circular shell.  Its Galactic latitude of just 2.2 degrees, unlike the other remnants discussed above, places it close to Galactic plane and in this nearly anti-center direction. Its highly circular appearance might indicate a relatively small and fairly young SNR. If that is correct, it would have a physical radius greater than 10 pc but less than 30 pc which would imply distances between 1 and 3 kpc
which would place it behind and farther away than the projected nearby large H~II region NGC~2264 which lies distance of 0.7 - 0.8 kpc. Due to its anticenter location, distances much greater than 3 kpc would place it on the very outer edge of the Milky Way. 

Low dispersion optical spectra was obtained on only three of its optical filaments, all selected based on their brightness in H$\alpha$.
These filaments are located along or near the remnant's northwestern rim with only one showing strong \O3 emission.
All three spectra (see Fig.~\ref{G195_fig3}) are quite noisy but appear to 
indicate [\ion{S}{2}] 
6716/6731 ratios around 1.0 - 1.2  which, if true, would imply fairly high electron densities of between 250 and 550 cm$^{-3}$ and a preshock filament density between 5 and 10 cm$^{-3}$. 
Using equation 5, we find an
$E(B-V) \simeq 0.5$ which agrees with \citet{Amores2005} for their models for a distance of 2 kpc and would help explain our lack of H$\beta$ detection. Finally, adopting an $N_{\rm H}$ = $5.7 \times 10^{21}$ cm$^{-2}$
we estimate a preshock ISM density of 0.9 cm$^{-3}$ for a distance of 2 kpc.
Finally, assuming pressure equilibrium 
and an average filament density of 7.5 cm$^{-3}$, 
$V_{s}$ of 320 km s$^{-1}$, a remnant energy of $\simeq 0.8 \times 10^{51}$  erg and hence an age of around 21 kyr.

\subsubsection{G239.9+7.0} 
This nebula is an example of an \O3 emission dominated remnant
with only a small section of its western rim displaying any substantial H$\alpha$ bright filaments.
As can be seen in Figure~\ref{G239_1}, the remnant could be easily missed in even deep H$\alpha$ SNR surveys (see right lower panel), whereas, in contrast, it is prominent and easily seen in  \O3 images. The remnant exhibits a fairly complete shell but with its southeast rim  missing in  \O3.

Its distance is completely unknown. If we limit's its distance above the Galactic plane to 500 pc, then its maximum distance is around 4 kpc. However, at such a distance it's diameter would be $\sim140$  pc ranking above even the largest confirmed Galactic SNRs and surprising given its overall strong \O3 emission. Thus, we adopted a much shorter distance of 
$1.0\pm0.5$ kpc which yields remnant diameters of $35\pm 17$ pc. 

We obtained spectra at three locations (P1, P2, and P3) selected based on their H$\alpha$ emission visibility. Although this allowed us to investigate the nebula's H$\alpha$ to [\ion{S}{2}] ratio, these spectra give a completely false impression as to 
the remnant's overall spectral properties. Indeed, its strong \O3 emission and near absence of detected 
H$\alpha$ for most  of the remnant's optical features indicates filament shock velocities  above 90 km s$^{-1}$ and more in the range of 110 - 130 km s$^{-1}$. 

All three spectra are of low S/N making estimates of filament densities highly uncertain. However if we take the [\ion{S}{2}] ratio seen at P4 at face value then
at that location, the very end of
an otherwise \O3 only filament
(Fig.~\ref{G239_slits}) the density is $225\pm70$ cm$^{-3}$ which assuming a filament shock velocity of 90 km s$^{-1}$ give a preshock density estimate $\simeq$ 6 cm$^{-3}$. 
Using these values and a 1 kpc distance yields an estimate remnant energy $\sim$ 
$0.6 \times 10^{51}$ erg and an age of $\sim$ 18 kyr, making it perhaps the youngest remnant in our 
group of five.

The significance of H$\beta$ not being detected at either P2 or P4 
may be due to both the remnant's faintness and a  moderate amount of extinction at a minimum despite a fairly high Galactic latitude. The extinction models of \citet{Amores2005} point to a modest $A_{V}$ $\sim$0.45 mag. Both reddening as well a filament densities could well be investigated further with much higher S/N spectra.

\subsection{The Monogem SNR's Optical Emissions}

The emission-line images presented above on the huge Monogem SNR reveal a far more extensive and relatively bright \O3 and H$\alpha$ emission filaments around much of the Monogem's X-ray periphery than previously 
realized. This is a significant improvement from the three Monogem related optical filaments currently in the
astrophysics literature.

We are not the first to identify many of these optical emission features as being ones associated with the Monogem SNR.
Recent images taken by several other amateurs have detected many of these filaments and have even commented on them as likely related to the Monogem Ring\footnote{https://stargazerslounge.com/topic/420847-help-me-identify-some-nebulae/}.
Plus, nearly a dozen amateurs have 
posted excellent images of the twisted H$\alpha$ filaments situated along the remnant's northwestern rim where it meets the Gemini H$\alpha$ Ring and 
shown in Figure~\ref{Gemini} and Figure~\ref{MG_Sean} and nicknamed the DNA Filaments\footnote{https://app.astrobin.com/}.

What we have presented, however, is a coherent picture of the Monogem SNR's optical emissions, linking individual and widely separated \O3 and H$\alpha$ emission filaments into a picture of a nearly complete
ring of optical emissions that surrounds and coincides with the remnant's X-ray boundaries.

We have also shown why the remnant's optical emission are located where they are.
For example.
Monogem's northwest expansion into the southern portion of the Gemini H$\alpha$ Ring and a collision with an interstellar cloud off to its northeast give rise to bright H$\alpha$ filaments in both places. While it is likely that these filaments are shocked and parts of the Monogem remnant, spectroscopic data on some or all of these filaments will be needed to confirm their shock nature to firmly establish their membership in the Monogem SNR. 

This picture of these separate filaments being parts of a large but single SNR, namely the Monogem Ring SNR, is different from some proposed origins. 
\citet{Ziegenbalg2025} views the very long \O3 filaments along Monogem's eastern limb which he calls ``the Gemini OIII Arc'' may be a 
fragment of the supernova remnant associated with the pulsar Geminga which lies several degrees to the northwest and well outside of the Monogem Ring.
He viewed the shorter, fainter, and more eastern \O3 filaments are the only ones related to the Monogem SNR, suggesting the brighter eastern \O3 filaments are instead part of a much larger (dia. $\sim55$ pc) emission ring.

Lastly, many of the remnants discussed above that are either near or projected inside the Monogem Ring make it clear this region's soft X-ray emission first detected more than half a century ago
is far more complex than anyone had realized.
There are currently  
four known SNRs plus several new candidate SNRs with projected locations near or within the remnant's X-ray boundaries (Fig.~\ref{MG1}).
Optical emissions from these remnants as well as from the Monogem remnant itself
may help better understand the small-scale X-ray features of this very large object deceptively called simply the Monogem  Ring. 

\section{Conclusions}

We have presented optical images on 11 Galactic SNR nebulae taken by amateur
astrophotographers through literally thousands of individual exposures totaling over 2000 hr of open shutter time. 
We have also presented follow-up low-dispersion spectra of several of these nebulae.
Our major findings include:

\smallskip

1) We have identified five new SNRs which range in size from one to over eight degrees in angular size. 
Low-dispersion optical spectra of these nebulae show shock emission line emissions
and all exhibit filamentary line emission features consistent with the presence of interstellar shocks.
Radio maps do not reveal any associated radio emissions.
These new SNRs are:
G27.8-17.1, G115.6+9.2, G190.5+25.3, G195.9+2.2 and G239.9+7.0. 
All but one lie at relatively high Galactic latitudes, and all but one, G190.5+25.3,  show strong \O3 line emissions.

\smallskip
2) Images in \O3 and H$\alpha$ of three additional nebulae are presented which, due to their
filamentary appearance, strong \O3 emissions, and lack of any central ionizing source, are 
seen as new SNR candidates. These are: G191.3+11.1  which lies along the northern boundary of the Monogem Ring remnant, G205.7-1.7 which is located close to the Rosette Nebula, and G305.4-0.7 in the southern hemisphere.  We also present images showing optical filaments along the northeastern rim of the suspected 
X-ray SNR G190.4+12.4 based on eROSITA data.

\smallskip
3) Images and spectra of the large and \O3 emission dominate SNR 
SNR G206.6+6.1 seen toward Monogem SNR's center are also presented.
This large remnant exhibits a highly filamentary morphology not apparent in radio maps. Deep H$\alpha$ images also reveal a possible shock blowout along the remnant's southeast limb.

\smallskip
4) We also present and discuss \O3 and H$\alpha$ images of numerous optical filaments around much of the X-ray emission rim 
of the 25 degree diameter Monogem Ring SNR. The remnant's \O3 filaments surround its X-ray emission along the eastern, northwest and southern boundaries, while
H$\alpha$ filaments appear where Monogem's shock has collided  with denser neighboring
clouds such as the Gemini H$\alpha$ Ring in the northwest, and the western tip of a large ISM cloud seen in the northeast.

\smallskip
5) Additional findings include: a) deep optical surveys can 
sometimes detect SNRs 
better than current radio and X-ray SNR searches, b) Balmer nonradiative shocks are not uncommon in SNRs located in low density ISM regions at high Galactic latitudes, and c) some \O3 emission dominated SNRs may have been missed in prior deep H$\alpha$ emission only surveys.

\bigskip

Completion of this work was 
greatly aided by our attendance at 
the Galactic Frontiers meeting held in June 2026 at the Physikzentrum, Bad Honnef, Germany.
We also wish to acknowledge our use of Stefan Ziegenbalg's on-line Northern Sky Narrowband Survey images which was invaluable for confirmation of faint emission features and for giving us wider views of certain nebular features.
Thanks also to Eric Galayda and the entire MDM staff at Kitt Peak for making follow-up optical image and spectral observations possible, sometimes changing equipment on very short notice.
We have made extensive use of the Simbad database and NASA's Skyview online data archives. 
This work is part of R.A.F's cooperative amateur High Altitude Reconnaissance 
project as part of the broad Archangel III Astrophysics Research Program at Dartmouth. 

\facilities{MDM Observatory at Kitt Peak, Arizona USA;
Starfront Observatory, Texas USA; Oukaimeden Observatories,  Morocco;
Observatorio El Sauce, Chile;
 New Mexico Skies Observatory, USA; Sierra Remote Observatory USA,
 Heaven's Mirror Observatory, Manton, NSW, Australia}
\software{IRAF, DS9 fits viewer \citet{Joye2003}, WCSTools \citet{Laycock2010},
Photoshop, PixInsight, Astropixel Processor } 

\appendix


\begin{deluxetable*}{lllccl}[h]
\footnotesize
\tablecolumns{6}
\tablecaption{SNR Names, Observers \& Observing Sites, Equipment, Image   and Exposure Details }
\tablehead{  \colhead{SNR} &  \colhead{Observers \& Observing Sites} & \colhead{Optics} &
\colhead{FOV/Scale} & \colhead{Filters} & \colhead{Exposures (hr)} }
\startdata
G27.9-17.6    & B. Falls, M.\ Drechsler                   & Takahashi FSQ-106EDX & $3.5\degr \times 5.2\degr$/2.0$''$& R,G,B & 6.2 ($75 \times 300$ s) \\
              & Deep Sky Chile                            &                      &                      & H$\alpha$ & 31.3 ($188 \times 600$ s) \\
              & Rio Hurtado Valley, Chile                 &                      &                      & [O III]   & 30.8 ($185 \times 600$ s) \\
               \hline
 G115.6+9.2 & Y.\ Sainty,  N.\ Martino, M.\ Drechsler     & Takahashi FSQ-106EDX & 3.1$\degr$/3.8$''$ & [O III]     & 65.6 ($787 \times 300$ s) \\
              &   K.\ Aziz, A. Soto, L.\ Leroux:          & Takahashi FSQ-85ED   &  2.7$\degr$/4.6$''$  &  H$\alpha$& 49.1 ($589 \times 300$ s) \\
              &   Oukaimeden Obs., Morocco                &                      &                      & R,G,B     & ~ 8.5 ($510 \times 60$ s) \\
              & T.\ Schaeffer, C.\ Bj\"ork, S.\ Body, N.\ Puig, & Askar 151 PHQ &$1.9\degr \times 1.2\degr$/1.1$''$& [O III]   &  196 ($496 \times 1424$ s) \\
              & J.\ Capel, J.\ Dzibua, J.\ Matzger,       & CFF 155mm refractor  &                      & H$\alpha$ &  232 ($589 \times 1421$ s) \\
              & S.\ Guberski, A.\ Jain, L.\ Carpenter,    & Esprit 150ED         &                      & R         & 23.8 ($193 \times 443$ s)  \\
              & A.\ Zhang, J.\  Schilling                 & 180 mm APO Pro       &                      & B         & 16.2 ($144 \times 406$ s) \\
              & various sites in Europe                   & Takahashi FSQ-106EDX &                      & G         & 20.3 ($173 \times 421$ s) \\ 
              \hline
 G190.4+12.5  & M.\ Drechsler, T.\ Schaeffer, C.\ Bj\"ork, & Takahashi FSQ-106EDX & 3.1$\degr$/3.8$''$  & H$\alpha$ & 40 ($240 \times 600$s) \\  
             & S.\ Body, and the Deep Sky Collective      &  Askar 151PHQ   & $1.9\degr \times 1.27\degr$/ 1.1$''$ & [O III]   & 108 ($271 \times 1439$ ) \\   
             &  various European sites                     &                      &                      & R,G,B     & ~ 8.5 ($510 \times 60$ s) \\
              \hline
 G190.5-25.3  & M.\ Drechsler, X.\ Strottner, O.\ Sharpen & AG Optical FA14     &$1.1\degr \times 0.7\degr$/1.0$''$& [O III]   & 21.4 ($643 \times 120$ s) \\
              & L.\ Mulato:                              &                      &                      & H$\alpha$ & 35.9 ($718 \times 180$ s) \\
              &  SkyEyE Observatory, Orio, Spain         &                      &                      & R,G,B     &  ~ 2.8  ~  ($85 \times 120$ s) \\
     & S.\ Walker, D. Di Cicco: MDW Survey      & Astro-Physics 130mm  &$3.4\degr \times 3.4\degr$/3.2$''$ & H$\alpha$ & ~ 3.9  ~ ($7 \times 2000$ s) \\  
                \hline          
 G191.4+11.1  & T.\ Schaeffer, C.\ Bj\"ork, S.\ Body,    & Takahashi FSQ-106EDX & 3.1$\degr$/3.8$''$   & H$\alpha$ & 40 ($240 \times 600$s) \\
              & T.\ Kottary, D.\ Arora, \&           &                      &                      & [O III]   & 60 ($360 \times 600$s ) \\
              & P.\ Sparkman various European sites      &                      &                      &           &                       \\
              \hline
 G195.9+2.2   & T.\ Schaeffer, C.\ Bj\"ork, S.\ Body, and& Askar 130PHQ         &$1.9\degr \times 1.27\degr$/ 1.1$''$ & H$\alpha$ & 139 ($348 \times 1439$s) \\
              & the Deep Sky Collective, and the New     & Askar 151PHQ         &                      & [O III]   & 108 ($271 \times 1439$ ) \\
              & Horizon Project; various sites in Europe &                      &                      &           &             \\
              \hline             
 G205.7-1.7   & T.\ Schaeffer, C.\ Bj\"ork, S.\ Body, and& Takahashi FSQ-106EDX & 3.1$\degr$/3.8$''$   & H$\alpha$ & 11.2 ($67 \times 600$s) \\
              & the Deep Sky Collective, and the New     &                      &                      & [O III]   & 10.5 ($63 \times 600$s ) \\
              & Horizon Project; various sites in Europe &                      &                      &           &                       \\
              \hline
 G203+12      & T.\ Schaeffer, C.\ Bj\"ork, S.\ Body,    & Takahashi FSQ-106EDX & 3.1$\degr$/1.9$''$   & H$\alpha$ & 216 ($1296 \times 600$s) \\
 (Monogem)    &  T. Kottary, D.\ Arora, P.\ Sparkman     &                      &                      & [O III]   & 216 ($1296 \times 600$s) \\ 
              & various sites in Europe                  &                      &                      &           &                          \\
              & S.\ Walker, D. Di Cicco: MDW Survey      & Astro-Physics 130mm  &$3.4\degr \times 3.4\degr$/3.2$''$ & H$\alpha$ & ~ 3.9  ~ ($7 \times 2000$ s) \\
              \hline
  G206.6+6.1   & B.\ Falls                                & RASA 8"              &                      & H$\alpha$ &  121.8                     \\
              & StarFront Observatories,                 & Takahashi FSQ-106EDX &                      & [O III]   & 157.2                      \\
              &  Rockwood, Texas USA                     &                      &                      & R,G,B     &  ~ ~ 4 ($80 \times 180$s) \\
              & T.\ Schaeffer, S.\ Body, T.\ Kottery,    & Takahashi FSQ-106EDX &$5.1\degr \times 6.0\degr$/$3.9''$ & H$\alpha$ & ~ 64 ($256 \times 900$s) \\
              & P.\ Sparkman                             &                      &                      & [O III]   & ~ 61 ($244 \times 900$ s) \\
              & various sites in Europe                  &                      &                      & [S II]    & ~ 35 ($140 \times 900$ a) \\
              &                                          &                      &                      & R,G,B     & 10.5 ($126 \times 300$ s) \\             
               \hline
  G239.9+7.0   &  B.\ Falls                               & Takahashi FSQ-106EDX & $3.5\degr \times 5.2\degr$/2.0$''$& [O III]   &  49.1 ($289 \times 600$ s) \\ 
              &  Deep Sky Chile                          &                      &                      & H$\alpha$ &  45.8 ($274 \times 660$ s)  \\
              & Rio Hurtado Valley, Chile                &                      &                      & R,G,B     &  ~ 1.8 ~ ($35 \times 180$ s)  \\ 
                            \hline
 G305.4-0.7   & M.\ Peitsch, J.\ Rodrigues     &Planewave CDK14    & $0.80\degr \times 0.53\deg$/$0.61''$&H$\alpha$ & 17.7 ($53 \times 1200$ s) \\
              & Heaven's Mirror Observatory    &                   &                                   & [S II]   & 34.3 ($103 \times 1200$ s) \\
              & Manton, NSW, Australia         &                   &                                   & [O III]  & 28.3 ($85 \times 1200$ s) \\
              &                                &                   &                                   & R,G,B    & ~1.0 ($60 \times 60$ s) \\
\enddata
\label{Equip}
\end{deluxetable*}




\begin{figure*}[t]
\begin{center}
\includegraphics[angle=0,width=16.0cm]{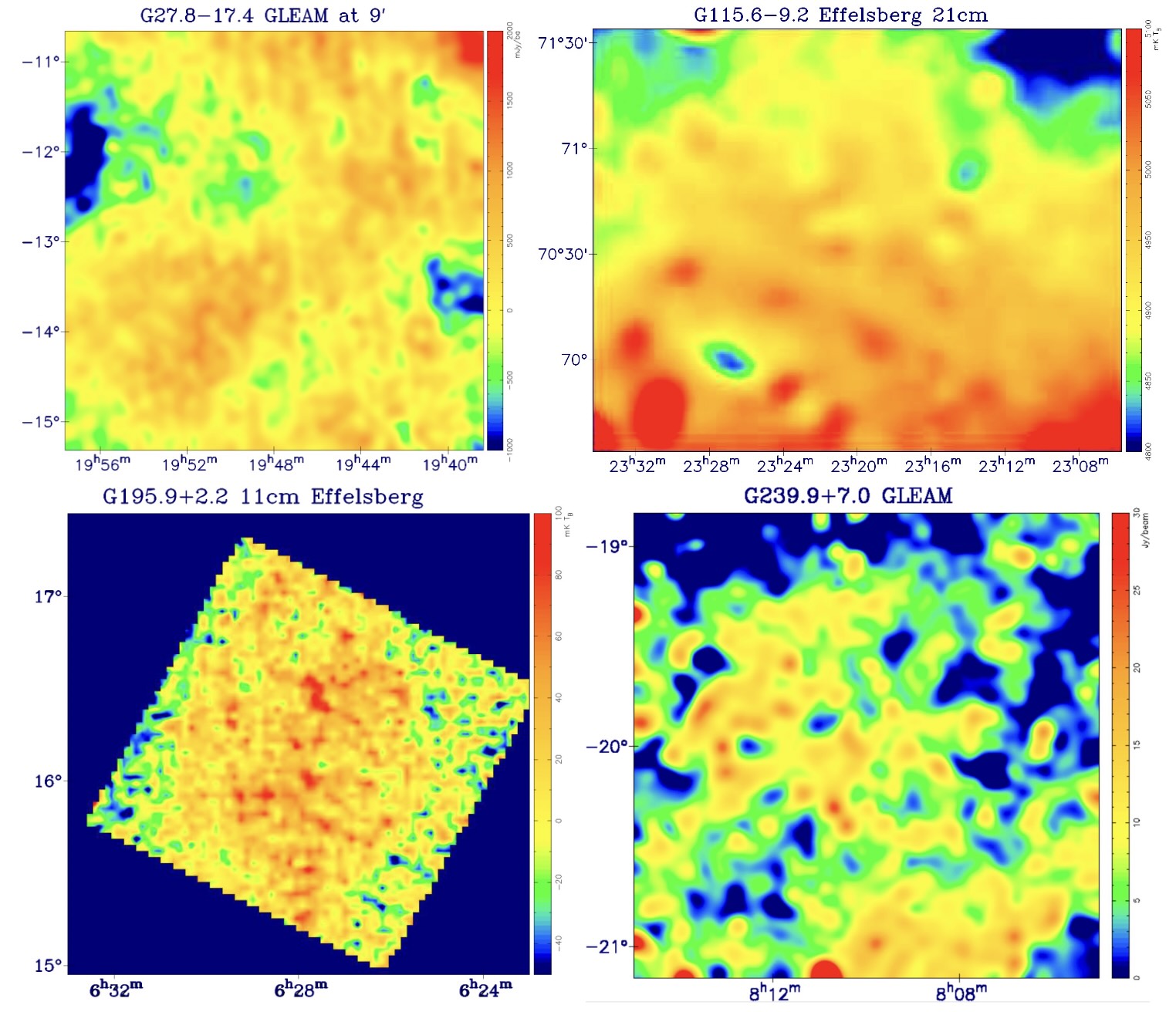} 
\caption{Radio images of four of our new Galactic SNRs showing no obvious 
related radio emission. The maps for G27.8-17.4 and G239.9+7.0
are from the GLEAM survey, band 170-230 MHz 
\citep{Hurley_Walker2019}.
The Effelsberg 11cm map of G195.2+2.2 is extracted from the Effelsberg
11cm-survey of the Galacic plane 
\citep{Furst1990}. The
Effelsberg 21cm map of G115.6-9.2 is from an unpublished section of
the EMLS \citep{Reich2004}.
Compact unresolved extragalactic sources 
have been subtracted from all maps to better shoe diffuse emission. 
The GLEAM maps have been additionally convolved to 5' for G239.9+7.0 
and 9' for G27.8-17.4.
\label{radio_images} 
} 
\end{center}
\end{figure*}

\clearpage
\newpage

\bibliography{AAS_ref.bib}{}

@ARTICLE{Aftab2026,
       author = {{Aftab}, Noor and {Zhang}, Xunhe (Andrew) and {Walker}, Sean and {di Cicco}, Dennis and {Mittelman}, David R. and {Gupta}, Sanya and {Saydjari}, Andrew K. and {Putman}, Mary and {Schiminovich}, David},
        title = "{The MDW H{\ensuremath{\alpha}} Sky Survey: Data Release 1}",
      journal = {\aj},
         year = 2026,
        month = jan,
       volume = {171},
       number = {1},
          eid = {17},
        pages = {17},
          doi = {10.3847/1538-3881/ae17b4},
archivePrefix = {arXiv},
       eprint = {2510.22900},
 primaryClass = {astro-ph.IM},
       adsurl = {https://ui.adsabs.harvard.edu/abs/2026AJ....171...17A}
}

@ARTICLE{Aktekin2025,
       author = {{Aktekin}, E. and {Bak{\i}{\textcommabelow s}}, H. and {Bak{\i}{\textcommabelow s}}, V. and {Sezer}, A.},
        title = "{Optical spectroscopy of the radio-identified supernova remnants G152.4{\ensuremath{-}}2.1 and G203.1 + 6.6}",
      journal = {\mnras},
         year = 2025,
        month = oct,
       volume = {543},
       number = {1},
        pages = {761-768},
          doi = {10.1093/mnras/staf1501},
       adsurl = {https://ui.adsabs.harvard.edu/abs/2025MNRAS.543..761A}
}

@ARTICLE{Aktekin2026,
       author = {{Aktekin}, E. and {Bak{\i}{\textcommabelow s}}, H. and {Bak{\i}{\textcommabelow s}}, V. and {Sezer}, A.},
        title = "{First detailed optical spectroscopic observations of the supernova remnants G107.7{\ensuremath{-}}5.1 and G150.3+4.5}",
      journal = {\mnras},
         year = 2026,
        month = apr,
       volume = {547},
       number = {4},
          eid = {stag530},
        pages = {stag530},
          doi = {10.1093/mnras/stag530},
archivePrefix = {arXiv},
       eprint = {2603.15815},
 primaryClass = {astro-ph.HE},
       adsurl = {https://ui.adsabs.harvard.edu/abs/2026MNRAS.547ag530A}
}

@ARTICLE{Amores2005,
       author = {{Am{\^o}res}, E.~B. and {L{\'e}pine}, J.~R.~D.},
        title = "{Models for Interstellar Extinction in the Galaxy}",
      journal = {\aj},
         year = 2005,
        month = aug,
       volume = {130},
       number = {2},
        pages = {659-673},
          doi = {10.1086/430957},
       adsurl = {https://ui.adsabs.harvard.edu/abs/2005AJ....130..659A}
}

@ARTICLE{2021Amores,
       author = {{Am{\^o}res}, Eduardo B. and {Jesus}, Ricardo M. and {Moitinho}, Andr{\'e} and {Arsenijevic}, Vladan and {Levenhagen}, Ronaldo S. and {Marshall}, Douglas J. and {Kerber}, Leandro O. and {K{\"u}nzel}, Roseli and {Moura}, Rodrigo A.},
        title = "{GALExtin: an alternative online tool to determine the interstellar extinction in the Milky Way}",
      journal = {\mnras},
         year = 2021,
        month = dec,
       volume = {508},
       number = {2},
        pages = {1788-1797},
          doi = {10.1093/mnras/stab2248},
archivePrefix = {arXiv},
       eprint = {2108.00561},
 primaryClass = {astro-ph.GA},
       adsurl = {https://ui.adsabs.harvard.edu/abs/2021MNRAS.508.1788A}
}

@ARTICLE{Araya2024,
       author = {{Araya}, Miguel},
        title = "{Nonthermal GeV emission from the Nereides nebula: Confirming the nature of the supernova remnant G107.7{\ensuremath{-}}5.1}",
      journal = {\aap},
         year = 2024,
        month = nov,
       volume = {691},
          eid = {A225},
        pages = {A225},
          doi = {10.1051/0004-6361/202451443},
archivePrefix = {arXiv},
       eprint = {2409.14006},
 primaryClass = {astro-ph.HE},
       adsurl = {https://ui.adsabs.harvard.edu/abs/2024A&A...691A.225A}
}

@ARTICLE{Bakis2023,
       author = {{Bak{\i}{\c{s}}}, H. and {Bulut}, G. and {Bak{\i}{\c{s}}}, V. and {Sano}, H. and {Sezer}, A.},
        title = "{Discovery of optical emission associated with the supernova remnant G107.5-1.5}",
      journal = {\mnras},
         year = 2023,
        month = may,
       volume = {521},
       number = {1},
        pages = {1099-1112},
          doi = {10.1093/mnras/stad576},
archivePrefix = {arXiv},
       eprint = {2302.10268},
 primaryClass = {astro-ph.HE},
       adsurl = {https://ui.adsabs.harvard.edu/abs/2023MNRAS.521.1099B}
}

@ARTICLE{Ball2023,
       author = {{Ball}, Brianna D. and {Kothes}, Roland and {Rosolowsky}, Erik and {West}, Jennifer and {Becker}, Werner and {Filipovi{\'c}}, Miroslav D. and {Gaensler}, B.~M. and {Hopkins}, Andrew M. and {Koribalski}, B{\"a}rbel and {Landecker}, Tom and {Leahy}, Denis and {Marvil}, Joshua and {Sun}, Xiaohui and {Bufano}, Filomena and {Carretti}, Ettore and {Ingallinera}, Adriano and {Van Eck}, Cameron L. and {Willis}, Tony},
        title = "{A catalogue of radio supernova remnants and candidate supernova remnants in the EMU/POSSUM Galactic pilot field}",
      journal = {\mnras},
         year = 2023,
        month = sep,
       volume = {524},
       number = {1},
        pages = {1396-1421},
          doi = {10.1093/mnras/stad1953},
archivePrefix = {arXiv},
       eprint = {2307.01948},
 primaryClass = {astro-ph.GA},
       adsurl = {https://ui.adsabs.harvard.edu/abs/2023MNRAS.524.1396B}
}

@ARTICLE{Ball2025,
       author = {{Ball}, B.~D. and {Kothes}, R. and {Rosolowsky}, E. and {Burger-Scheidlin}, C. and {Filipovi{\'c}}, M.~D. and {Lazarevi{\'c}}, S. and {Smeaton}, Z.~J. and {Becker}, W. and {Carretti}, E. and {Gaensler}, B.~M. and {Hopkins}, A.~M. and {Leahy}, D. and {Tahani}, M. and {West}, J.~L. and {Anderson}, C.~S. and {Loru}, S. and {Ma}, Y.~K. and {McClure-Griffiths}, N.~M. and {Micha{\l}owski}, M.~J.},
        title = "{A Catalog of Galactic Supernova Remnants and Supernova Remnant Candidates from the EMU/POSSUM Radio Sky Surveys. I.}",
      journal = {\apj},
         year = 2025,
        month = jul,
       volume = {988},
       number = {1},
          eid = {75},
        pages = {75},
          doi = {10.3847/1538-4357/addc63},
archivePrefix = {arXiv},
       eprint = {2507.19625},
 primaryClass = {astro-ph.GA},
       adsurl = {https://ui.adsabs.harvard.edu/abs/2025ApJ...988...75B}
}

@INCOLLECTION{Bally2008,
       author = {{Bally}, J.},
        title = "{Overview of the Orion Complex}",
    booktitle = {Handbook of Star Forming Regions, Volume I},
         year = 2008,
       editor = {{Reipurth}, B.},
       volume = {4},
        pages = {459},
          doi = {10.48550/arXiv.0812.0046},
       adsurl = {https://ui.adsabs.harvard.edu/abs/2008hsf1.book..459B}
}

@ARTICLE{Bakis2025a,
       author = {{Bak{\i}{\textcommabelow s}}, H. and {Aktekin}, E. and {Bak{\i}{\textcommabelow s}}, V. and {Sano}, H. and {Sezer}, A.},
        title = "{Optical investigation of supernova remnant G206.7+5.9}",
      journal = {\mnras},
         year = 2025,
        month = mar,
       volume = {537},
       number = {3},
        pages = {2412-2421},
          doi = {10.1093/mnras/staf172},
archivePrefix = {arXiv},
       eprint = {2501.16990},
 primaryClass = {astro-ph.HE},
       adsurl = {https://ui.adsabs.harvard.edu/abs/2025MNRAS.537.2412B}
}

@ARTICLE{Burrows1993,
       author = {{Burrows}, D.~N. and {Singh}, K.~P. and {Nousek}, J.~A. and {Garmire}, G.~P. and {Good}, J.},
        title = "{A Multiwavelength Study of the Eridanus Soft X-Ray Enhancement}",
      journal = {\apj},
         year = 1993,
        month = mar,
       volume = {406},
        pages = {97},
          doi = {10.1086/172423},
       adsurl = {https://ui.adsabs.harvard.edu/abs/1993ApJ...406...97B}
}

@ARTICLE{Boumis2002,
       author = {{Boumis}, P. and {Mavromatakis}, F. and {Paleologou}, E.~V. and
         {Becker}, W.},
        title = "{New optical filamentary structures in Pegasus}",
      journal = {\aap},
         year = 2002,
        month = dec,
       volume = {396},
        pages = {225-234},
          doi = {10.1051/0004-6361:20021365},
archivePrefix = {arXiv},
       eprint = {astro-ph/0209258},
 primaryClass = {astro-ph},
       adsurl = {https://ui.adsabs.harvard.edu/abs/2002A&A...396..225B}
}

@ARTICLE{Boumis2009,
       author = {{Boumis}, P. and {Xilouris}, E.~M. and {Alikakos}, J. and
         {Christopoulou}, P.~E. and {Mavromatakis}, F. and {Katsiyannis}, A.~C. and
         {Goudis}, C.~D.},
        title = "{Discovery of optical emission from the supernova remnant G 32.8-0.1 (Kes 78)}",
      journal = {\aap},
         year = 2009,
        month = jun,
       volume = {499},
       number = {3},
        pages = {789-797},
          doi = {10.1051/0004-6361/200811474},
archivePrefix = {arXiv},
       eprint = {0903.3124},
 primaryClass = {astro-ph.GA},
       adsurl = {https://ui.adsabs.harvard.edu/abs/2009A&A...499..789B}
}

@ARTICLE{Brown1995,
       author = {{Brown}, A.~G.~A. and {Hartmann}, D. and {Burton}, W.~B.},
        title = "{The Orion OB1 association. II. The Orion-Eridanus Bubble.}",
      journal = {\aap},
         year = 1995,
        month = aug,
       volume = {300},
        pages = {903},
          doi = {10.48550/arXiv.astro-ph/9503016},
archivePrefix = {arXiv},
       eprint = {astro-ph/9503016},
 primaryClass = {astro-ph},
       adsurl = {https://ui.adsabs.harvard.edu/abs/1995A&A...300..903B}
}

@ARTICLE{Bunner1971,
       author = {{Bunner}, A.~N. and {Coleman}, P.~L. and {Kraushaar}, W.~L. and {McCammon}, D.},
        title = "{Low-Energy Diffuse X-Rays}",
      journal = {\apjl},
         year = 1971,
        month = jul,
       volume = {167},
        pages = {L3},
          doi = {10.1086/180749},
       adsurl = {https://ui.adsabs.harvard.edu/abs/1971ApJ...167L...3B}
}

@ARTICLE{Bunner1973,
       author = {{Bunner}, A.~N. and {Coleman}, P.~L. and {Kraushaar}, W.~L. and {McCammon}, D. and {Williamson}, F.~O.},
        title = "{Observations of Spatial Structure in the Soft X-Ray Diffuse Flux}",
      journal = {\apj},
         year = 1973,
        month = feb,
       volume = {179},
        pages = {781-788},
          doi = {10.1086/151915},
       adsurl = {https://ui.adsabs.harvard.edu/abs/1973ApJ...179..781B}
}

@ARTICLE{Chevalier1978,
       author = {{Chevalier}, R.~A. and {Raymond}, J.~C.},
        title = "{Optical emission from a fast shock wave: the remnants of Tycho's supernova and SN 1006.}",
      journal = {\apjl},
         year = 1978,
        month = oct,
       volume = {225},
        pages = {L27-L30},
          doi = {10.1086/182785},
       adsurl = {https://ui.adsabs.harvard.edu/abs/1978ApJ...225L..27C}
}

@ARTICLE{Cowie1978,
       author = {{Cowie}, L.~L. and {York}, D.~G.},
        title = "{The velocity distribution of interstellar gas observed in strong UV absorption lines.}",
      journal = {\apj},
         year = 1978,
        month = aug,
       volume = {223},
        pages = {876-883},
          doi = {10.1086/156320},
       adsurl = {https://ui.adsabs.harvard.edu/abs/1978ApJ...223..876C}
}

@ARTICLE{Cowie1979,
       author = {{Cowie}, L.~L. and {Songaila}, A. and {York}, D.~G.},
        title = "{Orion's cloak: a rapidly expanding shell of gas centered on the Orion OB1 association.}",
      journal = {\apj},
         year = 1979,
        month = jun,
       volume = {230},
        pages = {469-484},
          doi = {10.1086/157103},
       adsurl = {https://ui.adsabs.harvard.edu/abs/1979ApJ...230..469C}
}

@ARTICLE{Cordova1989,
       author = {{Cordova}, F.~A. and {Hjellming}, R.~M. and {Mason}, K.~O. and {Middleditch}, J.},
        title = "{Soft X-Ray Emission from the Radio Pulsar PSR 0656+14}",
      journal = {\apj},
         year = 1989,
        month = oct,
       volume = {345},
        pages = {451},
          doi = {10.1086/167918},
       adsurl = {https://ui.adsabs.harvard.edu/abs/1989ApJ...345..451C}
}

@ARTICLE{Cox85,
       author = {{Cox}, D.~P. and {Raymond}, J.~C.},
        title = "{Preionization-dependent families of radiative shock waves.}",
      journal = {\apj},
         year = "1985",
        month = "Nov",
       volume = {298},
        pages = {651-659},
          doi = {10.1086/163649},
       adsurl = {https://ui.adsabs.harvard.edu/abs/1985ApJ...298..651C}
}

@ARTICLE{Davies1978,
       author = {{Davies}, R.~D. and {Elliott}, K.~H. and {Goudis}, C. and {Meaburn}, J. and {Tebbutt}, N.~J.},
        title = "{The Monoceros Supernova Remnant}",
      journal = {\aaps},
         year = 1978,
        month = feb,
       volume = {31},
        pages = {271-284},
       adsurl = {https://ui.adsabs.harvard.edu/abs/1978A&AS...31..271D}
}

@ARTICLE{Dennison1998,
       author = {{Dennison}, B. and {Simonetti}, J.~H. and {Topasna}, G.~A.},
        title = "{An imaging survey of northern galactic H{\ensuremath{\alpha}} emmission with arcminute resolution}",
      journal = {\pasa},
         year = 1998,
        month = apr,
       volume = {15},
       number = {1},
        pages = {147-48},
          doi = {10.1071/AS98147},
       adsurl = {https://ui.adsabs.harvard.edu/abs/1998PASA...15..147D}
}

@ARTICLE{Dokara2023,
       author = {{Dokara}, R. and {Gong}, Y. and {Reich}, W. and {Rugel}, M.~R. and {Brunthaler}, A. and {Menten}, K.~M. and {Cotton}, W.~D. and {Dzib}, S.~A. and {Khan}, S. and {Medina}, S.-N.~X. and {Nguyen}, H. and {Ortiz-Le{\'o}n}, G.~N. and {Urquhart}, J.~S. and {Wyrowski}, F. and {Yang}, A.~Y. and {Anderson}, L.~D. and {Beuther}, H. and {Csengeri}, T. and {M{\"u}ller}, P. and {Ott}, J. and {Pandian}, J.~D. and {Roy}, N.},
        title = "{A global view on star formation: The GLOSTAR Galactic plane survey. VII. Supernova remnants in the Galactic longitude range 28{\textdegree} < l < 36{\textdegree}}",
      journal = {\aap},
         year = 2023,
        month = mar,
       volume = {671},
          eid = {A145},
        pages = {A145},
          doi = {10.1051/0004-6361/202245339},
archivePrefix = {arXiv},
       eprint = {2211.13811},
 primaryClass = {astro-ph.GA},
       adsurl = {https://ui.adsabs.harvard.edu/abs/2023A&A...671A.145D}
}

@ARTICLE{Dordico1978,
       author = {{Dodorico}, S. and {Benvenuti}, P. and {Sabbadin}, F.},
        title = "{Supernova remnants in M33.}",
      journal = {\aap},
         year = 1978,
        month = feb,
       volume = {63},
        pages = {63-68},
       adsurl = {https://ui.adsabs.harvard.edu/abs/1978A&A....63...63D}
}

@ARTICLE{Dordico1980,
       author = {{Dodorico}, S. and {Dopita}, M.~A. and {Benvenuti}, P.},
        title = "{A catalogue of supernova remnant candidates in nearby galaxies.}",
      journal = {\aaps},
         year = 1980,
        month = apr,
       volume = {40},
        pages = {67-80},
       adsurl = {https://ui.adsabs.harvard.edu/abs/1980A&AS...40...67D}
}

@ARTICLE{Drew2005,
       author = {{Drew}, Janet E. and {Greimel}, R. and {Irwin}, M.~J. and
         {Aungwerojwit}, A. and {Barlow}, M.~J. and {Corradi}, R.~L.~M. and
         {Drake}, J.~J. and {G{\"a}nsicke}, B.~T. and {Groot}, P. and
         {Hales}, A. and {Hopewell}, E.~C. and {Irwin}, J. and {Knigge}, C. and
         {Leisy}, P. and {Lennon}, D.~J. and {Mampaso}, A. and
         {Masheder}, M.~R.~W. and {Matsuura}, M. and {Morales-Rueda}, L. and
         {Morris}, R.~A.~H. and {Parker}, Q.~A. and {Phillipps}, S. and
         {Rodriguez-Gil}, P. and {Roelofs}, G. and {Skillen}, I. and
         {Sokoloski}, J.~L. and {Steeghs}, D. and {Unruh}, Y.~C. and
         {Viironen}, K. and {Vink}, J.~S. and {Walton}, N.~A. and {Witham}, A. and
         {Wright}, N. and {Zijlstra}, A.~A. and {Zurita}, A.},
        title = "{The INT Photometric H{\ensuremath{\alpha}} Survey of the Northern Galactic Plane (IPHAS)}",
      journal = {\mnras},
         year = 2005,
        month = sep,
       volume = {362},
       number = {3},
        pages = {753-776},
          doi = {10.1111/j.1365-2966.2005.09330.x},
archivePrefix = {arXiv},
       eprint = {astro-ph/0506726},
 primaryClass = {astro-ph},
       adsurl = {https://ui.adsabs.harvard.edu/abs/2005MNRAS.362..753D}
}

@ARTICLE{Dubner2015,
       author = {{Dubner}, Gloria and {Giacani}, Elsa},
        title = "{Radio emission from supernova remnants}",
      journal = {\aapr},
         year = 2015,
        month = sep,
       volume = {23},
          eid = {3},
        pages = {3},
          doi = {10.1007/s00159-015-0083-5},
archivePrefix = {arXiv},
       eprint = {1508.07294},
 primaryClass = {astro-ph.HE},
       adsurl = {https://ui.adsabs.harvard.edu/abs/2015A&ARv..23....3D}
}

@INCOLLECTION{Dubner2017,
       author = {{Dubner}, Gloria},
        title = "{Radio Emission from Supernova Remnants}",
    booktitle = {Handbook of Supernovae},
         year = 2017,
       editor = {{Alsabti}, Athem W. and {Murdin}, Paul},
        pages = {2041},
          doi = {10.1007/978-3-319-21846-5_91},
       adsurl = {https://ui.adsabs.harvard.edu/abs/2017hsn..book.2041D}
}

@ARTICLE{Fesen1985,
       author = {{Fesen}, R.~A. and {Blair}, W.~P. and {Kirshner}, R.~P.},
        title = "{Optical emission-line properties of evolved galactic supernova remnants.}",
      journal = {\apj},
         year = "1985",
        month = "May",
       volume = {292},
        pages = {29-48},
          doi = {10.1086/163130},
       adsurl = {https://ui.adsabs.harvard.edu/\#abs/1985ApJ...292...29F}
}

@ARTICLE{Fesen2015,
       author = {{Fesen}, Robert A. and {Neustadt}, Jack M.~M. and {Black}, Christine S. and
         {Koeppel}, Ari H.~D.},
        title = "{Discovery of an Apparent High Latitude Galactic Supernova Remnant}",
      journal = {\apj},
         year = 2015,
        month = oct,
       volume = {812},
       number = {1},
          eid = {37},
        pages = {37},
          doi = {10.1088/0004-637X/812/1/37},
archivePrefix = {arXiv},
       eprint = {1508.06291},
 primaryClass = {astro-ph.HE},
       adsurl = {https://ui.adsabs.harvard.edu/abs/2015ApJ...812...37F}
}

@ARTICLE{Fesen2019,
       author = {{Fesen}, Robert A. and {Neustadt}, Jack M.~M. and {How}, Thomas G. and
         {Black}, Christine S.},
        title = "{Detection of extensive optical emission from the extremely radio faint Galactic supernova remnant G182.4+4.3}",
      journal = {\mnras},
         year = 2019,
        month = jul,
       volume = {486},
       number = {4},
        pages = {4701-4709},
          doi = {10.1093/mnras/stz1140},
archivePrefix = {arXiv},
       eprint = {1905.08901},
 primaryClass = {astro-ph.HE},
       adsurl = {https://ui.adsabs.harvard.edu/abs/2019MNRAS.486.4701F}
}

@ARTICLE{Fesen2020,
       author = {{Fesen}, Robert A. and {Weil}, Kathryn E. and {Raymond}, John C. and {Huet}, Laurent and {Rusterholz}, Martin and {di Cicco}, Dennis and {Mittelman}, David and {Walker}, Sean and {Drechsler}, Marcel and {Faworski}, Sheldon},
        title = "{G107.0+9.0: a new large optically bright, radio, and X-Ray faint galactic supernova remnant in Cepheus}",
      journal = {\mnras},
         year = 2020,
        month = nov,
       volume = {498},
       number = {4},
        pages = {5194-5206},
          doi = {10.1093/mnras/staa2765},
archivePrefix = {arXiv},
       eprint = {2008.05620},
 primaryClass = {astro-ph.HE},
       adsurl = {https://ui.adsabs.harvard.edu/abs/2020MNRAS.498.5194F}
}

@ARTICLE{Fesen2021,
       author = {{Fesen}, Robert A. and {Drechsler}, Marcel and {Weil}, Kathryn E. and {Strottner}, Xavier and {Raymond}, John C. and {Rupert}, Justin and {Milisavljevic}, Dan and {Subrayan}, Bhagya M. and {di Cicco}, Dennis and {Walker}, Sean and {Mittelman}, David and {Ludgate}, Mathew},
        title = "{Far-UV and Optical Emissions from Three Very Large Supernova Remnants Located at Unusually High Galactic Latitudes}",
      journal = {\apj},
         year = 2021,
        month = oct,
       volume = {920},
       number = {2},
          eid = {90},
        pages = {90},
          doi = {10.3847/1538-4357/ac0ada},
archivePrefix = {arXiv},
       eprint = {2102.12599},
 primaryClass = {astro-ph.HE},
       adsurl = {https://ui.adsabs.harvard.edu/abs/2021ApJ...920...90F}
}

@ARTICLE{Fesen2024,
       author = {{Fesen}, Robert A. and {Drechsler}, Marcel and {Strottner}, Xavier and {Falls}, Bray and {Sainty}, Yann and {Martino}, Nicolas and {Galli}, Richard and {Ludgate}, Mathew and {Blauensteiner}, Markus and {Reich}, Wolfgang and {Walker}, Sean and {di Cicco}, Dennis and {Mittelman}, David and {Morgan}, Curtis and {Kaeouach}, Aziz Ettahar and {Rupert}, Justin and {Benkhaldoun}, Zouhair},
        title = "{Deep Optical Emission-line Images of Nine Known and Three New Galactic Supernova Remnants}",
      journal = {\apjs},
         year = 2024,
        month = jun,
       volume = {272},
       number = {2},
          eid = {36},
        pages = {36},
          doi = {10.3847/1538-4365/ad410a},
archivePrefix = {arXiv},
       eprint = {2403.00317},
 primaryClass = {astro-ph.HE},
       adsurl = {https://ui.adsabs.harvard.edu/abs/2024ApJS..272...36F}
}

@ARTICLE{Filipovic2025,
       author = {{Filipovi{\'c}}, Miroslav D. and {Smeaton}, Zachary J. and {Kothes}, Roland and {Mantovanini}, Silvia and {Kosti{\'c}}, Petar and {Leahy}, Denis and {Ahmad}, Adeel and {Anderson}, Gemma and {Araya}, Miguel and {Ball}, Brianna D. and {Becker}, Werner and {Bordiu}, Cristobal and {Bradley}, Aaron C. and {Brose}, Robert and {Burger-Scheidlin}, Christopher and {Dai}, Shi and {Duchesne}, Stefan and {Galvin}, Timothy J. and {Hopkins}, Andrew M. and {Hurley-Walker}, Natasha and {Koribalski}, B{\"a}rbel S. and {Lazarevi{\'c}}, Sanja and {Lundqvist}, Peter and {Mackey}, Jonathan and {Martin}, Pierrick and {McGee}, Padric and {Mitra{\v{s}}inovi{\'c}}, Ana and {Payne}, Jeffrey L. and {Riggi}, Simone and {Ross}, Kathryn and {Rowell}, Gavin and {Rudnick}, Lawrence and {Sano}, Hidetoshi and {Sasaki}, Manami and {Roberto}, Soria and {Uro{\v{s}}evi{\'c}}, Dejan and {Vukoti{\'c}}, Branislav and {West}, Jennifer},
        title = "{Teleios (G305.4─2.2) ─ the mystery of a perfectly shaped new galactic supernova remnant}",
      journal = {\pasa},
         year = 2025,
        month = aug,
       volume = {42},
          eid = {e104},
        pages = {e104},
          doi = {10.1017/pasa.2025.10045},
archivePrefix = {arXiv},
       eprint = {2505.04041},
 primaryClass = {astro-ph.HE},
       adsurl = {https://ui.adsabs.harvard.edu/abs/2025PASA...42..104F}
}

@ARTICLE{Furst1986,
       author = {{F\"urst}, E. and {Reich}, W.},
        title = "{Multifrequency radio observations of S 147.}",
      journal = {\aap},
         year = 1986,
        month = jul,
       volume = {163},
        pages = {185-193},
       adsurl = {https://ui.adsabs.harvard.edu/abs/1986A&A...163..185F}
}

@ARTICLE{Furst1990,
       author = {{F\"urst}, E. and {Reich}, W. and {Reich}, P. and {Reif}, K.},
        title = "{A radio continuum survey of the Galactic Plane at 11 cmwavelength. III. The area 76deg <= L <= 240deg, -5deg <= B <= 5deg}",
      journal = {\aaps},
         year = 1990,
        month = oct,
       volume = {85},
        pages = {691-803},
       adsurl = {https://ui.adsabs.harvard.edu/abs/1990A&AS...85..691F}
}

@ARTICLE{Gaustad2001,
       author = {{Gaustad}, John E. and {McCullough}, Peter R. and {Rosing}, Wayne and {Van Buren}, Dave},
        title = "{A Robotic Wide-Angle H{\ensuremath{\alpha}} Survey of the Southern Sky}",
      journal = {\pasp},
         year = 2001,
        month = nov,
       volume = {113},
       number = {789},
        pages = {1326-1348},
          doi = {10.1086/323969},
archivePrefix = {arXiv},
       eprint = {astro-ph/0108518},
 primaryClass = {astro-ph},
       adsurl = {https://ui.adsabs.harvard.edu/abs/2001PASP..113.1326G}
}

@ARTICLE{Gao2022,
       author = {{Gao}, XuYang and {Reich}, Wolfgang and {Sun}, XiaoHui and {Zhao}, He and {Hong}, Tao and {Yuan}, ZhongSheng and {Reich}, Patricia and {Han}, JinLin},
        title = "{Peering into the Milky Way by FAST: IV. Identification of two new Galactic supernova remnants G203.1+6.6 and G206.7+5.9}",
      journal = {Science China Physics, Mechanics, and Astronomy},
         year = 2022,
        month = dec,
       volume = {65},
       number = {12},
          eid = {129705},
        pages = {129705},
          doi = {10.1007/s11433-022-2031-7},
archivePrefix = {arXiv},
       eprint = {2211.11408},
 primaryClass = {astro-ph.GA},
       adsurl = {https://ui.adsabs.harvard.edu/abs/2022SCPMA..6529705G}
}

@ARTICLE{Ghavamian2002,
       author = {{Ghavamian}, Parviz and {Winkler}, P. Frank and {Raymond}, John C. and {Long}, Knox S.},
        title = "{The Optical Spectrum of the SN 1006 Supernova Remnant Revisited}",
      journal = {\apj},
         year = 2002,
        month = jun,
       volume = {572},
       number = {2},
        pages = {888-896},
          doi = {10.1086/340437},
archivePrefix = {arXiv},
       eprint = {astro-ph/0202487},
 primaryClass = {astro-ph},
       adsurl = {https://ui.adsabs.harvard.edu/abs/2002ApJ...572..888G}
}

@ARTICLE{Goedhart2024,
       author = {{Goedhart}, S. and {Cotton}, W.~D. and {Camilo}, F. and {Thompson}, M.~A. and {Umana}, G. and {Bietenholz}, M. and {Woudt}, P.~A. and {Anderson}, L.~D. and {Bordiu}, C. and {Buckley}, D.~A.~H. and et al.},
        title = "{The SARAO MeerKAT 1.3 GHz Galactic Plane Survey}",
      journal = {\mnras},
         year = 2024,
        month = jun,
       volume = {531},
       number = {1},
        pages = {649-681},
          doi = {10.1093/mnras/stae1166},
archivePrefix = {arXiv},
       eprint = {2312.07275},
 primaryClass = {astro-ph.GA},
       adsurl = {https://ui.adsabs.harvard.edu/abs/2024MNRAS.531..649G}
}

@ARTICLE{Graham1982,
       author = {{Graham}, D.~A. and {Haslam}, C.~G.~T. and {Salter}, C.~J. and {Wilson}, W.~E.},
        title = "{A continuum study of galactic radio sources in the constellation of Monoceros}",
      journal = {\aap},
         year = 1982,
        month = may,
       volume = {109},
       number = {1},
        pages = {145-154},
       adsurl = {https://ui.adsabs.harvard.edu/abs/1982A&A...109..145G}
}

@ARTICLE{Green2025,
       author = {{Green}, D.~A.},
        title = "{An updated catalogue of 310 Galactic supernova remnants and their statistical properties}",
      journal = {Journal of Astrophysics and Astronomy},
         year = 2025,
        month = jan,
       volume = {46},
       number = {1},
          eid = {14},
        pages = {14},
          doi = {10.1007/s12036-024-10038-4},
archivePrefix = {arXiv},
       eprint = {2411.03367},
 primaryClass = {astro-ph.GA},
       adsurl = {https://ui.adsabs.harvard.edu/abs/2025JApA...46...14G}
}

@ARTICLE{D_Green2014,
       author = {{Green}, D.~A.},
        title = "{A catalogue of 294 Galactic supernova remnants}",
      journal = {Bulletin of the Astronomical Society of India},
         year = 2014,
        month = jun,
       volume = {42},
       number = {2},
        pages = {47-58},
          doi = {10.48550/arXiv.1409.0637},
archivePrefix = {arXiv},
       eprint = {1409.0637},
 primaryClass = {astro-ph.HE},
       adsurl = {https://ui.adsabs.harvard.edu/abs/2014BASI...42...47G}
}

@ARTICLE{D_Green2019,
       author = {{Green}, D.~A.},
        title = "{A revised catalogue of 294 Galactic supernova remnants}",
      journal = {Journal of Astrophysics and Astronomy},
         year = 2019,
        month = aug,
       volume = {40},
       number = {4},
          eid = {36},
        pages = {36},
          doi = {10.1007/s12036-019-9601-6},
archivePrefix = {arXiv},
       eprint = {1907.02638},
 primaryClass = {astro-ph.GA},
       adsurl = {https://ui.adsabs.harvard.edu/abs/2019JApA...40...36G}
}

@ARTICLE{Greimel2021,
       author = {{Greimel}, R. and {Drew}, J.~E. and {Mongui{\'o}}, M. and {Ashley}, R.~P. and {Barentsen}, G. and {Eisl{\"o}ffel}, J. and {Mampaso}, A. and {Morris}, R.~A.~H. and {Naylor}, T. and {Roe}, C. and {Sabin}, L. and {Stecklum}, B. and {Wright}, N.~J. and {Groot}, P.~J. and {Irwin}, M.~J. and {Barlow}, M.~J. and {Fari{\~n}a}, C. and {Fern{\'a}ndez-Mart{\'\i}n}, A. and {Parker}, Q.~A. and {Phillipps}, S. and {Scaringi}, S. and {Zijlstra}, A.~A.},
        title = "{High-resolution H{\ensuremath{\alpha}} imaging of the northern Galactic plane and the IGAPS image database}",
      journal = {\aap},
         year = 2021,
        month = nov,
       volume = {655},
          eid = {A49},
        pages = {A49},
          doi = {10.1051/0004-6361/202140950},
archivePrefix = {arXiv},
       eprint = {2107.12897},
 primaryClass = {astro-ph.GA},
       adsurl = {https://ui.adsabs.harvard.edu/abs/2021A&A...655A..49G}
}

@ARTICLE{Haffner2003,
       author = {{Haffner}, L.~M. and {Reynolds}, R.~J. and {Tufte}, S.~L. and {Madsen}, G.~J. and {Jaehnig}, K.~P. and {Percival}, J.~W.},
        title = "{The Wisconsin H{\ensuremath{\alpha}} Mapper Northern Sky Survey}",
      journal = {\apjs},
         year = 2003,
        month = dec,
       volume = {149},
       number = {2},
        pages = {405-422},
          doi = {10.1086/378850},
archivePrefix = {arXiv},
       eprint = {astro-ph/0309117},
 primaryClass = {astro-ph},
       adsurl = {https://ui.adsabs.harvard.edu/abs/2003ApJS..149..405H}
}

@ARTICLE{Heiles1976,
       author = {{Heiles}, C. and {Jenkins}, E.~B.},
        title = "{An almost complete survey of 21-cm line radiation for}",
      journal = {\aap},
         year = 1976,
        month = feb,
       volume = {46},
       number = {3},
        pages = {333-360},
       adsurl = {https://ui.adsabs.harvard.edu/abs/1976A&A....46..333H}
}

@ARTICLE{Heiles1979,
       author = {{Heiles}, C.},
        title = "{H I shells and supershells}",
      journal = {\apj},
         year = 1979,
        month = apr,
       volume = {229},
        pages = {533-537},
          doi = {10.1086/156986},
       adsurl = {https://ui.adsabs.harvard.edu/abs/1979ApJ...229..533H}
}

@ARTICLE{Heywood2022,
       author = {{Heywood}, I. and {Rammala}, I. and {Camilo}, F. and {Cotton}, W.~D. and {Yusef-Zadeh}, F. and {Abbott}, T.~D. and {Adam}, R.~M. and {Adams}, G. and {Aldera}, M.~A. and {Asad}, K.~M.~B. and et al.},
        title = "{The 1.28 GHz MeerKAT Galactic Center Mosaic}",
      journal = {\apj},
         year = 2022,
        month = feb,
       volume = {925},
       number = {2},
          eid = {165},
        pages = {165},
          doi = {10.3847/1538-4357/ac449a},
archivePrefix = {arXiv},
       eprint = {2201.10541},
 primaryClass = {astro-ph.GA},
       adsurl = {https://ui.adsabs.harvard.edu/abs/2022ApJ...925..165H}
}

@ARTICLE{HI4PI2016,
       author = {{HI4PI Collaboration} and {Ben Bekhti}, N. and {Fl{\"o}er}, L. and {Keller}, R. and {Kerp}, J. and {Lenz}, D. and {Winkel}, B. and {Bailin}, J. and {Calabretta}, M.~R. and {Dedes}, L. and {Ford}, H.~A. and {Gibson}, B.~K. and {Haud}, U. and {Janowiecki}, S. and {Kalberla}, P.~M.~W. and {Lockman}, F.~J. and {McClure-Griffiths}, N.~M. and {Murphy}, T. and {Nakanishi}, H. and {Pisano}, D.~J. and {Staveley-Smith}, L.},
        title = "{HI4PI: A full-sky H I survey based on EBHIS and GASS}",
      journal = {\aap},
         year = 2016,
        month = oct,
       volume = {594},
          eid = {A116},
        pages = {A116},
          doi = {10.1051/0004-6361/201629178},
archivePrefix = {arXiv},
       eprint = {1610.06175},
 primaryClass = {astro-ph.GA},
       adsurl = {https://ui.adsabs.harvard.edu/abs/2016A&A...594A.116H}
}

@ARTICLE{Hartigan1987,
       author = {{Hartigan}, Patrick and {Raymond}, John and {Hartmann}, Lee},
        title = "{Radiative Bow Shock Models of Herbig-Haro Objects}",
      journal = {\apj},
         year = 1987,
        month = may,
       volume = {316},
        pages = {323},
          doi = {10.1086/165204},
       adsurl = {https://ui.adsabs.harvard.edu/abs/1987ApJ...316..323H}
}

@ARTICLE{Heng2007,
       author = {{Heng}, Kevin and {McCray}, Richard},
        title = "{Balmer-dominated Shocks Revisited}",
      journal = {\apj},
         year = 2007,
        month = jan,
       volume = {654},
       number = {2},
        pages = {923-937},
          doi = {10.1086/509601},
archivePrefix = {arXiv},
       eprint = {astro-ph/0609331},
 primaryClass = {astro-ph},
       adsurl = {https://ui.adsabs.harvard.edu/abs/2007ApJ...654..923H}
}

@ARTICLE{Heng2010,
       author = {{Heng}, Kevin},
        title = "{Balmer-Dominated Shocks: A Concise Review}",
      journal = {\pasa},
         year = 2010,
        month = mar,
       volume = {27},
       number = {1},
        pages = {23-44},
          doi = {10.1071/AS09057},
archivePrefix = {arXiv},
       eprint = {0908.4080},
 primaryClass = {astro-ph.GA},
       adsurl = {https://ui.adsabs.harvard.edu/abs/2010PASA...27...23H}
}

@ARTICLE{How2018,
       author = {{How}, Thomas G. and {Fesen}, Robert A. and {Neustadt}, Jack M.~M. and
         {Black}, Christine S. and {Outters}, Nicolas},
        title = "{Optical emission associated with the Galactic supernova remnant G179.0+2.6}",
      journal = {\mnras},
         year = 2018,
        month = aug,
       volume = {478},
       number = {2},
        pages = {1987-1993},
          doi = {10.1093/mnras/sty1007},
archivePrefix = {arXiv},
       eprint = {1804.07403},
 primaryClass = {astro-ph.HE},
       adsurl = {https://ui.adsabs.harvard.edu/abs/2018MNRAS.478.1987H}
}

@ARTICLE{Hurley_Walker2019,
       author = {{Hurley-Walker}, N. and {Gaensler}, B.~M. and {Leahy}, D.~A. and {Filipovi{\'c}}, M.~D. and {Hancock}, P.~J. and {Franzen}, T.~M.~O. and {Offringa}, A.~R. and {Callingham}, J.~R. and {Hindson}, L. and {Wu}, C. and {Bell}, M.~E. and {For}, B.-Q. and {Johnston-Hollitt}, M. and {Kapi{\'n}ska}, A.~D. and {Morgan}, J. and {Murphy}, T. and {McKinley}, B. and {Procopio}, P. and {Staveley-Smith}, L. and {Wayth}, R.~B. and {Zheng}, Q.},
        title = "{Candidate radio supernova remnants observed by the GLEAM survey over 345{\textdegree} < l < 60{\textdegree} and 180{\textdegree} < l < 240{\textdegree}}",
      journal = {\pasa},
         year = 2019,
        month = nov,
       volume = {36},
          eid = {e048},
        pages = {e048},
          doi = {10.1017/pasa.2019.33},
archivePrefix = {arXiv},
       eprint = {1911.08124},
 primaryClass = {astro-ph.HE},
       adsurl = {https://ui.adsabs.harvard.edu/abs/2019PASA...36...48H}
}

@ARTICLE{Jenkins1974,
       author = {{Jenkins}, E.~B. and {Savage}, B.~D.},
        title = "{Ultraviolet photometry from the Orbiting Astronomical Observatory. XIV. An extension of the survey of Lyman-alpha absorption from interstellar hydrogen.}",
      journal = {\apj},
         year = 1974,
        month = jan,
       volume = {187},
        pages = {243-255},
          doi = {10.1086/152620},
       adsurl = {https://ui.adsabs.harvard.edu/abs/1974ApJ...187..243J}
}

@ARTICLE{Joubaud2019,
       author = {{Joubaud}, T. and {Grenier}, I.~A. and {Ballet}, J. and {Soler}, J.~D.},
        title = "{Gas shells and magnetic fields in the Orion-Eridanus superbubble}",
      journal = {\aap},
         year = 2019,
        month = nov,
       volume = {631},
          eid = {A52},
        pages = {A52},
          doi = {10.1051/0004-6361/201936239},
archivePrefix = {arXiv},
       eprint = {1909.10083},
 primaryClass = {astro-ph.HE},
       adsurl = {https://ui.adsabs.harvard.edu/abs/2019A&A...631A..52J}
}

@ARTICLE{Jin2025,
       author = {{Jin}, Yifei and {Raymond}, John},
        title = "{Dialog Concerning the Two Shock Codes}",
      journal = {\apj},
         year = 2025,
        month = aug,
       volume = {989},
       number = {2},
          eid = {203},
        pages = {203},
          doi = {10.3847/1538-4357/adeca2},
archivePrefix = {arXiv},
       eprint = {2507.03225},
 primaryClass = {astro-ph.GA},
       adsurl = {https://ui.adsabs.harvard.edu/abs/2025ApJ...989..203J}
}

@INPROCEEDINGS{Joye2003,
       author = {{Joye}, W.~A. and {Mandel}, E.},
        title = "{New Features of SAOImage DS9}",
    booktitle = {Astronomical Data Analysis Software and Systems XII},
         year = 2003,
       editor = {{Payne}, H.~E. and {Jedrzejewski}, R.~I. and {Hook}, R.~N.},
       series = {Astronomical Society of the Pacific Conference Series},
       volume = {295},
        month = jan,
        pages = {489},
       adsurl = {https://ui.adsabs.harvard.edu/abs/2003ASPC..295..489J}
}

@ARTICLE{Kim2007,
       author = {{Kim}, I.-J. and {Min}, K.-W. and {Seon}, K.-I. and {Park}, J.-W. and {Han}, W. and {Park}, J.-H. and {Nam}, U.-W. and {Edelstein}, J. and {Sankrit}, R. and {Korpela}, E.~J.},
        title = "{Far-Ultraviolet Observations of the Monogem Ring}",
      journal = {\apjl},
         year = 2007,
        month = aug,
       volume = {665},
       number = {2},
        pages = {L139-L142},
          doi = {10.1086/521441},
       adsurl = {https://ui.adsabs.harvard.edu/abs/2007ApJ...665L.139K}
}

@ARTICLE{Kop2020,
       author = {{Kopsacheili}, M. and {Zezas}, A. and {Leonidaki}, I.},
        title = "{A diagnostic tool for the identification of supernova remnants}",
      journal = {\mnras},
         year = "2020",
        month = "Jan",
       volume = {491},
       number = {1},
        pages = {889-902},
          doi = {10.1093/mnras/stz2594},
       adsurl = {https://ui.adsabs.harvard.edu/abs/2020MNRAS.491..889K}
}

@ARTICLE{Kop2021,
       author = {{Kopsacheili}, M. and {Zezas}, A. and {Leonidaki}, I. and {Boumis}, P.},
        title = "{The supernova remnant populations of the galaxies NGC 45, NGC 55, NGC 1313, NGC 7793: luminosity and excitation functions}",
      journal = {\mnras},
         year = 2021,
        month = nov,
       volume = {507},
       number = {4},
        pages = {6020-6036},
          doi = {10.1093/mnras/stab2395},
archivePrefix = {arXiv},
       eprint = {2108.07819},
 primaryClass = {astro-ph.GA},
       adsurl = {https://ui.adsabs.harvard.edu/abs/2021MNRAS.507.6020K}
}

@ARTICLE{Knies2018,
       author = {{Knies}, Jonathan R. and {Sasaki}, Manami and {Plucinsky}, Paul P.},
        title = "{Suzaku observations of the Monogem Ring and the origin of the Gemini H {\ensuremath{\alpha}} ring}",
      journal = {\mnras},
         year = 2018,
        month = jul,
       volume = {477},
       number = {4},
        pages = {4414-4422},
          doi = {10.1093/mnras/sty915},
       adsurl = {https://ui.adsabs.harvard.edu/abs/2018MNRAS.477.4414K}
}

@ARTICLE{Knies2024,
       author = {{Knies}, J.~R. and {Sasaki}, M. and {Becker}, W. and {Liu}, T. and {Ponti}, G. and {Plucinsky}, P.~P.},
        title = "{A new understanding of the Gemini-Monoceros X-ray enhancement from discoveries with eROSITA}",
      journal = {\aap},
         year = 2024,
        month = aug,
       volume = {688},
          eid = {A90},
        pages = {A90},
          doi = {10.1051/0004-6361/202348834},
archivePrefix = {arXiv},
       eprint = {2401.17289},
 primaryClass = {astro-ph.HE},
       adsurl = {https://ui.adsabs.harvard.edu/abs/2024A&A...688A..90K}
}

@ARTICLE{Laycock2010,
       author = {{Laycock}, S. and {Tang}, S. and {Grindlay}, J. and {Los}, E. and {Simcoe}, R. and {Mink}, D.},
        title = "{Digital Access to a Sky Century at Harvard: Initial Photometry and Astrometry}",
      journal = {\aj},
         year = 2010,
        month = oct,
       volume = {140},
       number = {4},
        pages = {1062-1077},
          doi = {10.1088/0004-6256/140/4/1062},
       adsurl = {https://ui.adsabs.harvard.edu/abs/2010AJ....140.1062L}
}

@ARTICLE{Lee2009,
       author = {{Lee}, Hsu-Tai and {Chen}, W.~P.},
        title = "{Triggered Star Formation on the Border of the Orion-Eridanus Superbubble}",
      journal = {\apj},
         year = 2009,
        month = apr,
       volume = {694},
       number = {2},
        pages = {1423-1434},
          doi = {10.1088/0004-637X/694/2/1423},
archivePrefix = {arXiv},
       eprint = {astro-ph/0608216},
 primaryClass = {astro-ph},
       adsurl = {https://ui.adsabs.harvard.edu/abs/2009ApJ...694.1423L}
}

@ARTICLE{Levenson1998,
       author = {{Levenson}, N.~A. and {Graham}, James R. and {Keller}, Luke D. and {Richter}, Matthew J.},
        title = "{Panoramic Views of the Cygnus Loop}",
      journal = {\apjs},
         year = 1998,
        month = oct,
       volume = {118},
       number = {2},
        pages = {541-561},
          doi = {10.1086/313136},
archivePrefix = {arXiv},
       eprint = {astro-ph/9805008},
 primaryClass = {astro-ph},
       adsurl = {https://ui.adsabs.harvard.edu/abs/1998ApJS..118..541L}
}

@ARTICLE{Long1977,
       author = {{Long}, K.~S. and {Patterson}, J.~R. and {Moore}, W.~E. and {Garmire}, G.~P.},
        title = "{A study of four soft X-ray enhancements.}",
      journal = {\apj},
         year = 1977,
        month = mar,
       volume = {212},
        pages = {427-437},
          doi = {10.1086/155061},
       adsurl = {https://ui.adsabs.harvard.edu/abs/1977ApJ...212..427L}
}

@INCOLLECTION{Long2017,
       author = {{Long}, Knox S.},
        title = "{Galactic and Extragalactic Samples of Supernova Remnants: How They Are Identified and What They Tell Us}",
    booktitle = {Handbook of Supernovae},
         year = 2017,
       editor = {{Alsabti}, Athem W. and {Murdin}, Paul},
        pages = {2005},
          doi = {10.1007/978-3-319-21846-5_90},
       adsurl = {https://ui.adsabs.harvard.edu/abs/2017hsn..book.2005L}
}

@ARTICLE{Long1990,
       author = {{Long}, Knox S. and {Blair}, William P. and {Kirshner}, Robert P. and {Winkler}, P. Frank},
        title = "{An Atlas of Confirmed and Candidate Supernova Remnants in M33}",
      journal = {\apjs},
         year = 1990,
        month = jan,
       volume = {72},
        pages = {61},
          doi = {10.1086/191409},
       adsurl = {https://ui.adsabs.harvard.edu/abs/1990ApJS...72...61L}
}

@ARTICLE{Lozinskaya1976,
       author = {{Lozinskaya}, T.~A.},
        title = "{Optical observations of supernova remnants: the filamentary nebula Simeiz 147}",
      journal = {\sovast},
         year = 1976,
        month = feb,
       volume = {20},
        pages = {19},
       adsurl = {https://ui.adsabs.harvard.edu/abs/1976SvA....20...19L}
}

@ARTICLE{Martini2011,
       author = {{Martini}, Paul and {Stoll}, Rebecca and {Derwent}, M.~A. and
         {Zhelem}, R. and {Atwood}, B. and {Gonzalez}, R. and {Mason}, J.~A. and
         {O'Brien}, T.~P. and {Pappalardo}, D.~P. and {Pogge}, Richard W.},
        title = "{The Ohio State Multi-Object Spectrograph}",
      journal = {\pasp},
         year = "2011",
        month = "Feb",
       volume = {123},
       number = {900},
        pages = {187},
          doi = {10.1086/658357},
       adsurl = {https://ui.adsabs.harvard.edu/abs/2011PASP..123..187M}
}

@ARTICLE{Madsen2006,
       author = {{Madsen}, G.~J. and {Reynolds}, R.~J. and {Haffner}, L.~M.},
        title = "{A Multiwavelength Optical Emission Line Survey of Warm Ionized Gas in the Galaxy}",
      journal = {\apj},
         year = 2006,
        month = nov,
       volume = {652},
       number = {1},
        pages = {401-425},
          doi = {10.1086/508441},
archivePrefix = {arXiv},
       eprint = {astro-ph/0609558},
 primaryClass = {astro-ph},
       adsurl = {https://ui.adsabs.harvard.edu/abs/2006ApJ...652..401M}
}

@ARTICLE{Mathewson1972,
       author = {{Mathewson}, D.~S. and {Clarke}, J.~N.},
        title = "{A Supernova Remnant in the Small Magellanic Cloud}",
      journal = {\apjl},
         year = 1972,
        month = dec,
       volume = {178},
        pages = {L105},
          doi = {10.1086/181095},
       adsurl = {https://ui.adsabs.harvard.edu/abs/1972ApJ...178L.105M}
}

@ARTICLE{Mathewson1973,
       author = {{Mathewson}, D.~S. and {Clarke}, J.~N.},
        title = "{Supernova Remnants in the Magellanic Clouds}",
      journal = {\apj},
         year = 1973,
        month = jun,
       volume = {182},
        pages = {697-698},
          doi = {10.1086/152177},
       adsurl = {https://ui.adsabs.harvard.edu/abs/1973ApJ...182..697M}
}

@ARTICLE{Massey1990,
       author = {{Massey}, Philip and {Gronwall}, Caryl},
        title = "{The Kitt Peak Spectrophotometric Standards: Extension to 1 Micron}",
      journal = {\apj},
         year = 1990,
        month = jul,
       volume = {358},
        pages = {344},
          doi = {10.1086/168991},
       adsurl = {https://ui.adsabs.harvard.edu/abs/1990ApJ...358..344M}
}

@ARTICLE{MC1973,
       author = {{Mathewson}, D.~S. and {Clarke}, J.~N.},
        title = "{Supernova remnants in the Large Magellanic Cloud.}",
      journal = {\apj},
         year = 1973,
        month = mar,
       volume = {180},
        pages = {725-738},
          doi = {10.1086/152002},
       adsurl = {https://ui.adsabs.harvard.edu/abs/1973ApJ...180..725M}
}

@ARTICLE{Mav2001,
       author = {{Mavromatakis}, F. and {Papamastorakis}, J. and {Ventura}, J. and {Becker}, W. and {Paleologou}, E.~V. and {Schaudel}, D.},
        title = "{The supernova remnants G 67.7+1.8, G 31.5-0.6 and G 49.2-0.7}",
      journal = {\aap},
         year = 2001,
        month = apr,
       volume = {370},
        pages = {265-272},
          doi = {10.1051/0004-6361:20010137},
archivePrefix = {arXiv},
       eprint = {astro-ph/0101198},
 primaryClass = {astro-ph},
       adsurl = {https://ui.adsabs.harvard.edu/abs/2001A&A...370..265M}
}

@ARTICLE{Mav2005,
       author = {{Mavromatakis}, F. and {Boumis}, P. and {Xilouris}, E. and
         {Papamastorakis}, J. and {Alikakos}, J.},
        title = "{The faint supernova remnant G 116.5+1.1 and the detection of a new candidate remnant}",
      journal = {\aap},
         year = 2005,
        month = may,
       volume = {435},
       number = {1},
        pages = {141-149},
          doi = {10.1051/0004-6361:20042187},
archivePrefix = {arXiv},
       eprint = {astro-ph/0502066},
 primaryClass = {astro-ph},
       adsurl = {https://ui.adsabs.harvard.edu/abs/2005A&A...435..141M}
}

@ARTICLE{Mav2009,
       author = {{Mavromatakis}, F. and {Boumis}, P. and {Meaburn}, J. and {Caulet}, A.},
        title = "{A new candidate supernova remnant G 70.5+1.9}",
      journal = {\aap},
         year = 2009,
        month = aug,
       volume = {503},
       number = {1},
        pages = {129-136},
          doi = {10.1051/0004-6361/200912211},
archivePrefix = {arXiv},
       eprint = {0905.3480},
 primaryClass = {astro-ph.GA},
       adsurl = {https://ui.adsabs.harvard.edu/abs/2009A&A...503..129M}
}

@ARTICLE{Nousek1981,
       author = {{Nousek}, J.~A. and {Cowie}, L.~L. and {Hu}, E. and {Lindblad}, C.~J. and {Garmire}, G.~P.},
        title = "{The Gem-Mon X-ray enhancement : a giant X-ray ring.}",
      journal = {\apj},
         year = 1981,
        month = aug,
       volume = {248},
        pages = {152-160},
          doi = {10.1086/159139},
       adsurl = {https://ui.adsabs.harvard.edu/abs/1981ApJ...248..152N}
}

@ARTICLE{Odegard1986,
       author = {{Odegard}, N.},
        title = "{Decameter Wavelength Observations of the Rosette Nebula and the Monoceros Loop Supernova Remnant}",
      journal = {\apj},
         year = 1986,
        month = feb,
       volume = {301},
        pages = {813},
          doi = {10.1086/163945},
       adsurl = {https://ui.adsabs.harvard.edu/abs/1986ApJ...301..813O}
}

@ARTICLE{Oke1974A,
       author = {{Oke}, J.~B.},
        title = "{Absolute Spectral Energy Distributions for White Dwarfs}",
      journal = {\apjs},
         year = 1974,
        month = feb,
       volume = {27},
        pages = {21},
          doi = {10.1086/190287},
       adsurl = {https://ui.adsabs.harvard.edu/abs/1974ApJS...27...21O}
}

@ARTICLE{Ochsendorf2015,
       author = {{Ochsendorf}, Bram B. and {Brown}, Anthony G.~A. and {Bally}, John and {Tielens}, Alexander G.~G.~M.},
        title = "{Nested Shells Reveal the Rejuvenation of the Orion-Eridanus Superbubble}",
      journal = {\apj},
         year = 2015,
        month = aug,
       volume = {808},
       number = {2},
          eid = {111},
        pages = {111},
          doi = {10.1088/0004-637X/808/2/111},
archivePrefix = {arXiv},
       eprint = {1506.02426},
 primaryClass = {astro-ph.GA},
       adsurl = {https://ui.adsabs.harvard.edu/abs/2015ApJ...808..111O}
}

@BOOK{Osterbrock2006,
       author = {{Osterbrock}, Donald E. and {Ferland}, Gary J.},
        title = "{Astrophysics of gaseous nebulae and active galactic nuclei}",
         year = 2006,
       adsurl = {https://ui.adsabs.harvard.edu/abs/2006agna.book.....O}
}

@ARTICLE{Parker2005,
       author = {{Parker}, Quentin A. and {Phillipps}, S. and {Pierce}, M.~J. and {Hartley}, M. and {Hambly}, N.~C. and {Read}, M.~A. and {MacGillivray}, H.~T. and {Tritton}, S.~B. and {Cass}, C.~P. and {Cannon}, R.~D. and {Cohen}, M. and {Drew}, J.~E. and {Frew}, D.~J. and {Hopewell}, E. and {Mader}, S. and {Malin}, D.~F. and {Masheder}, M.~R.~W. and {Morgan}, D.~H. and {Morris}, R.~A.~H. and {Russeil}, D. and {Russell}, K.~S. and {Walker}, R.~N.~F.},
        title = "{The AAO/UKST SuperCOSMOS H{\ensuremath{\alpha}} survey}",
      journal = {\mnras},
         year = 2005,
        month = sep,
       volume = {362},
       number = {2},
        pages = {689-710},
          doi = {10.1111/j.1365-2966.2005.09350.x},
archivePrefix = {arXiv},
       eprint = {astro-ph/0506599},
 primaryClass = {astro-ph},
       adsurl = {https://ui.adsabs.harvard.edu/abs/2005MNRAS.362..689P}
}

@ARTICLE{PlanckCollaboration2014P,
       author = {{Planck Collaboration} and {Abergel}, A. and {Ade}, P.~A.~R. and {Aghanim}, N. and {Alves}, M.~I.~R. and {Aniano}, G. and {Armitage-Caplan}, C. and {Arnaud}, M. and {Ashdown}, M. and {Atrio-Barandela}, F. and {Aumont}, J. and {Baccigalupi}, C. and {Banday}, A.~J. and {Barreiro}, R.~B. and {Bartlett}, J.~G. and {Battaner}, E. and {Benabed}, K. and {Beno{\^\i}t}, A. and {Benoit-L{\'e}vy}, A. and {Bernard}, J.-P. and {Bersanelli}, M. and {Bielewicz}, P. and {Bobin}, J. and {Bock}, J.~J. and {Bonaldi}, A. and {Bond}, J.~R. and {Borrill}, J. and {Bouchet}, F.~R. and {Boulanger}, F. and {Bridges}, M. and {Bucher}, M. and {Burigana}, C. and {Butler}, R.~C. and {Cardoso}, J.-F. and {Catalano}, A. and {Chamballu}, A. and {Chary}, R.-R. and {Chiang}, H.~C. and {Chiang}, L.-Y. and {Christensen}, P.~R. and {Church}, S. and {Clemens}, M. and {Clements}, D.~L. and {Colombi}, S. and {Colombo}, L.~P.~L. and {Combet}, C. and {Couchot}, F. and {Coulais}, A. and {Crill}, B.~P. and {Curto}, A. and {Cuttaia}, F. and {Danese}, L. and {Davies}, R.~D. and {Davis}, R.~J. and {de Bernardis}, P. and {de Rosa}, A. and {de Zotti}, G. and {Delabrouille}, J. and {Delouis}, J.-M. and {D{\'e}sert}, F.-X. and {Dickinson}, C. and {Diego}, J.~M. and {Dole}, H. and {Donzelli}, S. and {Dor{\'e}}, O. and {Douspis}, M. and {Draine}, B.~T. and {Dupac}, X. and {Efstathiou}, G. and {En{\ss}lin}, T.~A. and {Eriksen}, H.~K. and {Falgarone}, E. and {Finelli}, F. and {Forni}, O. and {Frailis}, M. and {Fraisse}, A.~A. and {Franceschi}, E. and {Galeotta}, S. and {Ganga}, K. and {Ghosh}, T. and {Giard}, M. and {Giardino}, G. and {Giraud-H{\'e}raud}, Y. and {Gonz{\'a}lez-Nuevo}, J. and {G{\'o}rski}, K.~M. and {Gratton}, S. and {Gregorio}, A. and {Grenier}, I.~A. and {Gruppuso}, A. and {Guillet}, V. and {Hansen}, F.~K. and {Hanson}, D. and {Harrison}, D.~L. and {Helou}, G. and {Henrot-Versill{\'e}}, S. and {Hern{\'a}ndez-Monteagudo}, C. and {Herranz}, D. and {Hildebrandt}, S.~R. and {Hivon}, E. and {Hobson}, M. and {Holmes}, W.~A. and {Hornstrup}, A. and {Hovest}, W. and {Huffenberger}, K.~M. and {Jaffe}, A.~H. and {Jaffe}, T.~R. and {Jewell}, J. and {Joncas}, G. and {Jones}, W.~C. and {Juvela}, M. and {Keih{\"a}nen}, E. and {Keskitalo}, R. and {Kisner}, T.~S. and {Knoche}, J. and {Knox}, L. and {Kunz}, M. and {Kurki-Suonio}, H. and {Lagache}, G. and {L{\"a}hteenm{\"a}ki}, A. and {Lamarre}, J.-M. and {Lasenby}, A. and {Laureijs}, R.~J. and {Lawrence}, C.~R. and {Leonardi}, R. and {Le{\'o}n-Tavares}, J. and {Lesgourgues}, J. and {Levrier}, F. and {Liguori}, M. and {Lilje}, P.~B. and {Linden-V{\o}rnle}, M. and {L{\'o}pez-Caniego}, M. and {Lubin}, P.~M. and {Mac{\'\i}as-P{\'e}rez}, J.~F. and {Maffei}, B. and {Maino}, D. and {Mandolesi}, N. and {Maris}, M. and {Marshall}, D.~J. and {Martin}, P.~G. and {Mart{\'\i}nez-Gonz{\'a}lez}, E. and {Masi}, S. and {Massardi}, M. and {Matarrese}, S. and {Matthai}, F. and {Mazzotta}, P. and {McGehee}, P. and {Melchiorri}, A. and {Mendes}, L. and {Mennella}, A. and {Migliaccio}, M. and {Mitra}, S. and {Miville-Desch{\^e}nes}, M.-A. and {Moneti}, A. and {Montier}, L. and {Morgante}, G. and {Mortlock}, D. and {Munshi}, D. and {Murphy}, J.~A. and {Naselsky}, P. and {Nati}, F. and {Natoli}, P. and {Netterfield}, C.~B. and {N{\o}rgaard-Nielsen}, H.~U. and {Noviello}, F. and {Novikov}, D. and {Novikov}, I. and {Osborne}, S. and {Oxborrow}, C.~A. and {Paci}, F. and {Pagano}, L. and {Pajot}, F. and {Paladini}, R. and {Paoletti}, D. and {Pasian}, F. and {Patanchon}, G. and {Perdereau}, O. and {Perotto}, L. and {Perrotta}, F. and {Piacentini}, F. and {Piat}, M. and {Pierpaoli}, E. and {Pietrobon}, D. and {Plaszczynski}, S. and {Pointecouteau}, E. and {Polenta}, G. and {Ponthieu}, N. and {Popa}, L. and {Poutanen}, T. and {Pratt}, G.~W. and {Pr{\'e}zeau}, G. and {Prunet}, S. and {Puget}, J.-L. and {Rachen}, J.~P. and {Reach}, W.~T. and {Rebolo}, R. and {Reinecke}, M. and {Remazeilles}, M. and {Renault}, C. and {Ricciardi}, S. and {Riller}, T.},
        title = "{Planck 2013 results. XI. All-sky model of thermal dust emission}",
      journal = {\aap},
         year = 2014,
        month = nov,
       volume = {571},
          eid = {A11},
        pages = {A11},
          doi = {10.1051/0004-6361/201323195},
archivePrefix = {arXiv},
       eprint = {1312.1300},
 primaryClass = {astro-ph.GA},
       adsurl = {https://ui.adsabs.harvard.edu/abs/2014A&A...571A..11P}
}

@ARTICLE{Plucinsky1996,
       author = {{Plucinsky}, Paul P. and {Snowden}, Steven L. and {Aschenbach}, Bernd and {Egger}, Roland and {Edgar}, Richard J. and {McCammon}, Dan},
        title = "{ROSAT Survey Observations of the Monogem Ring}",
      journal = {\apj},
         year = 1996,
        month = may,
       volume = {463},
        pages = {224},
          doi = {10.1086/177236},
       adsurl = {https://ui.adsabs.harvard.edu/abs/1996ApJ...463..224P}
}

@ARTICLE{Raymond79,
       author = {{Raymond}, J.~C.},
        title = "{Shock waves in the interstellar medium.}",
      journal = {The Astrophysical Journal Supplement Series},
         year = "1979",
        month = "Jan",
       volume = {39},
        pages = {1-27},
          doi = {10.1086/190562},
       adsurl = {https://ui.adsabs.harvard.edu/abs/1979ApJS...39....1R}
}

@ARTICLE{Reynolds1979,
       author = {{Reynolds}, R.~J. and {Ogden}, P.~M.},
        title = "{Optical evidence for a very large, expanding shell associated with the I Orion OB association, Barnard's loop, and the high galactic latitude Halpha filaments in Eridanus.}",
      journal = {\apj},
         year = 1979,
        month = may,
       volume = {229},
        pages = {942-953},
          doi = {10.1086/157028},
       adsurl = {https://ui.adsabs.harvard.edu/abs/1979ApJ...229..942R}
}

@ARTICLE{Reynolds2011,
       author = {{Reynolds}, S.~P.},
        title = "{Particle acceleration in supernova-remnant shocks}",
      journal = {\apss},
         year = 2011,
        month = nov,
       volume = {336},
       number = {1},
        pages = {257-262},
          doi = {10.1007/s10509-010-0559-8},
archivePrefix = {arXiv},
       eprint = {1012.1306},
 primaryClass = {astro-ph.HE},
       adsurl = {https://ui.adsabs.harvard.edu/abs/2011Ap&SS.336..257R}
}

@ARTICLE{Raymond1979,
       author = {{Raymond}, J.~C.},
        title = "{Shock waves in the interstellar medium.}",
      journal = {\apjs},
         year = 1979,
        month = jan,
       volume = {39},
        pages = {1-27},
          doi = {10.1086/190562},
       adsurl = {https://ui.adsabs.harvard.edu/abs/1979ApJS...39....1R}
}

@INPROCEEDINGS{Reich2002,
       author = {{Reich}, W.},
        title = "{Radio Observations of Supernova Remnants}",
    booktitle = {Neutron Stars, Pulsars, and Supernova Remnants},
         year = 2002,
       editor = {{Becker}, W. and {Lesch}, H. and {Tr{\"u}mper}, J.},
        month = jan,
        pages = {1},
          doi = {10.48550/arXiv.astro-ph/0208498},
archivePrefix = {arXiv},
       eprint = {astro-ph/0208498},
 primaryClass = {astro-ph},
       adsurl = {https://ui.adsabs.harvard.edu/abs/2002nsps.conf....1R}
}

@INPROCEEDINGS{Reich2004,
       author = {{Reich}, W. and {F{\"u}rst}, E. and {Reich}, P. and {Uyaniker}, B. and {Wielebinski}, R. and {Wolleben}, M.},
        title = "{The Effelsberg 1.4 GHz Medium Galactic Latitude Survey (EMLS)}",
    booktitle = {The Magnetized Interstellar Medium},
         year = 2004,
       editor = {{Uyaniker}, B. and {Reich}, W. and {Wielebinski}, R.},
        month = feb,
        pages = {45-50},
       adsurl = {https://ui.adsabs.harvard.edu/abs/2004mim..proc...45R}
}

@ARTICLE{Reich2020,
       author = {{Reich}, Wolfgang and {Reich}, Patricia and {Sun}, Xiaohui},
        title = "{Long, depolarising H{\ensuremath{\alpha}}-filament towards the Monogem ring}",
      journal = {\aap},
         year = 2020,
        month = sep,
       volume = {641},
          eid = {A121},
        pages = {A121},
          doi = {10.1051/0004-6361/202038349},
archivePrefix = {arXiv},
       eprint = {2007.08221},
 primaryClass = {astro-ph.GA},
       adsurl = {https://ui.adsabs.harvard.edu/abs/2020A&A...641A.121R}
}

@ARTICLE{Reich2021,
       author = {{Reich}, Wolfgang and {Gao}, Xuyang and {Reich}, Patricia},
        title = "{Radio properties of the optically identified supernova remnant G107.0+9.0}",
      journal = {\aap},
         year = 2021,
        month = nov,
       volume = {655},
          eid = {A10},
        pages = {A10},
          doi = {10.1051/0004-6361/202140844},
archivePrefix = {arXiv},
       eprint = {2108.08575},
 primaryClass = {astro-ph.GA},
       adsurl = {https://ui.adsabs.harvard.edu/abs/2021A&A...655A..10R}
}

@ARTICLE{Ren2018,
       author = {{Ren}, Juan-Juan and {Liu}, Xiao-Wei and {Chen}, Bing-Qiu and {Xiang}, Mao-Sheng and {Yuan}, Hai-Bo and {Huang}, Yang and {Zhang}, Hua-Wei and {Wang}, Chun and {Tian}, Zhi-Jia and {Liu}, Gao-Chao and {Wu}, Hong},
        title = "{Mapping the emission line strengths and kinematics of supernova remnant S147 with extensive LAMOST spectroscopic observations}",
      journal = {Research in Astronomy and Astrophysics},
         year = 2018,
        month = aug,
       volume = {18},
       number = {9},
          eid = {111},
        pages = {111},
          doi = {10.1088/1674-4527/18/9/111},
archivePrefix = {arXiv},
       eprint = {1804.10989},
 primaryClass = {astro-ph.GA},
       adsurl = {https://ui.adsabs.harvard.edu/abs/2018RAA....18..111R}
}

@ARTICLE{Reimers1984,
       author = {{Reimers}, D. and {Wendker}, H.~J.},
        title = "{Possible optical detection of a filament from the Monogem ring.}",
      journal = {\aap},
         year = 1984,
        month = feb,
       volume = {131},
        pages = {375-377},
       adsurl = {https://ui.adsabs.harvard.edu/abs/1984A&A...131..375R}
}

@ARTICLE{Russell1990,
       author = {{Russell}, Stephen C. and {Dopita}, Michael A.},
        title = "{Abundances of the Heavy Elements in the Magellanic Clouds. II. H II Regions and Supernova Remnants}",
      journal = {\apjs},
         year = 1990,
        month = sep,
       volume = {74},
        pages = {93},
          doi = {10.1086/191494},
       adsurl = {https://ui.adsabs.harvard.edu/abs/1990ApJS...74...93R}
}

@ARTICLE{Sabin2013,
       author = {{Sabin}, L. and {Parker}, Q.~A. and {Contreras}, M.~E. and {Olgu{\'\i}n}, L. and {Frew}, D.~J. and {Stupar}, M. and {V{\'a}zquez}, R. and {Wright}, N.~J. and {Corradi}, R.~L.~M. and {Morris}, R.~A.~H.},
        title = "{New Galactic supernova remnants discovered with IPHAS}",
      journal = {\mnras},
         year = 2013,
        month = may,
       volume = {431},
       number = {1},
        pages = {279-291},
          doi = {10.1093/mnras/stt160},
archivePrefix = {arXiv},
       eprint = {1301.6416},
 primaryClass = {astro-ph.SR},
       adsurl = {https://ui.adsabs.harvard.edu/abs/2013MNRAS.431..279S}
}

@ARTICLE{Sanduleak1975,
       author = {{Sanduleak}, N.},
        title = "{A new planetary nebula in Puppis.}",
      journal = {\pasp},
         year = 1975,
        month = oct,
       volume = {87},
        pages = {705},
          doi = {10.1086/129831},
       adsurl = {https://ui.adsabs.harvard.edu/abs/1975PASP...87..705S}
}

@ARTICLE{Sankrit2023,
       author = {{Sankrit}, Ravi and {Blair}, William P. and {Raymond}, John C.},
        title = "{Third Epoch HST Imaging of a Nonradiative Shock in the Cygnus Loop Supernova Remnant}",
      journal = {\apj},
         year = 2023,
        month = may,
       volume = {948},
       number = {2},
          eid = {97},
        pages = {97},
          doi = {10.3847/1538-4357/acc860},
       adsurl = {https://ui.adsabs.harvard.edu/abs/2023ApJ...948...97S}
}

@ARTICLE{Sasaki2020,
       author = {{Sasaki}, Manami},
        title = "{Supernova remnants in nearby galaxies}",
      journal = {Astronomische Nachrichten},
         year = 2020,
        month = feb,
       volume = {341},
       number = {2},
        pages = {156-162},
          doi = {10.1002/asna.202023772},
       adsurl = {https://ui.adsabs.harvard.edu/abs/2020AN....341..156S}
}

@INPROCEEDINGS{Safi-Harb2019,
       author = {{Safi-Harb}, Samar and {Ramsay}, Michael and {Ferrand}, G. and {West}, Jennifer},
        title = "{A New Version of SNRcat: the High Energy Catalogue of Supernova Remnants}",
    booktitle = {Supernova Remnants: An Odyssey in Space after Stellar Death II},
         year = 2019,
        month = jun,
          eid = {61},
        pages = {61},
       adsurl = {https://ui.adsabs.harvard.edu/abs/2019sros.confE..61S}
}

@ARTICLE{Schlafly2014,
       author = {{Schlafly}, E.~F. and {Green}, G. and {Finkbeiner}, D.~P. and {Juri{\'c}}, M. and {Rix}, H.-W. and {Martin}, N.~F. and {Burgett}, W.~S. and {Chambers}, K.~C. and {Draper}, P.~W. and {Hodapp}, K.~W. and {Kaiser}, N. and {Kudritzki}, R.-P. and {Magnier}, E.~A. and {Metcalfe}, N. and {Morgan}, J.~S. and {Price}, P.~A. and {Stubbs}, C.~W. and {Tonry}, J.~L. and {Wainscoat}, R.~J. and {Waters}, C.},
        title = "{A Map of Dust Reddening to 4.5 kpc from Pan-STARRS1}",
      journal = {\apj},
         year = 2014,
        month = jul,
       volume = {789},
       number = {1},
          eid = {15},
        pages = {15},
          doi = {10.1088/0004-637X/789/1/15},
archivePrefix = {arXiv},
       eprint = {1405.2922},
 primaryClass = {astro-ph.GA},
       adsurl = {https://ui.adsabs.harvard.edu/abs/2014ApJ...789...15S}
}

@ARTICLE{Sezer2012,
       author = {{Sezer}, A. and {G{\"o}k}, F. and {Aktekin}, E.},
        title = "{The first optical light from the supernova remnant G182.4+4.3 located in the Galactic anticentre region}",
      journal = {\mnras},
         year = 2012,
        month = dec,
       volume = {427},
       number = {2},
        pages = {1168-1174},
          doi = {10.1111/j.1365-2966.2012.22015.x},
archivePrefix = {arXiv},
       eprint = {1208.5990},
 primaryClass = {astro-ph.HE},
       adsurl = {https://ui.adsabs.harvard.edu/abs/2012MNRAS.427.1168S}
}

@ARTICLE{Shull1979,
       author = {{Shull}, J.~M. and {McKee}, C.~F.},
        title = "{Theoretical models of interstellar shocks. I.}",
      journal = {\apj},
         year = 1979,
        month = jan,
       volume = {227},
        pages = {131-149},
          doi = {10.1086/156712},
       adsurl = {https://ui.adsabs.harvard.edu/abs/1979ApJ...227..131S}
}

@ARTICLE{Smith1993,
       author = {{Smith}, R.~C. and {Kirshner}, Robert P. and {Blair}, William P. and {Long}, Knox S. and {Winkler}, P.~F.},
        title = "{Optical Emission-Line Properties of M33 Supernova Remnants}",
      journal = {\apj},
         year = 1993,
        month = apr,
       volume = {407},
        pages = {564},
          doi = {10.1086/172538},
       adsurl = {https://ui.adsabs.harvard.edu/abs/1993ApJ...407..564S}
}

@ARTICLE{Stupar2007,
       author = {{Stupar}, M. and {Parker}, Q.~A. and {Filipovi{\'c}}, M.~D. and
         {Frew}, D.~J. and {Boji{\v{c}}i{\'c}}, I. and {Aschenbach}, B.},
        title = "{Multiwavelength study of a new Galactic SNR G332.5-5.6}",
      journal = {\mnras},
         year = 2007,
        month = oct,
       volume = {381},
       number = {1},
        pages = {377-388},
          doi = {10.1111/j.1365-2966.2007.12296.x},
archivePrefix = {arXiv},
       eprint = {0708.0615},
 primaryClass = {astro-ph},
       adsurl = {https://ui.adsabs.harvard.edu/abs/2007MNRAS.381..377S}
}

@ARTICLE{Stupar2008,
       author = {{Stupar}, M. and {Parker}, Q.~A. and {Filipovi{\'c}}, M.~D.},
        title = "{Newly confirmed and candidate Galactic SNRs uncovered from the AAO/UKST H{\ensuremath{\alpha}} survey}",
      journal = {\mnras},
         year = 2008,
        month = nov,
       volume = {390},
       number = {3},
        pages = {1037-1054},
          doi = {10.1111/j.1365-2966.2008.13761.x},
archivePrefix = {arXiv},
       eprint = {0807.5004},
 primaryClass = {astro-ph},
       adsurl = {https://ui.adsabs.harvard.edu/abs/2008MNRAS.390.1037S}
}

@ARTICLE{Stupar2011,
       author = {{Stupar}, M. and {Parker}, Q.~A.},
        title = "{Catalogue of known Galactic SNRs uncovered in H{\ensuremath{\alpha}} light}",
      journal = {\mnras},
         year = 2011,
        month = jul,
       volume = {414},
       number = {3},
        pages = {2282-2296},
          doi = {10.1111/j.1365-2966.2011.18547.x},
archivePrefix = {arXiv},
       eprint = {1102.4453},
 primaryClass = {astro-ph.GA},
       adsurl = {https://ui.adsabs.harvard.edu/abs/2011MNRAS.414.2282S}
}

@ARTICLE{Stupar2012,
       author = {{Stupar}, M. and {Parker}, Q.~A.},
        title = "{Optical detection and spectroscopic confirmation of supernova remnant G213.0-0.6 (now redesignated as G213.3-0.4)}",
      journal = {\mnras},
         year = 2012,
        month = jan,
       volume = {419},
       number = {2},
        pages = {1413-1420},
          doi = {10.1111/j.1365-2966.2011.19797.x},
archivePrefix = {arXiv},
       eprint = {1110.1827},
 primaryClass = {astro-ph.GA},
       adsurl = {https://ui.adsabs.harvard.edu/abs/2012MNRAS.419.1413S}
}

@ARTICLE{Stupar2010,
       author = {{Stupar}, M. and {Parker}, Q.~A. and {Filipovi{\'c}}, M.~D.},
        title = "{The optical emission nebulae in the vicinity of WR 48 ({\ensuremath{\Theta}} Mus): true Wolf-Rayet ejecta or unconnected supernova remnant?}",
      journal = {\mnras},
         year = 2010,
        month = jan,
       volume = {401},
       number = {3},
        pages = {1760-1769},
          doi = {10.1111/j.1365-2966.2009.15814.x},
archivePrefix = {arXiv},
       eprint = {0910.1546},
 primaryClass = {astro-ph.GA},
       adsurl = {https://ui.adsabs.harvard.edu/abs/2010MNRAS.401.1760S}
}

@ARTICLE{Stupar2018,
       author = {{Stupar}, M. and {Parker}, Q.~A. and {Frew}, D.~J.},
        title = "{Confirmation of G6.31+0.54 as a part of a Galactic supernova remnant}",
      journal = {\mnras},
         year = 2018,
        month = oct,
       volume = {479},
       number = {4},
        pages = {4432-4439},
          doi = {10.1093/mnras/sty1684},
archivePrefix = {arXiv},
       eprint = {1806.09745},
 primaryClass = {astro-ph.GA},
       adsurl = {https://ui.adsabs.harvard.edu/abs/2018MNRAS.479.4432S}
}

@ARTICLE{vdb1973,
       author = {{van den Bergh}, Sidney and {Marscher}, Alan P. and {Terzian}, Yervant},
        title = "{An Optical Atlas of Galactic Supernova Remnants}",
      journal = {\apjs},
         year = 1973,
        month = aug,
       volume = {26},
        pages = {19},
          doi = {10.1086/190278},
       adsurl = {https://ui.adsabs.harvard.edu/abs/1973ApJS...26...19V}
}

@ARTICLE{vanDokkum01,
       author = {{van Dokkum}, Pieter G.},
        title = "{Cosmic-Ray Rejection by Laplacian Edge Detection}",
      journal = {\pasp},
         year = 2001,
        month = Nov,
       volume = {113},
        pages = {1420-1427},
          doi = {10.1086/323894},
archivePrefix = {arXiv},
       eprint = {astro-ph/0108003},
 primaryClass = {astro-ph},
       adsurl = {https://ui.adsabs.harvard.edu/#abs/2001PASP..113.1420V}
}

@ARTICLE{Vuceti2023,
       author = {{Vu{\v{c}}eti{\'c}}, M. and {Milanovi{\'c}}, N. and {Uro{\v{s}}evi{\'c}}, D. and {Raymond}, J. and {Oni{\'c}}, D. and {Milo{\v{s}}evi{\'c}}, S. and {Petrov}, N.},
        title = "{Proper Motion of Cygnus Loop Shock Filaments}",
      journal = {Serbian Astronomical Journal},
         year = 2023,
        month = dec,
       volume = {207},
        pages = {9-19},
          doi = {10.2298/SAJ2307009V},
archivePrefix = {arXiv},
       eprint = {2403.05215},
 primaryClass = {astro-ph.HE},
       adsurl = {https://ui.adsabs.harvard.edu/abs/2023SerAJ.207....9V}
}

@ARTICLE{Villa2025,
       author = {{Villa-Durango}, M{\'o}nica A. and {Barrera-Ballesteros}, Jorge and {Rom{\'a}n-Z{\'u}{\~n}iga}, Carlos G. and {Moran}, Emma R. and {Ybarra}, Jason E. and {M{\'e}ndez-Delgado}, J. Eduardo and {Drory}, Niv and {Kreckel}, Kathryn and {Ibarra-Medel}, Hector and {S{\'a}nchez}, S.~F. and {Johnston}, Evelyn J. and {Roman-Lopes}, A. and {Hernandez}, Jes{\'u}s and {Fern{\'a}ndez-Trincado}, Jos{\'e} G. and {Stutz}, Amelia M. and {Henney}, William J. and {Ghosh}, A. and {Sarbadhicary}, Sumit K. and {Lugo-Aranda}, A.~Z. and {Bizyaev}, Dmitry and {Jones}, Amy M. and {Blanc}, Guillermo A.},
        title = "{SDSS-V Local Volume Mapper (LVM): revealing the structure of the Rosette Nebula}",
      journal = {\mnras},
         year = 2025,
        month = oct,
       volume = {543},
       number = {2},
        pages = {1196-1213},
          doi = {10.1093/mnras/staf1530},
archivePrefix = {arXiv},
       eprint = {2509.10665},
 primaryClass = {astro-ph.GA},
       adsurl = {https://ui.adsabs.harvard.edu/abs/2025MNRAS.543.1196V}
}

@ARTICLE{Walker2001,
       author = {{Walker}, A.~J. and {Zealey}, W.~J. and {Parker}, Q.~A.},
        title = "{Filamentary Shell Structures from the AAO/UKST H{\ensuremath{\alpha}} Survey}",
      journal = {\pasa},
         year = 2001,
        month = jan,
       volume = {18},
       number = {3},
        pages = {259-266},
          doi = {10.1071/AS01063},
       adsurl = {https://ui.adsabs.harvard.edu/abs/2001PASA...18..259W}
}

@ARTICLE{Welty2002,
       author = {{Welty}, Daniel E. and {Jenkins}, Edward B. and {Raymond}, John C. and {Mallouris}, Christoforos and {York}, Donald G.},
        title = "{Intermediate- and High-Velocity Ionized Gas toward {\ensuremath{\zeta}} Orionis}",
      journal = {\apj},
         year = 2002,
        month = nov,
       volume = {579},
       number = {1},
        pages = {304-326},
          doi = {10.1086/342755},
archivePrefix = {arXiv},
       eprint = {astro-ph/0208374},
 primaryClass = {astro-ph},
       adsurl = {https://ui.adsabs.harvard.edu/abs/2002ApJ...579..304W}
}

@ARTICLE{Weinberger2006,
       author = {{Weinberger}, R. and {Temporin}, S. and {Stecklum}, B.},
        title = "{Detection of an optical filament in the Monogem Ring}",
      journal = {\aap},
         year = 2006,
        month = mar,
       volume = {448},
       number = {3},
        pages = {1095-1100},
          doi = {10.1051/0004-6361:20054183},
archivePrefix = {arXiv},
       eprint = {astro-ph/0511546},
 primaryClass = {astro-ph},
       adsurl = {https://ui.adsabs.harvard.edu/abs/2006A&A...448.1095W}
}

@ARTICLE{Xiao201,
       author = {{Xiao}, L. and {Zhu}, M.},
        title = "{Radio perspectives on the Monoceros SNR G205.5+0.5}",
      journal = {\aap},
         year = 2012,
        month = sep,
       volume = {545},
          eid = {A86},
        pages = {A86},
          doi = {10.1051/0004-6361/201218938},
archivePrefix = {arXiv},
       eprint = {1207.4873},
 primaryClass = {astro-ph.GA},
       adsurl = {https://ui.adsabs.harvard.edu/abs/2012A&A...545A..86X}
}

@ARTICLE{ZW1997,
       author = {{Zanin}, C. and {Weinberger}, R.},
        title = "{The ``Criss-Cross Nebula'': an interaction of the Orion-Eridanus Bubble with a small interstellar cloud.}",
      journal = {\aap},
         year = 1997,
        month = aug,
       volume = {324},
        pages = {1165-1169},
       adsurl = {https://ui.adsabs.harvard.edu/abs/1997A&A...324.1165Z}
}

@ARTICLE{Zhao2018,
       author = {{Zhao}, He and {Jiang}, Biwei and {Gao}, Shuang and {Li}, Jun and {Sun}, Mingxu},
        title = "{The Distance to and the Near-infrared Extinction of the Monoceros Supernova Remnant}",
      journal = {\apj},
         year = 2018,
        month = mar,
       volume = {855},
       number = {1},
          eid = {12},
        pages = {12},
          doi = {10.3847/1538-4357/aaacd0},
archivePrefix = {arXiv},
       eprint = {1802.01069},
 primaryClass = {astro-ph.GA},
       adsurl = {https://ui.adsabs.harvard.edu/abs/2018ApJ...855...12Z}
}

@ARTICLE{Ziegenbalg2025,
       author = {{Ziegenbalg}, Stefan},
        title = "{Northern Sky Narrowband Survey DR0.2: First [O III] and [S II] Results}",
      journal = {Research Notes of the American Astronomical Society},
         year = 2025,
        month = aug,
       volume = {9},
       number = {8},
          eid = {227},
        pages = {227},
          doi = {10.3847/2515-5172/adfec7},
       adsurl = {https://ui.adsabs.harvard.edu/abs/2025RNAAS...9..227Z}
}
\bibliographystyle{aasjournal}

\end{document}